\documentclass[11pt,eprint,nofootinbib]{revtex4-2}
\usepackage[margin=1 in]{geometry}
\usepackage[utf8]{inputenc}
\usepackage{relsize}
\usepackage{
physics, bbold, mathrsfs, amssymb, simpler-wick, slashed, float}
\usepackage{cancel}
\usepackage{graphicx}
\usepackage{bm}
\usepackage[colorlinks=true,unicode,citecolor=magenta]{hyperref}
\usepackage{amsmath}
\usepackage{amsthm}
\usepackage{amssymb}
\usepackage{eucal}
\usepackage{wasysym}
\usepackage{tipx}
\usepackage{scalerel}
\usepackage{mathalpha}
\usepackage{bbold}
\usepackage{pifont}
\usepackage{slashed}
\usepackage[export]{adjustbox}
\usepackage{textcomp}
\usepackage{gensymb}
\usepackage{BOONDOX-uprscr}
\usepackage{multirow}
\usepackage[normalem]{ulem}
\usepackage{comment}
\usepackage{tikz}
\usetikzlibrary{decorations.pathmorphing} 
\usetikzlibrary{matrix} 
\usetikzlibrary{arrows} 
\usetikzlibrary{calc} 

\tikzstyle{block} = [draw,rectangle,thick,minimum height=2em,minimum width=2em]
\tikzstyle{sum} = [draw,circle,inner sep=0mm,minimum size=2mm]
\tikzstyle{connector} = [->,thick]
\tikzstyle{line} = [thick]
\tikzstyle{branch} = [circle,inner sep=0pt,minimum     size=1mm,fill=black,draw=black]
\tikzstyle{guide} = []
\tikzstyle{snakeline} = [connector, decorate, decoration={pre length=0.2cm,
                         post length=0.2cm, snake, amplitude=.4mm,
                         segment length=2mm},thick, blue, ->]
\tikzset{graviton/.style={decorate, decoration={snake, amplitude=.3mm, segment length=1.3mm, pre length=0mm, post length=0mm}, double}}
\usepackage[compat=1.1.0]{tikz-feynman}
\usepackage[compat=1.1.0]{tikz-feynhand}
\usepackage{color,lipsum}
\usepackage{mathtools}
\numberwithin{equation}{section}
\newcommand{\intDx}[0]{\int_{\scriptscriptstyle{x(-T/2)=x_a}}^{\scriptscriptstyle{x(T/2)=x_b}}\mathcal{D}x}
\newcommand{\intDxi}[0]{\int_{\scriptscriptstyle{\xi(0)=\eta_a}}^{\scriptscriptstyle{\bar \xi(T)=\bar \eta_b}}\mathcal{D}\bar \xi \mathcal{D}\xi}
\newcommand{\Fint}[1]{\int \mathcal{D}#1 \, }
\newcommand{\Fiint}[2]{\int \mathcal{D}#1 \mathcal{D}#2 \, }
\newcommand{\dfourk}[1]{\frac{d^4 #1}{(2\pi)^4}}
\newcommand{\dthreek}[1]{\frac{d^3 #1}{(2\pi)^3}}
\newcommand{\dtwok}[1]{\frac{d^2 #1}{(2\pi)^2}}
\newcommand{\FTk}[2]{\int \dfourk{#1}e^{i#1 \cdot #2}}
\newcommand{\vol}[0]{\text{Vol}}
\newcommand{\suchthat}[0]{\, \Big|\, }
\newcommand{\fourdot}[2]{#1 \cdot #2}
\newcommand{\intdx}[0]{\int_0^1 dx \,}
\newcommand{\intoinf}[1]{\int_0^\infty d#1 \,}
\newcommand{\intinf}[1]{\int_{-\infty}^\infty d#1 \,}
\newcommand{\intinfT}[0]{\int_0^\infty d T\, }
\newcommand{\nn}[0]{\nonumber \\}
\newcommand{\rr}[1]{\textcolor{red}{\it{\textbf{[rr: #1]}}} }
\newcommand{\dva}[1]{\textcolor{blue}{\it{\textbf{[dv: #1]}}} }
\newcommand{\yd}[1]{\textcolor{teal}{\it{\textbf{[yd: #1]}}} }
\newcommand{\sa}[1]{\textcolor{magenta}{\it{\textbf{[sa: #1]}}} }
\newcommand{\sid}[1]{\textcolor{magenta}{\textbf{#1}}}
\newcommand\be{\begin{equation}}
\newcommand\ba{\begin{eqnarray}}
\newcommand\ee{\end{equation}}
\newcommand\ea{\end{eqnarray}}
\newcommand\bw{\begin{widetext}}
\newcommand\ew{\end{widetext}}
\usepackage{hyperref}
\usepackage{tabularx}
\usepackage{epstopdf}

\begin{document}

\title{Worldline Formalism,  Eikonal Expansion and the Classical Limit of Scattering Amplitudes}
\author{Siddarth Ajith${}^*$, Yuchen Du${}^{\dag, *}$, Ravisankar Rajagopal${}^*$, Diana Vaman${}^*$}
\affiliation{
${}^*$ Department of Physics, University of Virginia, Charlottesville, Virginia 22904, USA}
\affiliation{
${}^\dag$ School of Physics and Shing-Tung Yau Center, Southeast University, Nanjing 210018, China}

\date{\today}
\begin{abstract}

We revisit the fundamentals of two different methods for calculating classical observables: the eikonal method, which is a scattering amplitude-based method, and the worldline quantum field theory (WQFT) method. The latter has been considered an extension of the worldline effective field theory. We show that the eikonal and WQFT methods are equivalent and that calculations can be translated freely between them. We also show how the scattering amplitude can serve as the generating function of all the WQFT diagrams. This offers another computational avenue for the  WQFT diagrams, which can be useful at higher loop orders.

Concretely,  we focus on  2-into-2 scattering processes mediated by massless force carriers. On one hand, taking the classical limit of the QFT scattering amplitude leads to the eikonal method. On the other hand, since in the classical limit the scattering particles are almost on-shell throughout the scattering process, the worldline, a first quantized formalism, is the most efficient framework to study the scattering amplitude. This is an alternate but equivalent formalism to the quantum field theoretical (QFT) framework. By taking the classical limit of the scattering amplitude computed in the worldline, we can derive the WQFT rules of Mogull, Plefka and Steinhoff \cite{Mogull:2020sak}. 
In WQFT, the Feynman diagrams are reorganized into a new set of diagrams that facilitate the $\hbar$ expansion.  Unlike the QFT eikonal method, which identifies the eikonal phase recursively, the worldline-based computation allows to target and systematically extract the classical contributions directly through a specific set of WQFT diagrams. In worldline formalism the perturbative expansion of the scattering amplitude is naturally organized
 in diagrams which factorize (reducible) and diagrams
which are new to that order (irreducible), in a one-to-one map with the structure of the amplitude in the eikonal method. This opened up the possibility to investigate and prove the conjectured exponentiation of the eikonal phase in \cite{ADRV1}.

\end{abstract}

\maketitle

  \hypersetup{linkcolor=blue}
   \tableofcontents

\section{Introduction}\label{sect:introduction}

The detection of gravitational waves (GWs) by the LIGO/Virgo/KAGRA collaboration has heralded the age of gravitational wave astronomy. These detectors allow  probing black holes and/or neutron stars in binaries, and future detectors like the space-based LISA will offer even more sources like white dwarf stars or extreme mass ratio inspirals. However, to fully take advantage of this new window into the universe, we must improve our understanding of gravitational theory; GW detection depends heavily on theoretical modeling, and as GW detectors improve their sensitivity, one must also extend the accuracy of the theoretical calculations.

Motivated by this,  the last few years have brought general relativity (GR) and high energy communities ever closer.  Scattering amplitudes, which are the bread and butter of high energy physicists, through their classical limit,   hold useful information for the GR community.   In particular, an integral part of GW physics is understanding gravitation in the weak field limit of binary sources.   When approached with conventional GR methods, such as solving the Einstein field equations order-by-order, this problem becomes increasingly intractable, and finding efficient alternatives is desirable. This new way of doing the calculation has been achieved through quantum field theory techniques.

In this work, we consider the field theoretical 2-body scattering amplitudes. The scattering amplitudes can be used to extract classical observables, such as the deflection angle of the scattering particles (or massive bodies in GR), or extract the effective one-body gravitational potential of a binary.  It turns out that the classical limit of certain loop diagrams 
can be mapped to the post-Minkowskian GR expansion, providing a completely different method to advance the post-Minkowskian formalism.  By phrasing this problem as a field theory amplitudes calculation, we can import many innovative techniques that save considerable labor.
In particular, the classical observables can be derived in QFT through the eikonal expansion~\cite{Damgaard:2023vnx,DiVecchia:2020ymx,DiVecchia:2021ndb,DiVecchia:2021bdo,Bjerrum-Bohr:2021din,Bjerrum-Bohr:2021vuf,DiVecchia:2019myk,DiVecchia:2019kta,DiVecchia:2022owy,DiVecchia:2022piu,Bellazzini:2022wzv,Cristofoli:2020uzm,KoemansCollado:2018hss,KoemansCollado:2019ggb,Parra-Martinez:2020dzs,DiVecchia:2019kta,DiVecchia:2019myk}
or KMOC formalism \cite{Kosower:2018adc,Damgaard:2023vnx,Maybee:2019jus,Herrmann:2021lqe,Herrmann:2021tct,Cristofoli:2021vyo,Georgoudis:2023lgf}.
Another approach is to construct effective field theories (EFT) \cite{Goldberger:2004jt,Porto:2005ac, Rothstein:2014sra, Porto:2016pyg, Goldberger:2022ebt,porto-main,porto-eft-radiation-reaction,porto-bootstraping} which give the sought-after classical information, as long as one can recover GR in some limit of a fundamental scale of the EFT. A particularly relevant EFT is the worldline quantum field theory (WQFT) \cite{Mogull:2020sak, Jakobsen:2021lvp, Jakobsen:2021smu, Jakobsen:2021zvh, Jakobsen:2022fcj, Jakobsen:2023oow},
 which 
 has also been used to perform post-Minkowskian calculations in a remarkably efficient way \cite{porto-conservative-dynamics-third,porto-conservative-dynamics-fourth,dlapa2024localintimeconservativebinarydynamics}.

 In this paper, we show how the 2-body scattering amplitude can be computed either through Feynman diagrams in a QFT (second quantized) framework or in the worldline (first quantized) framework. 
Taking the classical limit ($\hbar\to0$) of the QFT amplitude casts the calculation as an eikonal expansion, where the scattering particles' momenta are of the order $\hbar^0$, while the momenta of all the virtual mediators are soft, of the order $\hbar^1$.  A consequence of this limit is that the scattering particles are almost on-shell throughout the scattering process. This suggests that a first quantized approach for the scattering particles can be more efficient. As we will briefly advertise, that is indeed the case.

Of course, if instead of the QFT starting point, we work with the equivalent worldline formalism, we do end up with the same expression for the scattering amplitude. We find the operator approach (as opposed to the path integral one) quite useful,  particularly when it comes to implementing the LSZ reduction. In the operator approach this is done by creating the on-shell asymptotic scattering states through the action of vertex operators.
With the QFT and worldline computations agreeing, we can take the classical limit in both cases and arrive at the eikonal expansion. On the worldline formalism side, we show that the classical limit leads to the WQFT diagrammatic expansion. The WQFT rules of Mogull, Plefka and Steinhoff \cite{Mogull:2020sak} are one way to perform the calculation, though we choose not to go to energy space. The contractions between the vertex operators creating the asymptotic states with the  vertex operators corresponding to the mediators have the net effect of setting up an expansion around the classical trajectory parametrized in terms of the in-out momenta $p$ and $p'$ as $x(\tau)=(p+p') \tau/2 + \delta x(\tau)=v\tau+\delta x(\tau)$. 
After Fourier transforming the scattering amplitude to impact parameter space $b$,
the worldline effective field theory (WEFT) perspective emerges naturally from the amplitude
with each worldline parametrized as $x=b+v\tau +\delta x(\tau)$. We also provide a way to systematically generate and calculate the WQFT diagrams, with the scattering amplitude in the worldline formalism being promoted to a generating function. With this alternative way, we show with a side-by-side analysis how the QFT eikonal expansion of each Feynman diagram maps into a particular set of WQFT diagrams. 
Concretely, we review  the classical (eikonal) limit  of 2-body scattering in scalar Yukawa theory, first in its non-relativistic version, then in the relativistic theory. We then perform the same calculation in the worldline framework. The WQFT method efficiency is clearly visible. 
We also discuss the same problem in scalar QED and gravity.
We consider not only the contribution coming from the conservative sector, but we account for the radiative corrections as well. 

\begin{figure}[h!]
\includegraphics[width=\textwidth]{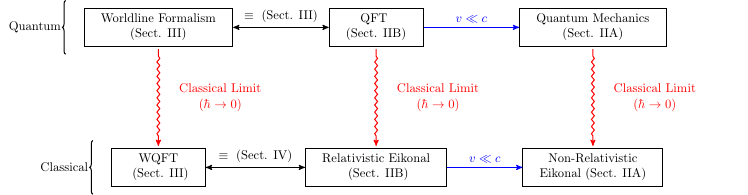}
\caption{A flowchart showing the relationships elaborated on throughout this work.}
\label{fig:eikonalW-QFT}
\end{figure}

{The classical terms arise at $\order{\hbar^{-1}}$ in the scattering amplitude expressed in impact parameter space.} The more singular terms are called superclassical, and the less singular terms are called quantum. A given Feynman loop diagram could receive all types of contributions. We perform a detailed $\hbar$ counting analysis, allowing us to map the various terms in the eikonal expansion into specific WQFT diagrams. A theory with a ``good" classical limit is such that all superclassical terms can be factorized into products of lower order (in perturbation theory) classical and quantum contributions. Scalar QED and scalar gravity are such theories. This factorizable nature of the superclassical terms is behind the exponentiation of the leading order eikonal. The exponentiation conjecture of the scattering amplitude in impact parameter space suggests a similar factorization at play. On the QFT side, the velocity cuts are the counterpart of the factorization \cite{Bjerrum-Bohr:2021din}{, but they are not immediately visible in a QFT construction}. WQFT diagrams are manifestly displaying this factorization structure, and once more the operator approach is well suited to show the factorization in mathematical terms.  In a separate work \cite{ADRV1} we employed this observation to prove the exponentiation of the eikonal phase to all loop orders. Here, we lay the foundation by giving an in-depth look at the relationship between QFT, worldline, and their classical limits, the eikonal and  WQFT, as shown in Figure 
\ref{fig:eikonalW-QFT}.

 The rest of the paper is organized as follows. In Section \ref{sect:eikonal-qm-qft}, we review the classical limit of scattering amplitudes in both quantum mechanics and quantum field theory using scalar Yukawa theory as a simple example, and review the connection between the classical limit and the eikonal expansion. In Section~\ref{sect:wordline}, we review the worldline formalism 
 for scalar particles in a 
background field (in turn scalar, electromagnetic and gravitational). We discuss how to compute 2-body scattering amplitudes, by integrating out the background fields and effectively gluing two scalar worldlines with mediator $n$-point functions. We show how the classical limit yields the WQFT diagrammatic expansion.
  In Section~\ref{sect:wqft-eikonal}, we  review the  QFT eikonal method and show the equivalence between the classical limit of QFT and WQFT computations. {We also discuss an alternative way to generate and calculate WQFT diagrams based on the worldline formalism, which maintains the relation to the scattering amplitude.} In Section~\ref{sect:wqft-sQED}, we go up to the sub-sub-leading eikonal order in scalar QED, and finally, in Section~\ref{sect:wqft-sGravity} we make some comments on the relevant WQFT diagrams for scalars interacting  gravitationally.  In Appendix~\ref{app:sQED} we give the conventional QFT classical limit treatment to the 2-body scattering amplitudes in  scalar QED for comparison purposes and to show the advantages of WQFT. 
 
 Regarding conventions, we use the mostly plus signature for the metric. Barred quantities are wave numbers or dimensionless coupling constants whose  $\hbar$ dependence can be made explicit. The notation for the scattering amplitudes in momentum space is $\mathcal A ^{(j)}_{i}$ where $i$ gives the order of perturbation (essentially PM order) and $j$ counts the orders of the eikonal expansion (related to $\hbar$); note that our convention is such that $j=0$ is the classical limit, $j>0$ is quantum (and thus not considered here), and $j<0$ gives super${}^{-j}$classical terms. Throughout the work, we use the TikZ-FeynHand package for Feynman diagrams~\cite{Ellis:2016jkw,Dohse:2018vqo}.


\section{The Eikonal Expansion in QM and QFT}\label{sect:eikonal-qm-qft}

In this section, we review the eikonal expansion, the effective one-body potential, and the leading exponentiation of the scattering amplitude.
We consider two-into-two scattering processes, which can be phrased as one particle scattering off an effective potential. This potential will be a function of the relative distance between the particles or, equivalently, the transferred momentum in Fourier space. Additionally, there can be dependence on kinematic parameters such as the velocity and spin of the scattering particles, as well as the nature of their interactions. 
In both quantum mechanics (QM) and its relativistic version, quantum field theory (QFT),  the scattering is expressed through the $S$-matrix
\begin{equation}
    \hat S = \hat U(\infty,-\infty)=1+ i\hat { T},
\end{equation}
where $\hat U$ is the time-evolution operator and ${ \hat T}$ is the {transfer matrix}. The QM transfer matrix $\hat T$ can be formally written as an infinite series in terms of the effective potential $V$
\begin{equation}
    \hat T = \hat V + \hat V \hat G_0\hat V + \hat V\hat G_0\hat V\hat G_0\hat V + \cdots,
\end{equation}
where $\hat G_0$ the Green's function of the free particle.  In QFT,  the 2-into-2 scattering is mediated by virtual particle exchanges, and the scattering amplitude is evaluated using Feynman rules.  The one-body effective potential $V$ can be extracted from the QFT scattering amplitude in the eikonal limit (i.e., large center of mass energy, small scattering angle). We take the limit by performing a power expansion in $\hbar$, so in this sense, the eikonal limit yields the classical limit of the scattering amplitudes. For the sake of completeness, we review side-by-side the steps taken in  QM and QFT calculations, order by order in perturbation theory.  

\subsection{Non-relativistic eikonal}\label{subsect:eikonal-qm}

We begin with the non-relativistic (QM) problem.  The probability amplitude for a particle with momentum $\vb p$ in the infinite past to be deflected to momentum $\vb {p'}$ in the infinite future,  due to scattering off a potential $V(\vb r)$, is derived from the $S$-matrix element:
\begin{equation}
\langle \vb{p'}|\hat S|\vb{p}\rangle=    \langle \vb {p'} | {\cal T}\exp(-\frac i{\hbar} 
    \int_{-\infty}^\infty dt \; V_I(\hat {\vb r},t))| \vb p \rangle\,,
\end{equation}
where $ V_I$ is the interaction-picture potential, and we marked the operators with a hat symbol. 
In perturbation theory, the amplitude is given by:
\begin{equation}
i\mathcal A= \langle \vb {p'} | \left[  \frac {-i}{\hbar} 
    \int_{-\infty}^\infty dt V_I(\hat {\vb r},t)) +\frac 1{2!} \frac{(-i)^2}{\hbar^2} \,\mathcal T \left(\int_{-\infty}^\infty dt_1 \int_{-\infty}^\infty dt_2
     V_I(\hat{\vb r},t_1)   V_I(\hat{\vb r},t_2) +\dots \right)\right]|\vb p \rangle{.}\label{qmamp}
\end{equation}
For comparison with the QFT amplitude, and for the purpose of taking the classical limit through the eikonal expansion, we will pay attention to the units (dimensions) of the various expressions.
Working in units where the speed of light is $c=1$, the dimension of the amplitude in \eqref{qmamp} is $M^{-3}$, where $M$ is some mass scale, since the units are the same as for $\langle \vb {p'}| \hat 1 |\vb p \rangle=\delta^3{(\vb{p'}-\vb p})$.
\bigskip \par To first order in perturbation theory, the amplitude is given by
\begin{equation}
 i\mathcal A_{0}  =  -\frac i{\hbar} 
    \int_{-\infty}^\infty dt\, \langle \vb {p'} |V_I(\hat {\vb r},t)| \vb p \rangle=- \frac{i}{\hbar}
   \frac{ 2\pi\hbar  \delta({E_{p'}-E_p}) }{(2\pi\hbar)^3}\int d^3 \vb r \,V(\vb r)
   \exp(i (\vb {\bar p}-\vb{\bar{p}}')\cdot \vb r)\,,
\end{equation} 
where $E_{p'}$ and $E_p$ are the energies of the free particle states $|\vb {p'}\rangle $ and respectively $|\vb p\rangle$, e.g.
\begin{equation}
    E_p=\frac{\vb p^2}{2m}\,,
\end{equation}
and where we introduced the barred notation
\begin{equation}
    \vb {\bar p}\equiv \frac{\vb p}{\hbar}\,.
\end{equation}
In other words, the barred momenta are wave vectors and have dimensions of inverse length.
Finally, denoting the momentum transferred by $\vb q$,
\begin{equation}
    \vb q\equiv \vb{p'}-\vb p\,,
\end{equation}
to first order in perturbation theory, one recovers the Born approximation
\begin{equation}
  \mathcal A_{0}=   
   -\frac{1}{(2\pi \hbar)^3}\,V(\vb {\bar q})\,2\pi   \delta({E_{p'}-E_p})\,,\label{born 1}
\end{equation}
where $V({\vb { \bar q}})$ is the Fourier-transform of the scattering potential.  
As an example, consider a central Coulomb-type potential 
\begin{equation}
    V(r)=\frac {{\cal Q}^2}{4\pi r}\,,
\end{equation}
where ${\cal Q} $ is the  charge (with ${\cal Q}^2/\hbar$ dimensionless, which can be inferred from requiring that $(1/\hbar) \int dt V(r)$ be dimensionless). Its Fourier-transform is
\begin{equation}
    V(\vb {\bar q})=\int d^3 \vb r\ V(r)e^{i\bar{\vb q}\cdot \vb r}=\frac{{\cal Q}^2}{\bar {\vb q}^2}\,.
\end{equation}
Putting everything together, the scattering amplitude to first order in perturbation theory  is 
\begin{equation}
\mathcal A_{0}=
    -\frac{ 1}{(2\pi \hbar)^3} \frac{{\cal Q}^2}{\bar {\vb q}^2}2\pi\delta({E_{p'}-E_p})\,.\label{born 2}
\end{equation}
 \par Next, we  move on to  second order in perturbation theory:
\begin{equation}
i\mathcal A_{1}=  \frac 1{2!} \left(\frac{-i}  {\hbar}\right)^2\langle \vb {p'} | {\cal T}\left(
  \int_{-\infty}^\infty dt_1 \int_{-\infty}^\infty dt_2 V_I(\hat {\vb r},t_1) V_I(\hat {\vb r},t_2)\right) | \vb p \rangle\,.
  \end{equation}
Inserting complete sets of states yields
\begin{eqnarray}
i\mathcal A_{1}&=&\left(\frac{-i}  {\hbar}\right)^2 \frac{1}{(2\pi \hbar)^6}\int_{-\infty}^\infty  d t_1 \int_{-\infty}^{t_1} d t_2 \exp\left[\frac i\hbar (E_{p'}t_1-E_p t_2)\right]
\int d^3\vb{r_1} \int d^3 \vb{r_2 }\int d^3 \vb k\nonumber \\ 
&& \times\exp\left[-\frac i\hbar (E_{k}t_1-E_k t_2)\right]\exp\left[i(\vb{\bar p}-\vb{\bar k})\cdot \vb{r_1}\right]
\exp\left[i(\vb{\bar k}-\vb{\bar p'})\cdot \vb{r_2}\right]
V(r_1)V(r_2)\,.
\end{eqnarray}
We trade off the $t_2$ integral for an integral over $\Delta t=t_1-t_2$. The integral over $t_1$ yields a delta-function expressing the energy conservation $\delta(E_{p'}-E_p)$. The integral over $\Delta t$ is regularized by adding a convergence factor $\exp(-\epsilon\Delta t)$, and performing the $\Delta t$ integral yields the (non-relativistic) free particle propagator
\begin{equation}
    G_0(k)=\frac{-i{\hbar}}{E_p-E_k-i\epsilon}\,.
\end{equation}
Lastly, each of the potential factors $V(r_1)$, $V(r_2)$ can be written in terms of their Fourier-transforms $V(\bar{\vb {k}}_1)$, $V(\bar{\vb {k}}_2)$, and the $\vb {r_1}$, $\vb r_2$ integrals can be performed. 
This leads to the following expression
\begin{equation}
i\mathcal  A_{1}={\cal Q}^4 \left(\frac{-i}{\hbar}\right)^2 \frac{1}{\hbar^3} \,   
2\pi \hbar\delta(E_{p'}-E_p)
\int \frac{d^3\bar{\vb {k}}_1}{(2\pi)^3}\int \frac{d^3\bar{\vb {k}}_2}{(2\pi)^3}
\delta^3(\bar{\vb{k}}_1+\vb{\bar{k}_2}-\bar{\vb q})\frac{1}{\bar{\vb{k}}_1{}^2
}\frac{1}{\bar{\vb{k}}_2{}^2}\frac{-i{\hbar}}{E_p-E_{p+k_1}-i\epsilon}\label{nonrel2nd}\,.
\end{equation}
In the { small scattering angle limit}, this expression can be further manipulated as follows.
Under the integral sign,   the non-relativistic particle propagator can be symmetrized under a $\vb{k}_1, \vb{k}_2$ interchange
\begin{eqnarray}
&&\frac 12\left(\frac{-i{\hbar}}{E_p-E_{p+k_1}-i\epsilon}+\frac{-i{\hbar}}{E_p-E_{p+k_2}-i\epsilon}\right)=\frac 12\left(\frac{-2im\hbar}{\vb p^2-(\vb{p}+\vb{k_1})^2-i\epsilon}+\frac{-2im\hbar}{\vb p^2-(\vb{p}+\vb{k_2})^2-i\epsilon}\right)
\nonumber\\
&&=im\left(\frac{1}{2\vb p\cdot\bar{\vb {k}}_1+\hbar \bar{\vb{k}}_1^2+i\epsilon}
+\frac{1}{2\vb p\cdot\vb{\bar {k}_2} +\hbar \bar {\vb k}_2^2+i\epsilon}
\right)\,.
\end{eqnarray}

Next we substitute $\vb p\cdot\bar {\vb{k}}_2=\vb p\cdot(\vb{\bar q}-\vb{\bar {k}}_1)$ and use that
\begin{equation}
\vb p\cdot \vb{\bar q}=\frac12\left[(\vb p+\vb {p'})\cdot \vb{\bar q}-(\vb {p'}-\vb p)\cdot \vb{\bar q}\right]=-\frac 12 \hbar \vb{\bar q}^2\, ,
\end{equation}
where in the last step we used that $\vb p^2=\vb {p'}^2$.  The small scattering angle assumption is being made by treating $\bar{\vb  q}$ finite, which in turn implies that $\vb q=\hbar \bar {\vb q}$ is small,  of the order $\hbar$.
Expanding the denominator of the non-relativistic propagator in powers of $\hbar $
leads to 
\begin{equation}
im \left[ \frac{1}{2\vb p\cdot\bar{\vb {k}}_1+i\epsilon}\left(1-\frac{\hbar {\bar {\vb{k}}_1^2}}{2\vb p\cdot\bar{\vb{k}}_1+i\epsilon}\right)+ \frac{1}{-2\vb p\cdot\bar{\vb {k}}_1+i\epsilon}\left(1-\frac{\hbar ({\bar{\vb  {k}}_2^2}-\vb{\bar q}^2)}{-2\vb p\cdot\bar{\vb {k}}_1+i\epsilon}\right)\right]+{\cal O}(\hbar^2) \label{nonrelpropexp}\,.
\end{equation}
This is the eikonal approximation:   each scattering on the potential $V$ imparts a small 
momentum kick. 
The leading order terms  in $\hbar $  yield a delta-function
\begin{equation}
    im\left( \frac{1}{2\vb p\cdot\bar{\vb {k}}_1+i\epsilon}-\frac{1}{2\vb p\cdot\bar{\vb {k}}_1-i\epsilon}
    \right)=2\pi  m\,\delta(2
    \vb p\cdot\bar{\vb {k}}_1)\,.
\end{equation}
Substituting this into Eq.~\eqref{nonrel2nd} leads to the so-called leading eikonal contribution to the scattering amplitude 
\begin{eqnarray}
i\mathcal  A_{1}^{(-1)}&=&    {\cal Q}^4\left(\frac{-i}{\hbar}\right)^2 \frac{1}{\hbar^3} \,   
2\pi \hbar\,\delta(E_{p'}-E_p)
\int \frac{d^3\bar{\vb {k}}_1}{(2\pi)^3}\int \frac{d^3\bar{\vb {k}}_2}{(2\pi)^3}
\delta^3(\bar{\vb{k}}_1+\bar{\vb{k}}_2-\bar{\vb q})\frac{1}{\bar{\vb{k}}_1{}^2
}\frac{1}{\bar{\vb{k}}_2{}^2} \, 2\pi  m\, \delta(2
    \vb p\cdot\vb{\bar {\vb{k}}_1})\nonumber\\
    &=&-\frac{1}{\hbar^2}\frac{\mathcal Q^4}{\hbar^2}
    \int \frac{d^3\bar{\vb {k}}_1}{(2\pi)^3}\int \frac{d^3\bar{\vb {k}}_2}{(2\pi)^3}
\delta^3(\bar{\vb{k}}_1+\bar{\vb{k}}_2-\bar{\vb q})\frac{1}{\bar{\vb{k}}_1{}^2
}\frac{1}{\bar{\vb{k}}_2{}^2} 
   2\pi  m\, \delta(2
    \vb p\cdot\vb{\bar {\vb{k}}_1}) 2\pi \delta(E_{p'}-E_p)\,.\label{qm2nd}
\end{eqnarray}
In Eq.~\eqref{qm2nd}, we introduced the following notation: the superscript $\mathcal A^{(-1)}$ denotes that this term is one order higher in $\hbar^{-1}$ relative to the classical terms, which we henceforth denote with a superscript $(0)$. The subscript  $\mathcal A_1$ denotes the order in perturbation theory. For example $\mathcal A_n^{(m)}$ is proportional to $\mathcal Q^{2n+2} \hbar^{-3+m}$. The terms with negative $m$ labels are called super${}^{-m}$classical. The terms  $\mathcal A_n^{(m)}$ with $m$ positive are higher order in $\hbar$ relative to the classical ones and yield quantum corrections to the scattering amplitude.  
 
 In a coordinate frame such that the momentum of the incoming particle is   
\begin{equation}
    \vb p=(0,0,p_z)\,,
\end{equation} we have 
\begin{equation}
    \vb q=(q_x,q_y,q_z)=\vb q^\perp+\frac{\vb p\cdot \vb q}{|\vb p|}= (\hbar \bar q_x, \hbar \bar q_y, \mathcal O(\hbar^2))\,,
\end{equation}
 and the 3-dimensional integrals are reduced to 2-dimensional integrals (transverse to $\vb p$, due to the delta-function $\delta(2
    \vb p\cdot\vb{\bar {k}}_1)$). The amplitude given in Eq.~\eqref{qm2nd}  can be written as
    \begin{eqnarray}   
i\mathcal A_{1}^{(-1)}&=&
-\frac{|\vb p|}{(2\pi)^3 \hbar^2 m}\int d^2\vb b^\perp \ \frac 12 e^{i \vb q\cdot \vb b^\perp} \,\delta^2(\vb b^\perp)
    \ 2\pi \delta(E_{p'}-E_p)  \,,
    \end{eqnarray}
    where $\vb b^\perp$ is the impact parameter (which is transverse to the incoming momentum $\vb p$) and where 
    \begin{equation}
        \delta(\vb b^\perp) = \frac{m}{\hbar|\vb p|}\int_{-\infty}^\infty d\xi \ V\qty(\vb b^\perp+ \xi \frac{\vb p}{|\vb p|})
    \end{equation}
is the eikonal \cite{Levy:1969cr}. 
This, we shall see, matches the leading order superclassical contribution of the corresponding QFT scattering amplitude to second order in perturbation theory, in the eikonal limit. 
The leading order in perturbation theory amplitude given in ~\eqref{born 1} can also be written in terms of the eikonal as
\begin{equation}
    i\mathcal A_0=-i\frac{|\vb p|}{(2\pi)^3 \hbar^2 m}\int d^2\vb b^\perp e^{i \vb q\cdot \vb b^\perp}\delta(\vb b^\perp)\  2\pi \delta(E_{p'}-E_p) \,. 
\end{equation}
The resummation of these leading orders in $1/\hbar $ terms in each order in perturbation theory yields  the leading-order eikonal phase
\cite{Levy:1969cr}
\begin{equation}
    i\mathcal A= \frac{|\vb p|}{(2\pi)^3 \hbar^2 m}\int d^2\vb b^\perp  e^{i \vb q\cdot \vb b^\perp} \bigg(e^{-i\delta(\vb b^\perp)}-1\bigg)\ 2\pi \delta(E_{p'}-E_p)  \,.
\end{equation}

\subsection{Relativistic eikonal for scalar interactions}\label{sect:eikonal-qft}
For simplicity, we will consider a scalar Yukawa QFT, and a 2-body scattering problem with the interaction between two massive, charged scalar $\phi$ particles mediated by a massless scalar $\varphi$ particle. The interaction potential is then $V= \bar g \phi \phi^* \varphi$. We continue to work in units where the speed of light is $c=1$, but we do not set $\hbar$ to 1, as it is customary in QFT calculations.   Our  Minkowski metric convention is mostly plus.
The  QFT action is normalized such that the action is dimensionless:
\begin{equation}
{\cal S}=\int d^4 x\bigg[  -\partial_\mu \phi  \partial^\mu\phi^* -\bar m^2 \phi \phi^*-\frac 12\partial_\mu\varphi\partial^\mu\varphi-\bar g\phi\phi^* \varphi\bigg]\,,
\end{equation}
where we introduced the barred mass notation
\begin{equation}
\bar m^2\equiv \frac{m^2}{\hbar^2}\,.
\end{equation}
The fields $\phi ,\varphi$ and the coupling constant $\bar g$ have dimension of inverse length $[\phi]=[\varphi]=[\bar g]=[L^{-1}].$  We also recall that $\hbar$ has units of length times mass: $[\hbar]=L M$ and so $[m^2/\hbar^2]=L^{-2}$, as needed.
The Feynman propagators (with dimensions of length squared) are
\begin{equation}
G_\phi(\bar k)=\frac{-i}{{\bar k}^2+\bar m^2-i\epsilon}, \qquad G_\varphi=\frac{-i}{\bar k^2-i\epsilon}\,.
\end{equation}
The cubic vertices are $-i \bar g$.  We define a scalar charge ${\cal Q}$ (such that ${\cal Q}^2/\hbar$ is dimensionless) using  
\begin{equation}
\bar g^2\equiv \frac{{\cal Q}^2}{\hbar} \bar m^2= \frac{{\cal Q}^2 m^2}{\hbar^3}\,.
\end{equation}
We compute the  $S$ matrix element corresponding to a scattering process of two $\phi$-particles into two $\phi$-particles
\begin{equation}
    \langle p_1, p_2|\hat S| p'_1, p'_2\rangle =  \langle p_1, p_2|1+ i\hat T | p'_1, p'_2\rangle\label{s}\,.
\end{equation}
Its units are  $M^{-4}$, which can be inferred by thinking about the dimensions of the matrix element given in Eq.~\eqref{s}  with the interactions turned off.
To get the scattering amplitude, we should sum up the Feynman diagrams
\be 
\includegraphics[width=\textwidth]{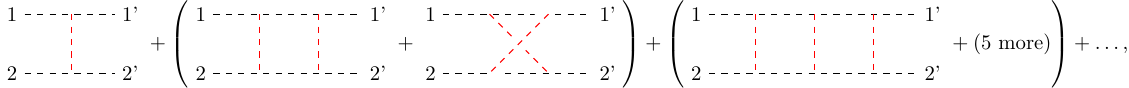}
\ee
where the red lines indicate the massless scalar field $\varphi$ which  mediates the interaction between the $\phi$ particles.

 \par We work in the same {\it small scattering angle limit} as in the corresponding QM scattering problem. For the QFT scattering amplitude,  this means small $t$, with the Mandelstam variable defined as usual: $t=-(p_1-p'_1)^2$.
We also take the limit of large center of mass energy $s=-(p_1+p_2)^2$, since we will be interested in the classical limit of the scattering amplitude.  Put together, this means we work in the eikonal limit.
Furthermore, we  assume that each mediator exchange imparts a very soft kick to the scattering particles. 
The red color used for the $\varphi$ propagator in the Feynman diagrams above is meant as a visual reminder that the virtual $\varphi$ particles are soft.

To first order in perturbation theory, Eq.~\eqref{s}  yields the scattering amplitude \footnote{We might expect contributions from all $s,t,u$ channels. Instead we included the $t$-channel contribution only. The other channels are suppressed for the scattering process (large $s$, small $t$) we are studying, with the 
mediators imparting only small kicks to the scattering particles. 
Alternatively, we could have devised a scenario where particles of 1 and 2 are quanta of different fields, and each of those fields couples with the mediator $\varphi$ as in \cite{Levy:1969cr}.  In that case,  only the $t$ channel exchange is non-zero.}
\begin{equation}\label{amplmom}
i\mathcal A_0=(-i\bar g)^2 \frac{-i}{\bar q^2} (2\pi)^4\delta^4(p_1+p_2-p'_1-p'_2).
\end{equation}
Note that the units are $M^{-4}$,  as expected. For simplicity of notation, we will no longer explicitly write the energy-momentum delta function and  will keep it as an implicit factor in the scattering amplitude.

One subtlety in matching the scattering amplitudes is that we also need to consider the corresponding QM 2-into-2 scattering process. The motion separates into the uniform motion of the center of mass and the motion of a particle of reduced mass moving in a potential $V(r)$ where $\vb r=\vb r_1-\vb r_2$ is the relative distance between the scattering particles. Then, the QM scattering amplitude will include a delta-function enforcing the CM (total) momentum conservation: $\delta^3 (\vb p_1+\vb p_2 -\vb p'_1-\vb p'_2)$. This is matched by the corresponding total momentum conservation of the QFT scattering amplitude: $
\delta^3(\vb p_1+\vb p_2-\vb p'_1-\vb p'_2)$. Note that since we defined the QFT asymptotic states to be eigenstates of momentum $|\vb p\rangle$ (and not wavenumber eigenstates $|\bar{\vb p}\rangle$), the QFT amplitude contains the factor $\delta^4(p_1+p_2-{p'}_1-{p'}_2)$ (and not a delta-function of the barred momenta). After factoring out the momentum conservation delta functions, the resulting amplitude has mass dimension units $M^{-1}$. Lastly, we recall that the relativistic free particle momentum eigenstates have a different normalization than the non-relativistic (QM) eigenstates
\begin{eqnarray}
    \langle \vb {p'} | \vb p\rangle_{QFT} = \sqrt{2(\vb p^2 +m^2)} (2\pi)^3\delta^3(\vb {p'}-\vb p),\\
    \langle \vb {p'} | \vb p\rangle_{QM} =  \delta^3(\vb {p'}-\vb p)\,.
\end{eqnarray}
Accounting for this $M^2$ normalization factor (each particle normalization yields one $M$ factor), and recalling that we factored out the total momentum conservation delta-function, the dimensions of the QFT scattering amplitude now match mass units ($M$) of the QM amplitude in Eq.~\eqref{born 2}.  

We now go to the center of mass frame,  to extract the effective one-body potential.\footnote{Alternatively, we could start with two massive fields, with $m_1\ll m_2$. Then the $m_2$ particle could be approximated as a heavy source,  essentially static, and we can extract the effective one-body potential in the lab frame.} The relativistic $S$-matrix element still contains a delta function enforcing the conservation of the total relativistic energy
$
2 \sqrt{m^2+\vb p^2}= 2\sqrt{m^2+\vb {p'}{}^2}.$ In the non-relativistic limit this becomes $2\pi \delta(2 E_p-2 E_{p'})$. 
Putting everything together,  the non-relativistic limit of the QFT scalar Yukawa amplitude is
\begin{equation}
 i\mathcal A_{0} =i   \frac{{\cal Q}^2 m^2}{(2\pi \hbar)^3}\frac{1}{\bar {\vb q}^2} 2\pi \delta(2(E_{p'}-E_p)) \frac{1}{ (2m)^2}\,.\label{Yqftnonrel}
\end{equation}
where the factor of $1/(2 m)^2$ is the non-relativistic limit of the normalization factors $1/(\sqrt{2(m^2+\vb p^2)})^2$ $ 1/(\sqrt{2(m^2+\vb {p'}{}^2)})^2$. In the last step, we replaced the relativistic  $1/q^2$ propagator with $1/{\vb q}^2$ by enforcing the energy conservation condition, which implies $q^0=0$.  In order to extract the one-body effective potential from the 2-body interaction, we need to account for the relative motion of a particle of reduced mass $m_1 m_2/(m_1+m_2)=m/2$. The relative momentum transfer in the CM frame is $2(\vb {p'}-\vb p)=2\vb q$,
so we need to match Eq.~\eqref{Yqftnonrel} with the appropriately modified Eqs.~\eqref{born 1} and \eqref{born 2}: $-i \ V(2 \bar{\vb q}) \ 2\pi \delta(2(E_{p'}-E_p)) / (2\pi \hbar)^3$.
We recover the {\it massless} scalar Yukawa theory potential (massless since the mediator $\varphi$  is massless),  which is an attractive potential: 
\begin{equation}
V(\bar{\vb q})=-\frac{ {\cal Q}^2}{ \bar {\vb  q}^2}.
\end{equation}

To second order in perturbation theory,  the QFT scattering amplitude in our toy-model scalar Yukawa theory is given by the  two boson exchange diagrams 
\be
\includegraphics
{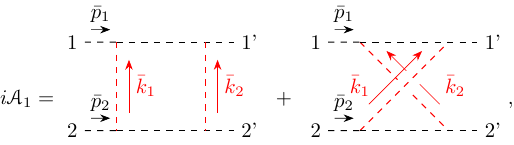}
\ee
where each loop barred momentum is integrated over.  We introduce the short-hand notation
\begin{equation}
    \int_{\bar k}\equiv \int \frac{d^4 \bar k }{(2 \pi)^4} 
    \,.
\end{equation}
Under the integral sign, we can exchange $\bar k_1$ and $\bar k_2$:
\be 
\includegraphics[width=\textwidth]{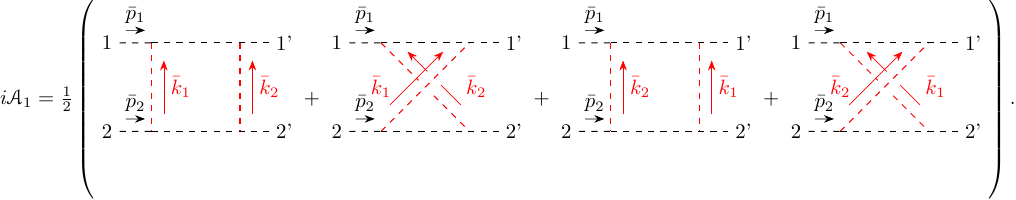}
\ee 
These diagrams sum up to
\begin{multline}
    i\mathcal{A}_{1} = \frac{(i\bar g)^4}{2 i^4}\int_{\bar k_1, \bar k_2} (2\pi)^4 {\delta}^4(\bar k_1 + \bar k_2 - \bar q) \frac{1}{\bar k_1^2-i\epsilon}\,\frac{1}{\bar k_2^2-i\epsilon} \\ \times\qty(\frac{1}{(\bar p_1 + \bar k_1)^2 + \bar m_1^2-i\epsilon} + \frac{1}{(\bar p_1 + \bar k_2)^2 + \bar m_1^2-i\epsilon})\qty(\frac{1}{(\bar p_2 - \bar k_1)^2 + \bar m_2^2-i\epsilon} + \frac{1}{(\bar p_2 - \bar k_2)^2 + \bar m_2^2-i\epsilon}),
\end{multline}
where we labeled the two scattering particles' masses by $m_1$ and $m_2$. We do this so we can consider the case when these particles are distinguishable, though, for the purpose of this section, we can set the masses equal $m_1=m_2=m$. 

We recall that we work in the eikonal limit, and allow the massive scattering particles to be only slightly off-shell throughout. 
Then, each mediator exchange imparts a very soft kick to the scattering particles. 
We implement this condition by keeping track of factors of  $\hbar$: the loop momenta are all of the order $\hbar$ since $k=\hbar\bar k$, but the external momenta $p$ are finite and of order $\hbar^0$. In this way, the eikonal limit is the classical limit.

The massive particle propagators can be expanded in $\hbar$ to yield\footnote{In the more recent literature (see for example \cite{Akhoury:2013yua}), the eikonal expansion is done in the ``eikonal frame" with the external momenta for the incoming/outgoing particle of type 1 given by $$p_1=\left(E_1, -\frac{\vb q}2, p_{1z}\right),\qquad p_1'=\left(E_1, \frac{\vb q}2,p_{1z}\right)$$ and similarly for the other scattering particles. This will lead to a reshuffling of order $\hbar$ terms. Here, as in \cite{Levy:1969cr}, we use instead $$p_1=(E_1,0,0,p_{1z}).$$
One advantage of the eikonal frame is that the transferred momentum $\vb q$ is ``transverse": $\vb q=(q_x,q_y,0)$, though in the same eikonal frame, neither $\vb p$ nor $\vb p'$ are exactly longitudinal. 
}
\begin{equation}
    \frac{1}{(\bar p_1 + \bar k_1)^2 + \bar m_1^2 - i\epsilon} = \frac{1}{2 \bar p_1 \cdot \bar k_1 + \bar k_1^2 - i\epsilon} = \frac{\hbar}{2 p_1 \cdot \bar k_1 + \hbar \bar k_1^2 - i\epsilon}=
    \frac{\hbar}{2 p_1 \cdot \bar k_1 -i\epsilon}\sum_{n=0}^\infty\left(\frac{-\hbar \bar k_1^2}{2 p_1 \cdot \bar k_1 -i\epsilon}\right)^n\,.
\end{equation}
We notice that in the eikonal/classical limit, we keep only one of the two poles of the relativistic propagator. Essentially, we are focusing only on the  particle pole 
$-p_1^0+\sqrt{(p_1^0)^2-\vb k_1^2 -2 \vb k_1\cdot \vb p_1} $ and discard the anti-particle one $-p_1^0-\sqrt{(p_1^0)^2-\vb k_1^2 -2 \vb k_1\cdot \vb p_1} $. This is mathematically consistent with the physical picture of the massive particles being only slightly off-shell during the scattering process. For this reason, we will see that the worldline formalism is especially well-suited to capture the eikonal/classical limit: the worldline formalism is the first quantized version of the quantum field theory, and each worldline describes a massive scattering particle interacting with the soft mediators.

The on-shell condition for the outgoing states
\be 
    0 = p'_{a}{}^2 + m_{a}^2 = (p_{a} + \eta_a \ \hbar \bar q)^2 + m_a^2 = 2\eta_a \ \hbar p_a \cdot \bar q + \hbar^2 \bar q^2\,,
\ee 
\begin{equation}
     2 p_a \cdot \bar q = - \eta_a \ \hbar \bar q^2\,,\label{qpishbar2}
\end{equation}
\be
    2 p_a \cdot \bar k_2 = 2 p_a \cdot (\bar q - \bar k_1) = -2 p_a \cdot \bar k_1 +\eta_a \ \hbar \bar q^2\,,
\ee
where $\eta_1=1$ and $\eta_2=-1$. Note that  $p_{1,2} \cdot \bar q$ are actually of order $\hbar$.

Then, the two-particle exchange scattering amplitude can be written as
\begin{multline}
    i\mathcal{A}_{1} = \frac{(i\bar g)^4 \hbar^2}{2 i^4 }\int_{\bar k_1, \bar k_2} (2\pi)^4 {\delta}^4(\bar k_1 + \bar k_2 - \bar q) \frac{1}{\bar k_1^2-i\epsilon}\frac{1}{\bar k_2^2-i\epsilon} \\
     \times \qty(\frac{1}{2 p_1 \cdot \bar k_1 + \hbar \bar k_1^2 - i\epsilon} + \frac{1}{-2 p_1 \cdot \bar k_1 + \hbar (\bar k_2^2 - \bar q^2) - i\epsilon})
     \\\times
     \qty(\frac{1}{-2p_2 \cdot \bar k_1 + \hbar \bar k_1^2 - i\epsilon} + \frac{1}{2p_2 \cdot \bar k_1 + \hbar (\bar k_2^2 - \bar q^2) -i\epsilon}).\label{m200}
\end{multline}
At this point, we can proceed with the $\hbar$ expansion of the massive propagators
\begin{multline}
    \qty[\frac{1}{2p_1 \cdot \bar k_1 - i\epsilon}\sum_{n=0}^\infty \qty(\frac{-\hbar \bar k_1^2}{2p_1 \cdot \bar k_1})^n - \frac{1}{2p_1 \cdot \bar k_1 + i\epsilon}\sum_{n=0}^\infty
    \qty(\frac{\hbar ( \bar k_2^2 - \bar q^2)}{2p_1 \cdot \bar k_1 })^n] \\ 
    \times\qty[-\frac{1}{2p_2 \cdot \bar k_1 + i\epsilon}\sum_{n=0}^\infty \qty(\frac{\hbar \bar k_1^2}{2p_2 \cdot \bar k_1})^n + \frac{1}{2p_2 \cdot \bar k_1 - i\epsilon}\sum_{n=0}^\infty \qty(\frac{-\hbar (\bar k_2^2 - \bar q^2)}{2p_2 \cdot \bar k_1})^n]\,.\label{secondsubleading2}
\end{multline}
The  leading order terms in the $\hbar$  expansion are:
\begin{equation}
    \qty[\qty(\frac{1}{2p_1 \cdot \bar k_1 - i\epsilon} - \frac{1}{2p_1 \cdot \bar k_1 + i\epsilon}) \qty(\frac{1}{2p_2 \cdot \bar k_1 - i\epsilon} - \frac{1}{2p_2 \cdot \bar k_1 + i\epsilon})] = (2 \pi i) \delta(2 p_2 \cdot\bar k_1) (2 \pi i) \delta(2 p_1 \cdot \bar k_1)\,.\label{m2001}
\end{equation}
After factoring out the delta functions enforcing the total momentum conservation (i.e. the center of mass motion), and including the difference between the normalization factors of QFT and QM eigenstates of momentum, the leading eikonal amplitude
\begin{equation}
    i \mathcal  A_1^{(-1)}=\frac{\mathcal Q^4 m^4}{2\hbar^4}\frac{2\pi\delta(2({p'}{}^0-p^0))}{(2\pi)^3(\sqrt{2p^0})^2\ (\sqrt{2p'{}^0})^2}\int_{\bar k_1, \bar k_2} (2\pi)^4{\delta}^4(\bar k_1 + \bar k_2 - \bar q) \frac{1}{\bar k_1^2}\frac{1}{\bar k_2^2}
(2 \pi i) \delta(2 p_2 \cdot\bar k_1) (2 \pi i) \delta(2 p_1 \cdot \bar k_1)\,.\label{qft2nd0}
\end{equation}
Note that $\mathcal A_1^{(-1)}$ is of order $1/\hbar^4$. As discussed earlier, such contributions that are higher order in $1/\hbar$ than the classical one (which is of order $1/\hbar^3$) are called superclassical. To match with Eq.~\eqref{qm2nd} we take the non-relativistic limit $p^0\approx m +E_p$, etc in Eq.~\eqref{qft2nd0}.
Using one of the delta functions, we can integrate over the temporal component of the loop barred-momentum ${\bar k}_1^0$ and substitute ${\bar k}_1^0=\vb{p}_a\cdot\bar{\vb{k}}_1/p_a^0$, where the label $a=1,2$ indicates the incoming momenta.
In the non-relativistic limit, ${\bar k}_1^0$ becomes negligible.  Since $q^0=k_1^0+k_2^0=0$, that means that ${\bar k}_2^0$ is negligible as well. This allows us to write
\begin{eqnarray}
    i\mathcal  A_1^{(-1)}&=&-\frac{\mathcal Q^4  m^4}{2(2\pi)^3\hbar^4}\frac{2\pi\delta(2(E_{p'}-E_p))}{(2m)^3}\int\frac{d^3\bar{\vb k}_1 }{(2\pi)^3} \int\frac{d^3\bar{\vb k}_2 }{(2\pi)^3}  \delta^3(\bar {\vb k}_1 + \bar {\vb k}_2 - \bar {\vb q}) \frac{1}{\bar{\vb k}_1^2}\frac{1}{\bar {\vb k}_2^2}(2\pi)^4\delta(2\vb p\cdot \bar{\vb k}_1)\nonumber\\
   &=&-\frac{\mathcal Q^4\, 2\pi m}{2\hbar^4}2\pi\delta((E_{p'}-E_p)
   \int\frac{d^3\bar{\vb k}_1 }{(2\pi)^3} \int\frac{d^3\bar{\vb k}_2 }{(2\pi)^3}  \delta^3(\bar {\vb k}_1 + \bar {\vb k}_2 - \bar {\vb q}) \frac{1}{4\bar{\vb k}_1^2}\frac{1}{4\bar {\vb k}_2^2}\delta(2\vb p\cdot \bar{\vb k}_1)
 \,.
\end{eqnarray}
After recalling the earlier comments that we need to match with the QM expression for a Coulomb-type potential $V(2\bar{\vb k})$ due to the choice of CM frame and using the reduced mass $m/2$, we see that this indeed matches the leading QM eikonal in Eq.~\eqref{qm2nd}.

The next order term in Eq.~\eqref{secondsubleading2} is the subleading eikonal. This term is one order in $\hbar $ higher than the leading order, and it will contribute at the classical level. Considering just the internal line propagators, we get
\begin{equation}
    \hbar \qty[\frac{2 \bar k_1 \cdot \bar k_2}{(2p_1 \cdot \bar k_1 + i \epsilon)^2} (2 \pi i) \delta(2 p_2 \cdot \bar k_1) + \frac{2 \bar k_1 \cdot\bar k_2}{(2p_2 \cdot \bar k_1 - i \epsilon)^2} (2 \pi i) \delta(2 p_1 \cdot \bar k_1)]\,.
\end{equation}
Using the delta functions, we can integrate over the temporal component of the loop barred-momentum $\bar k_1^0$ and substitute $\bar {k}_1^0=\vb{p}_a\cdot\bar{\vb{k}}_1/p_a^0$, where the label $a=1,2$ indicates the incoming momenta.
In the non-relativistic limit, $\bar {k}_1^0$ and $\bar {k}_2^0$ become negligible.   Then the non-relativistic limit of the relativistic propagators (after integrating out $\bar k_1^0$) reproduces each term of the non-relativistic scattering amplitude to second order in perturbation theory as derived earlier in Eq.~\eqref{nonrelpropexp}.  
From the QFT amplitude, we can further get the non-relativistic corrections to the one-body effective potential.

Without taking the non-relativistic limit, the resummation and exponentiation of the leading order $\hbar$ terms can be seen as follows. 
We first need to Fourier-transform the amplitude $\mathcal A$ to impact parameter space 
\be
\mathcal A(b)\equiv\int_{\bar q} e^{i\bar q\cdot b}\delta(2\bar q\cdot \bar v_1)\delta (2\bar q\cdot \bar v_2) \mathcal A =\int_{\bar q} e^{i\bar q\cdot b}\delta(2\bar q\cdot \bar v_1)\delta (2\bar q\cdot \bar v_2) \bigg(\mathcal A_0+\mathcal A_1^{(-1)}+\mathcal A_2^{(-2)}+\dots\bigg)\,.
\ee
The delta functions are enforcing the mass-shell conditions constraining $\bar q$:
\be
\bar q \cdot \bar v_a\equiv \bar q\cdot \frac{\bar p_a+\bar p'_a}2=\eta_a (\bar p'_a-\bar p_a)\cdot \frac{\bar p_a+\bar p'_a}2=0\,,
\ee
where there was no summation implied over the index $a=1,2$,  and we recall the convention $\eta_1=1=-\eta_2$. To leading $\hbar$ order in the eikonal expansion we can approximate $\delta(2 q\cdot \bar v_a)\approx \delta(2 q \cdot \bar p_a)$.
Then we can write
\begin{eqnarray}
i\mathcal A^{(0)}(b)&=&
\int_{\bar k_1} \frac{i \bar g^2 e^{i\bar k_1\cdot b}}{\bar k_1^2}\delta(2\bar k_1\cdot \bar p_1)\delta(2\bar k_1\cdot \bar p_2)\nonumber\\&+&\frac 12
\int_{\bar k_1}\frac{i \bar g^2 e^{i\bar k_1\cdot b}}{\bar k_1^2}\delta(2\bar k_1\cdot \bar p_1)\delta(2\bar k_1\cdot \bar p_2)
\int_{ \bar k_2}\frac{i \bar g^2 e^{i\bar k_2\cdot b}}{\bar k_2^2}\delta(2\bar k_2\cdot \bar p_1)\delta(2\bar k_2\cdot \bar p_2)+\dots\nonumber\\
&=&\exp(i\delta^{(0)}(b))-1\,,
\end{eqnarray}
where the leading QFT eikonal phase is
\be\label{lead eik}
\delta^{(0)}(b)= \bar g^2\int_{\bar k} \frac{ e^{i \bar k\cdot b}}{\bar k^2}\delta(2\bar k\cdot \bar p_1)\delta(2\bar k\cdot \bar p_2)=\frac{i \mathcal Q^2 m^2}{\hbar}\int_{\bar k} \frac{e^{i \bar k\cdot b} }{ \bar k^2}\delta(2\bar k\cdot p_1)\delta(2\bar k\cdot  p_2)\,.
\ee
Note that the classical eikonal phase $\delta(b)$ is of the order $1/\hbar$, which can be easily seen from \eqref{lead eik}. 

Similarly, the classical amplitude  in impact parameter space is of the order $1/\hbar$, as we can see from $i\mathcal A^{(0)}(b)$ above. Relative to the momentum space amplitude given in \eqref{amplmom}, which is of the order $1/\hbar^3$ (after substituting $\bar g^2=\mathcal Q^2 m^2/\hbar^3$), there is a difference of a factor of $\hbar^2$ between the momentum space and impact parameter space.

With this, we conclude our review of the relativistic and non-relativistic limits of the eikonal expansion.

\section{Worldline formalism}\label{sect:wordline}


In this section, we review the basics of the worldline formalism, which is a first-quantized version of QFT.  It is somewhat reminiscent of string theory, which is also formulated as a first quantized theory.  The worldline formalism begins with Bern and Kosower \cite{Bern:1990cu, Bern:1990ux} who, in 1990, considered the infinite tension limit of string amplitudes (which restricts the infinite tower of string states to massless states); they gave a master formula and a set of rules which allowed the calculation of one-loop N-point amplitudes efficiently.  Strassler \cite{Strassler:1992zr} derived the Bern and Kosower formula and rules, without using string theory or Feynman diagrams through a rewriting of the field theory loop amplitudes in terms of a 
path integral over point-particle coordinates, i.e. the worldline of a point-particle. See also \cite{Schubert:2001he, Bastianelli:2006rx}.

\subsection{Scalar Interactions}\label{subsect:scalar-wordline}

Consider a relativistic scalar particle which is interacting with other scalar particles. The position-space Feynman propagator can be expressed as a path integral over the  particle's trajectory (or worldline)
\begin{eqnarray}
G_0(x', x)&=&\left\langle x'\left|\frac{1}{\Box-\bar m^2+i\epsilon}\right|x\right\rangle=\int_0^\infty dT \langle x'|\exp(i T(\Box-\bar m^2+i\epsilon))|x\rangle\nonumber\\
&=&\int_0^\infty dT  \int_{-\infty}^\infty \frac{d^4\bar p}{(2\pi)^4} e^{i \bar p\cdot (x'-x)}e^{-iT(\bar p^2+\bar m^2-i\epsilon)}\nonumber\\
&=&\int _0^\infty dT \int^{x^\mu(T)= x'{}^\mu}_{ x^\mu(0)= x^\mu}{\cal D}x(\tau)\int {\cal D}\bar p(\tau)e^{-i \int_0^T d\tau( \bar p^2(\tau)+\bar m^2 - \bar p\cdot \dot x-i\epsilon)}\nonumber\\
&=&\int _0^\infty dT\int^{x^\mu(T)=x'{}^\mu}_{ x^\mu(0)= x^\mu} {\cal D}x(\tau)
\exp(i \int_0^T d\tau \mathcal{L}_0[x,\dot x])\,,\label{freepropx}
\end{eqnarray}
with $\mathcal L_0 =(\dot x^2)/4-\bar m^2$.  By rescaling the worldline (Schwinger) time $\tau\rightarrow \tau/2$ (and correspondingly $T\to T/2$), the free particle Lagrangian becomes canonically normalized 
\begin{equation}\label{freeL}
\mathcal L_0=\frac 12 (\dot x^2-\bar m^2)\,.
\end{equation}
Next, we turn on the interactions by coupling the worldline with background fields.
For scalar Yukawa interactions, these background fields will be scalars.
On the  worldline, we add a  potential term $\Phi(x)$ to the free Lagrangian
\begin{equation}
    \mathcal L =  \mathcal L_0-
\frac{\bar g}2\Phi(x(\tau))\,,
\end{equation}
where $\bar g$ is the coupling constant with the background field.  
This potential will be decomposed in Fourier modes (plane waves)
\be
\Phi(x)=\int_{\bar k} \Phi_{\bar k}e^{i \bar k\cdot x},
\ee
 and each $\exp(i \bar k\cdot x(\tau))$ factor will define an interaction of the worldline with a background quantum with wave vector $\bar k$.  We can think of 
\begin{equation}
 V_i(\tau)=-i \bar g\exp(i \bar k_i\cdot x)
\end{equation}
as a vertex operator for the emission or absorption of a background 
scalar particle with momentum $\hbar\bar k_i$. The dressed Feynman propagator,
\begin{equation}
G(x',x) = \int^{\infty}_0 \frac{dT}2\int^{x(T)=x'}_{x(0)=x} \mathcal{D}x(\tau)
\ e^{i \int^T_0 {d\tau} \mathcal{L}(x(\tau), \dot{x}(\tau))}
\end{equation}can be Taylor expanded and written in terms of 
\begin{equation}
G_N(x',x) =  \int^{\infty}_{0}\frac{dT}2\ \qty(\prod_{i=1}^{N}\int^{T}_{0} \frac{d\tau_{i}}2) \langle x, \tau_{N+1}{=}T|\mathcal{T}\{\hat V_{N}(\tau_{N}) \hat V_{N-1}(\tau_{N-1})...\hat V_{1}(\tau_{1})\}|x', \tau_0{=}0\rangle  \,,
\end{equation}
where $G_N$ describes the interaction with $N$ background particles. 

After Fourier-transforming the dressed propagator
\begin{eqnarray}
G_N(\bar p{}',\bar p) &=& \int d^4 x \int d^4 x' e^{i \bar p\cdot x} 
e^{-i \bar p' \cdot x'} G_N(x',x) 
=\frac{1}{i(\bar p^2 + \bar m^2-i\epsilon)} \mathcal{M}_N \frac{1}{i(\bar p'^2 + \bar m^2-i\epsilon)}\,,\label{gnan}
\end{eqnarray}
 as well as amputating the scalar propagators and setting the external momenta $p$ and $p'$ on shell, we obtain the scattering amplitude off $N$ background quanta.  We will refer to $\mathcal{M}_N$ as a worldline amplitude, despite the possible off-shellness of the background quanta.

This is easily illustrated with $N=1$:
\begin{align}
    G_1(\bar p', \bar p) &= \int_0^\infty  \frac{dT}2\int_{0}^{T} \frac{d\tau}2 \, \bra{\bar p', T}  \hat V(\tau) \ket{\bar p, 0} \nonumber \\
    &= \int_0^\infty \frac{dT}2 \int_{0}^{T} \frac{d\tau}2 \,\bra{\bar p'} e^{-i\hat H_0 T}  e^{i \hat H_0 \tau} \hat V(0) e^{-i\hat H_0 \tau} \ket{\bar p}\nonumber \\
    &= \frac 14\intinfT \int_{0}^{T} d\tau \, e^{-i \frac{\bar p'^2 + \bar m^2-i\epsilon}{2}\qty({T} - \tau)} e^{-i \frac{\bar p^2 + \bar m^2-i\epsilon}{2}\tau } \bra{\bar p{}'}\hat V(0) \ket{\bar p}\nonumber\\
    &= \frac{1}{i(\bar p^2 + \bar m^2 - i\epsilon)} \bra{\bar p'}\hat V(0) \ket{\bar p}\frac{1}{i(\bar p{}'^2 + \bar m^2 - i\epsilon)} \,.
\end{align}

This factoring of the external leg propagators happens in general for any $N$ \cite{Du:2023nzo}. 
Focusing on the scattering amplitude, we place the external legs on-shell and write 
\begin{equation}
    \ket{\bar p} = \ket{\bar p, \tau{=}\pm \infty} = e^{i \bar p \cdot \hat x(\tau{=}\pm \infty)} \ket{0}\,,
\end{equation}
where $\ket{0}$ is the vacuum state. The first equality is possible since the momenta are on-shell and hence $\hat H_0 \ket{\bar p} = 0$. Thus, by amputating the external legs from the dressed propagator and placing them on-shell, we extract the worldline scattering amplitude:
\be\label{An}
	\mathcal{M}_{N} =  \lim_{\substack{\tau_{N+1} \rightarrow +\infty,\\ \tau_0 \rightarrow -\infty}}\qty(\prod_{i=2}^{N}\int^{\infty}_{-\infty} \frac{d\tau_{i}}2e^{-\epsilon|\tau_i|}) \langle \mathcal{T}\{\hat V_{\text{out}}(\tau_{N+1})\hat V_{N}(\tau_{N}) \hat V_{N-2}(\tau_{N-2})...\hat V_{1}(0)\hat V_{\text{in}}(\tau_0)\}\rangle \,,
\ee
where $\hat V(\tau_0)$ and $\hat V(\tau_{N+1})$ are vertex operators creating the in and out states given by
\begin{equation}
\hat V_{\text{in}}(\tau_0)=\exp( i\bar p\cdot \hat x(\tau_0)),  \,\,\,\, p^2+m^2=0, \qquad \hat V_{\text{out}}(\tau_{N+1})=\exp(-i \bar p'\cdot \hat x(\tau_{N+1})),  \,\,\,\,
p'{}^2+m^2=0,\label{inout}
\end{equation}
and where the angular brackets denote the vacuum expectation value. 
Notice that in the scattering amplitude describing an interaction with $N$ background quanta,  there are $N$ background vertex operators, two vertex operators that create the in and out worldline states, and $N-1$ worldline time integrals from minus infinity to plus infinity, with one background interaction arbitrarily fixed (accomplished by choosing $\tau_1=0$ in Eq.~\eqref{An}).

\subsubsection*{The Dressed Propagator and The On-Shell Limit}

Consider the fully dressed propagator in momentum space\footnote{We recall that we rescaled the Schwinger time $\tau\to\tau/2$ to get the canonically normalized Lagrangian in \eqref{freeL}. We rescaled similarly $T\to T/2$, so in terms of the rescaled variables, the boundary values of the path integral remain the same. Another consequence of these rescalings is that coupling constant $\bar g$ was rescaled to $\bar g/2$, in the same way the mass term in the original Lagrangian got rescaled by a factor of $1/2$.}
\be
G(p',p)=\int d^4 x \int d^4{x'} e^{i \bar p\cdot x} e^{-i \bar p'\cdot x'}
\int _0^\infty \frac{dT}2 \int^{ x^\mu(T)= x'{}^\mu}_{ x^\mu(0)= x^\mu} {\cal D}x(\tau)e^{i \int_0^T d\tau \mathcal L_0[x,\dot x]} e^{-i(\bar g/2 )\int_0^T d\tau  \Phi(x(\tau))}\,,
\ee
and rewrite the path integral in terms of fluctuations around the classical trajectory
\be
x^\mu(\tau)=x^\mu+\frac{x'{}^\mu-x^\mu}{T}\tau+\delta x^\mu(\tau)\,.
\ee
The  fluctuations $\delta x^\mu$,  which obey vanishing boundary conditions on an interval,  have the correlation function
\begin{equation}\label{eqn-finite-line-prop}
    \langle \mathcal T \{\delta x^{\mu}(\tau)\delta x^{\nu}(\tau')\}\rangle = -\frac i{T} \eta^{\mu\nu}\left[
\theta(\tau-\tau') \tau (T-\tau')+\theta(\tau'-\tau) \tau' (T-\tau)\right]\,.
\end{equation}
In particular,  we have
\be
\left\langle  e^{i \bar k \cdot \delta x(\tau)}\right\rangle = e^{\frac {i\bar k^2}{2T} (\tau^2-T\tau)}.
\ee
Next,  expand the background field in Fourier modes and perform the integration over $\Delta x=x'-x$. The dressed propagator becomes
\begin{eqnarray}
G(p',p)&=&\int_0^\infty \frac{dT}2 e^{-\frac i2 (\bar p'^2+\bar m^2-i\epsilon) T}\int  d^4x e^{-i (\bar p'-\bar 
p)\cdot x} \int^{ \delta x^\mu(T)= 0}_{ \delta x^\mu(0)= 0}
 {\cal D}\delta x(\tau) \exp(\frac i2\int_0^T d\tau  \delta \dot x^2)
\nonumber\\
&&
\bigg[1+\sum_{N=1}^\infty\frac{(-i\bar g )^N}{2^N N!}\prod_{i=1}^N\int_{\bar {k_i}} \int_0^T d\tau_i \Phi_{\bar k_i}  e^{i\bar k_i \cdot x} e^{i \bar k_i\cdot \delta x(\tau_i)} \exp(-i  (\bar p' \cdot \bar k_i \tau_i+\frac 1{2T} \bar k_i^2 \tau_i^2))\bigg]\nonumber\\
&&=\sum_{N=0}^\infty\frac{(-i\bar g)^N}{2^N N!}\bigg(\prod_{i=1}^N\int_{\bar {k_i}} \Phi_{\bar k_i} \bigg)G_N(p', p)(2\pi)^4\delta^4(\bar p'-\bar p-\sum_{i=1}^N \bar k_i)\,,
\end{eqnarray}
where  $\Phi_{\bar k_i}$ is the Fourier transformed background field.
The lowest-order terms in the perturbative expansion of the dressed  propagator are
\be
G_0 (p',p)=\frac{1}{i(\bar p^2+\bar m^2-i\epsilon)} \,,
\ee
and
\be
G_1(p',p)=(-i\bar g)\frac{1}{i(\bar p'{}^2+\bar m^2-i\epsilon)} \frac{1}{i((\bar p'-\bar q)^2+\bar m^2-i\epsilon)} \,,
\ee
as expected, and  coincide with the results we obtained  earlier using an operator approach.

It is interesting to notice that the Schwinger times $\tau_i$ map into Feynman parameters.  For any Feynman diagram corresponding to the probability amplitude to interact with $N$ background quanta, we can combine the particle propagators with the help of Feynman parameters:
\begin{eqnarray}
&&\frac{1}{(p'^2+m^2)}\frac{1}{(p'-k_1)^2+m^2)} \cdots \frac1{(p'+\sum_{i=1}^{N} k_i)^2+m^2}=\prod_{i=1}^{N+1}\int_0^1 dy_i \delta(1-\sum_{i=1}^{N+1} y_i)\nonumber\\
&&\frac{1}{(p'^2+m^2 + (k_1^2-2 p'\cdot k_1)  \sum_{i=1}^N y_i+(k_2^2-2 k_2\cdot p') \sum_{i=2}^N y_i + \dots (k_N^2-2 k_N\cdot p' ) y_N)^{N+1}}.\nonumber\\
\label{treeprop}
\end{eqnarray}
This structure is mapped into the worldline expression of the dressed propagator for a particular time ordering $0<\tau_1<\tau_2<\dots<\tau_N<T$ by defining $\tau_1=y_1 T,  \tau_2-\tau_1=y_2 T, \dots,  T-\tau_N=y_{N+1} T$. Then the constraint obeyed by the Feynman parameters $y_1+y_2+\dots y_{N+1}=1$ is algebraically satisfied.

The eikonal expansion of the propagator is relevant for a particle which remains almost on-shell as it interacts with the background field.  This means that we will keep $\bar k=k/\hbar$ and $p = \hbar \bar p$ finite as $\hbar \to0$, and we have $p^2+m^2\sim \mathcal O(\hbar)$ and $p'^2+m^2\sim \mathcal O(\hbar)$.
Then,  to leading order in the eikonal expansion, we can neglect the insertions $e^{i\bar k_i \cdot \delta x(\tau_i)}$ in the path integral,  perform the integrals over $\bar k_i$ and arrive at 
\be
\label{lssym}
G^{(0)}(p', p)=\int_0^\infty \frac{dT}2 \int d^4 x e^{-\frac i2 (\bar p'^2+\bar m^2-i\epsilon) T} e^{-i \bar q\cdot x} e^{-i(\bar g/2) \int_0^T d\tau \Phi(x+\bar p' \tau)}
\,,
\ee
where we recall that $\bar q=\bar p'-\bar p$. The advantage of the worldline formalism is immediate. We no longer need to sum up Feynman diagrams and manipulate the products of Feynman propagators through various algebraic identities, as Levy and Sucher \cite{Levy:1969cr} did,  in order  to arrive at the leading eikonal dressed propagator.  Instead, using the worldline formalism, this result is immediately obtained by expanding the interaction potential in fluctuations about the classical trajectory and keeping the leading order term.

For the purpose of taking the on-shell limit (for the eikonalized scalar particle line)
it is useful to consider the difference
with the free propagator. Then $G^{(0)}-G_0$
\be
\sum_{N=1}^\infty\frac{(-i\bar g)^N}{2^N N!}\int d^4x e^{-i \bar q\cdot x}\int_0^\infty \frac{dT}2 \int_0^T d\tau_1 e^{-\frac i2(\bar p'^2+\bar m^2-i\epsilon)T}
\int_{\bar k_1}\Phi_{\bar k_1}
e^{i \bar k_1\cdot x}e^{i\bar k_1\cdot p'\tau_1}
\prod_{i=2}^N
\int_{\bar k_i}
\int_0^T d\tau_i \Phi_{\bar k_i} e^{i \bar k_i\cdot x}e^{i \bar k_i\cdot p'\tau_i}
\ee
can be further 
manipulated by defining new variables $\sigma=\tau_1, \sigma'=T-\tau_1,\sigma_i=\tau_i-\tau_1$
for $i=2,3,\dots$ into
\begin{eqnarray}
&&\sum_{N=1}^\infty\frac{(-i\bar g)^{N-1}}{2^{N-1} N!}\int_0^\infty \frac{d\sigma' }2\int_0^\infty \frac{d\sigma}2 e^{-\frac i2(\bar p'{}^2+\bar m^2 -i\epsilon)\sigma'}e^{-\frac i2(\bar p^2+\bar m^2 -i\epsilon)\sigma}\int d^4 x\,(-i\bar g)\Phi(x )e^{-i\bar q\cdot x}\prod_{i=2}^N
\int_{-\sigma}^{\sigma'}d\tilde\sigma \Phi(x+\bar p'\tilde\sigma)\nonumber\\
&&=\int_0^\infty \frac{d\sigma}2 e^{-\frac i2(\bar p'{}^2+\bar m^2 -i\epsilon)\sigma}\int_0^\infty \frac{d\sigma'}2 e^{-\frac i2(\bar p^2+\bar m^2 -i\epsilon)\sigma}\int d^4 x\,(-\bar g) \Phi(x )\,e^{-i\bar q\cdot x} \,\frac{e^{i\delta_0(\sigma,\sigma')}-1}{\delta_0(\sigma,\sigma')}
\label{dressed0}\,,\end{eqnarray}
where in the exponents we dropped the quadratic terms $\bar k_i\cdot\bar k_j$ which are subleading. In the last step 
of \eqref{dressed0} we resumed the background fields into the exponent
\be
\delta_0(\sigma,\sigma')=-\frac{\bar g}2\int
_{-\sigma}^{\sigma'}d \tilde\sigma \Phi(x+\bar p'\tilde\sigma)\,.\ee
The on-shell limit is taken by amputating the external scalar legs in the usual fashion: multiplying \eqref{dressed0}
by $\bar p^2+\bar m^2$ and $\bar p'{}^2+\bar m^2$ and taking the limit $\bar p^2+\bar m^2 \to 0$ and $\bar p'{}^2+\bar m^2\to 0$.
This yields the familiar result \cite{Levy:1969cr} for the leading eikonal  of the scattering amplitude in a background field
\be
\int d^4 x\,(-\bar g)\Phi(x )e^{-i\bar q\cdot x} \frac{e^{i\delta_0}-1}{\delta_0}
\ee
where $\delta_0\equiv\delta_0(\infty,\infty)$ is now the (leading order) eikonal. 

We can even derive a form of the dressed propagator beyond the leading order in the eikonal expansion. Integrating out the $\delta x$ fluctuations using the 
finite interval correlation function \eqref{eqn-finite-line-prop}, and
performing the same changes of variable we used before (but without dropping any terms), the difference between the dressed propagator and the free propagator takes the form 
\begin{eqnarray}
\label{deltag}
    G(p,p')-G^{(0)}(p,p')&=&
\sum_{N=1}^\infty\frac{(-i\bar g)^{N-1}}{2^{N-1}N!}\!\int d^4x \,\Phi(x)\,e^{-i \bar q\cdot x}\int_0^\infty \frac{d\sigma }2\int_0^\infty \frac{d\sigma' }2
e^{-\frac i2(\bar p'{}^2+\bar m^2 -i\epsilon)\sigma'}e^{-\frac i2(\bar p^2+\bar m^2 -i\epsilon)\sigma}\nonumber\\
&&\prod_{i=2}^N\int_{\bar k_i}\int_{-\sigma}^{\sigma'} d\sigma_i \Phi_{\bar k_i}
e^{i\bar k_i\cdot x}
e^{i\bar k_i \cdot \bar v \sigma_i}\int {\cal D}\delta x(\tau) \,e^{i\bar k_i\cdot \delta x(\sigma_i)}
e^{\frac i2\int_{-\infty}^\infty d\tau  \delta \dot x^2}
\end{eqnarray}
where $\bar v$ is the average wave vector $\bar v= (\bar p+\bar p')/2$.
To arrive at the final expression given in \eqref{deltag}, we factored out  $\exp(-i(\bar p^2+\bar m^2-i\epsilon)\sigma/2)$. We recognized the remaining factors as correlations of $\langle \prod_i\exp(i\bar k_i \cdot\delta \hat x(\sigma_i))\rangle $, where the fluctuations $\delta x$ are integrated now over the whole line, and not over a finite interval anymore. The correlation function on the unrestricted line is given in \eqref{xxprop}. Taking the on-shell limit yields the worldline scattering amplitude $\mathcal M_N$. We amputate the external scalar legs as we did before; this removes the integrals over $\sigma$ and $\sigma'$, and extends the integration domain for the $\sigma_i$ integrals to the whole real line\footnote{We have derived here a more symmetric form of the dressed propagator than \cite{Mogull:2020sak}, one which allows us to amputate both legs in the same fashion.}. Instead of starting from the path integral of the dressed propagator and taking the on-shell limit, we will later reproduce these results in \eqref{newMn} from the operator approach more straightforwardly.

\subsection{Scalar Particle in a Background Maxwell Field}\label{subsect:maxwell-wordline}

Next, we review the worldline formalism for scalar Quantum Electrodynamics (scalar QED).  To be consistent with our earlier QFT conventions with a dimensionless action,  the dimension of the gauge field $A_\mu$ must be inverse length:
\begin{equation}
    [A^\mu] = L^{-1}\, ,
\end{equation}
and the coupling constant  is dimensionless
\begin{equation}
    [\bar e] = 1\,.
\end{equation}
The scalar electron charge is $e$ with $\bar e^2=e^2/\hbar.$
To facilitate the transition to the worldline formalism for scalar QED, we follow Strassler \cite{Strassler:1992zr} and re-write the scalar propagator (assuming a non-zero background gauge field) as
\begin{equation}
    G(x', x) = \left\langle x'\left| \frac{1}{-\mathscr{D}_\mu \mathscr{D}^\mu + \bar m^2 - i\epsilon}\right|x\right\rangle,
\end{equation}
where 
\begin{equation}
    \mathscr{D}_\mu = \partial_\mu - i \bar e A_\mu.
\end{equation}
The denominator can be exponentiated using the Schwinger proper time
\begin{equation}
    G(x',x) = \int_0^\infty dT \, \, \bra{x'} \exp{-iT\qty(-\mathscr{D}_\mu \mathscr{D}^\mu + \bar m^2-i\epsilon)}\ket{x}.
\end{equation}
The matrix element under the proper time integral looks like a Schrodinger picture transition amplitude, with the time evolution generated  by the Hamiltonian 
\begin{equation}
    \hat H = \frac{(\hat {\bar p}_\mu +  \bar e \hat A_\mu(x))^2 + \bar m^2}{2}\,,
\end{equation}
where we have used $\hat {\bar p}_\mu = -i \partial_\mu$. The factor of $1/2$ is for convenience and is compensated by a re-scaling of $T$. 
The corresponding Lagrangian $\mathcal L=p\cdot \dot x -H$ is given by
\begin{equation}
    \mathcal L = \frac{1}{2}(\dot x^2-\bar m^2) - \bar e \dot x \cdot A(x)\,.
\end{equation}
Therefore, we can write a path integral expression for the propagator,
\begin{equation}
    G(x', x) = \int_0^\infty \frac{dT}2\int^{x(T)=x'}_{x(0)=x} \mathcal{D}x(\tau)\ \exp{i \int_{0}^{T} d\tau \, \qty(\frac{1}{2} (\dot x^2 - \bar m^2) - \bar e \dot x \cdot A(x))}\,.
\end{equation}
To get the propagator dressed with $N$ distinct photons, we simply expand the part of the exponential with $A$ to the $N$th power. There are $N!$ permutations for $N$ distinct photons, hence the $1/N!$ from the expansion cancels. We are left with a time-ordered expectation value
\begin{equation}
    G_N(x, x') = \int_0^\infty \frac{dT}2\left\langle x',T
    \left|\mathcal{T}
    \left\{\prod_{j=1}^N 
    \int_{0}^{T} d\tau_j \hat V(\tau_j)\right\} \right|x,0\right\rangle\,,
\end{equation}
where the background vertex operator for the emission/absorption of a photon with momentum $k$ and polarization $\varepsilon$ is 
\begin{equation}
    \hat V(\tau) = (-i\bar e) \hat{\dot x}(\tau) \cdot \varepsilon \ e^{i \bar k\cdot \hat x(\tau)}\label{gaugevertex}\,.
\end{equation}
Repeating the same steps that led to Eq.~\eqref{An} (amputating the external legs and placing them on-shell ) yields a similar expression for the scattering amplitude 
\begin{align}
\label{worldlinescatteringampgeneralformula1}
	\mathcal{M}_{N} = 2\lim_{\substack{\tau_{N+1} \rightarrow +\infty\\ \tau_0 \rightarrow -\infty}}\qty(\prod_{i=2}^{N}\int^{\infty}_{-\infty} d\tau_{i}) \left\langle \mathcal{T}\left\{\hat V_{\text{out}}(\tau_{N+1})\hat V_{N}(\tau_{N}) \hat V_{N-2}(\tau_{N-2})...\hat V_{1}(0)\hat V_{\text{in}}(\tau_0)\right\}\right\rangle \,.
\end{align}
where the in and out vertex operators, responsible for creating the asymptotic scalar particle states, were defined in Eq.~\eqref{inout}.
The only difference with the earlier discussion of scalar interactions comes from the vertex operators, which are now given by Eq.~\eqref{gaugevertex}. For simplicity we are dropping the $\exp(-\epsilon|\tau_i|)$ regulators, though their effect is taken into account whenever we perform the time-ordered integrals.

\subsection{Scalar Particle in a Curved Background}\label{subsect:gravity-wordline}

As in the previous two sections, to arrive at the worldline expression describing the interactions of a scalar point-particle with a background metric,  we begin in a similar way. Consider the curved space propagator of a minimally coupled scalar field, given by
\be
    G(x',x)=\left\langle x'\left|\frac{1}{-\frac{1}{\sqrt{\hat g}}\partial_\mu(\sqrt{\hat g} \hat g^{\mu\nu}\partial_\nu)+\bar m^2-i\epsilon}\right|x\right\rangle\,.
\ee
This expression can be exponentiated as well, yielding
\be
G(x',x)=\int_0^\infty dT
\langle x' |
\exp[-i T(\hat H-i\epsilon)]|x\rangle\,,\label{csprop}
\ee
where the Hamiltonian $\hat H$ can be written in a manifestly  Hermitian form as
\be
    \hat H=\frac{1}{\hat g^{1/4}}\hat {\bar p}_\mu
    \hat g^{1/2} \hat g^{\mu\nu} \hat {\bar p}_\nu\frac{1}{\hat g^{1/4}}+\bar m^2\,, \label{csham}
\ee
and where the Hermitian barred momentum operators are defined through
\be
\hat {\bar p} = -i\frac{1}{\hat g^{1/4}}\partial_\mu
{\hat g^{1/4}}\,.
\ee
The position eigenstates 
form a complete basis
\be
\int d^d x \sqrt{g}|x \rangle \langle x| = 1\,.
\ee
They are normalized 
such that
\be
\langle x'|x\rangle=\frac 1{\sqrt{g(x)}}\delta^d(x'-x)\,.
\ee
Similarly, the barred-momentum eigenstates obey a completeness relation
\be
\int d^d \bar p |\bar p\rangle\langle \bar p|=1\,.
\ee
As a consequence, their inner product with the position eigenstates is given by
\be
\langle x| \bar p\rangle = \frac{
\exp\left(
i \bar p\cdot x\right)}{(2\pi )^{d/2} g^{1/4}(x)}\,.
\ee
Since the curved space Hamiltonian dependence on the $\hat{\bar p}$ and $\hat x$ operators is neither separable (as it was in flat space) nor symmetric (as it was in a background Maxwell field, where the mixed term $\hat {\bar p}_\mu \hat A^\mu(x) + \hat A^\mu(x) \hat{\bar p}_\mu$ was symmetric), turning the matrix element in Eq.~\eqref{csprop} into a path integral requires care. The issue is accounting for the specific ordering of $\hat x$ and $\hat{\bar p}$ operators present in the curved space Hamiltonian. Consider the case of a one-dimensional system with a toy Hamiltonian $\hat x\hat {\bar p}$ (for the purpose of this example, we ignore that this toy Hamiltonian is non-Hermitian). The  matrix element $\langle x'| \exp(i \varepsilon \hat x\hat{\bar p}) |x\rangle$ with $\varepsilon$ being a small parameter, is evaluated by first writing $\hat x \hat {\bar p}$ in a symmetric form plus commutators ($\hat x \hat{\bar p}=\frac 12(\hat x \hat {\bar p}+\hat{\bar p }\hat x) -\frac 12 [\hat {\bar p},\hat x]$), and then exponentiating back. The result is $\frac{1}{\sqrt{2 \pi}} \int d\bar p \exp\{i \varepsilon [(x+x')\bar p/2 +\frac i2] \} \exp(i \bar p\cdot (x-x'))$. 
Similarly,  the matrix element of another toy Hamiltonian, which in this example is both  Hermitian and symmetric, $\hat{\bar p}\hat x^2 \hat{\bar p}=\frac{1}{4} (2\hat x \hat{\bar p}^2 \hat x+\hat x^2 \hat{\bar p}^2+\hat{\bar p}^2 \hat x^2)$, evaluates to $\langle x'|\exp(i\varepsilon \hat {\bar p} \hat x^2 \hat{\bar p})|x\rangle  =\frac{1}{\sqrt{2 \pi}} \int d\bar p \exp[i \varepsilon (x+x')^2\bar p/4] \exp[i \bar p\cdot (x-x')]. $
In the case of the  scalar Hamiltonian in curved space \eqref{csham}, these considerations yield the following path integral form of propagator in \eqref{csprop}
\begin{eqnarray}
G(x',x)&=&\lim_{N\to\infty}\int_0^\infty
dT \prod_{i=1}^{N-1}\int_{-\infty}^\infty d^d x_i \frac{g^{1/4}(x')}{g^{1/4}(x)} 
\prod_{j=1}^N \int d^d p_j\nonumber\\
&&\times\exp\left\{i\varepsilon_N \left[ (x_j-x_{j-1}) \cdot \bar p_j   - \bar p_{\mu,j} \bar p_{\nu, j} g^{\mu\nu}\left(\tfrac{{x_j}+x_{j-1}}2\right)-\bar m^2\right] -V_{CT}(\tfrac{{x_j}+x_{j-1}}2)\right\},\nonumber\\
\end{eqnarray}
where $\varepsilon_N=T/N$ and $x_0=x, \ x_N=x'$. The contribution from the commutators needed to turn the Hamiltonian \eqref{csham}
into its symmetric form is responsible for the so-called ``counterterm" potential
\be
V_{CT}= \frac 18 R  + \frac 18 g^{\mu\nu} \Gamma_{\mu\rho}^\sigma \Gamma_{\nu\sigma}^\rho
\ee
Taking the continuum limit and performing the integration over the momenta leads to the fully dressed worldline propagator \cite{Bastianelli:2000pt,Bastianelli:2000nm,Bastianelli:2006rx}
\be
G(x',x)=\int_0^\infty\frac{dT}2 \int^{x(T)=x'}_{x(0)=x} \mathcal{D}x(\tau)
\exp\left\{i\int_0^T d\tau\left[   g_{\mu\nu} (\tfrac 12\dot x^\mu \dot x^\nu +\tfrac 12 a^\mu a^\nu +b^\mu c^\nu)-\tfrac 12 \bar m^2 -\tfrac 18 R \right]\right\}\,,
\label{cspi}
\ee
after a similar rescaling of the Schwinger time needed to achieve the canonical normalization of the Lagrangian. 
The $\sqrt{g(x)}$ factors obtained after integrating the momenta, which can be thought of as the product of two factors $g(x)\times 1/\sqrt{g(x)}$, are exponentiated as well, with the help of two sets of auxiliary fields (``ghosts"), one anticommuting pair ($b$ and $c$), and one commuting ($a$). The path integral expression of the dressed propagator given in \eqref{cspi} is a rather formal expression, and to properly define it, we need to regularize it. When regularizing the path integral by discretizing the time (``time slicing"), in addition to the Ricci scalar term, we must recall that we need to add the non-general covariant $\Gamma \Gamma$ terms derived earlier. Other modes of regularization (e.g. momentum cut-off) will require different, possibly non-general covariant terms  \cite{Bastianelli:2006rx}. In particular, a widely used regularization scheme, dimensional regularization, is general covariant and does not require any additional contribution to the path integral \eqref{cspi} \cite{Bastianelli:2000nm, Bastianelli:2000pt}.
The next steps towards writing a scattering amplitude formula are the same as in the previous two sections:  first, we expand the background metric around flat space $g_{\mu\nu}=\eta_{\mu\nu}+h_{\mu\nu}$. Then, we expand the metric fluctuations into plane waves. To describe the interaction with $N$ background gravitons with different momenta, we expand to order $ N$ in perturbation theory. Unlike the case of scalar Yukawa or Maxwell interactions, where the interactions of the worldline with the background fields were linear, the gravitational interactions are non-linear due to $V_{CT}$. Nonetheless, the same philosophy applies, and we collect the graviton vertex operators. In particular, the linearized vertex operator for the emission/absorption of a graviton with momentum $k$ and polarization tensor $\varepsilon_{\mu\nu}$ is 
\be
\hat V(\tau)|_{lin}=i\varepsilon_{\mu\nu} \left[\frac{1}2 \left(\hat{\dot  x}^\mu
\hat {\dot x}^\nu + \hat a^\mu \hat a^\nu + \hat b^\mu \hat c^\nu\right) -\frac 1{8} \left(\eta^{\mu\nu} \bar k^2-\bar k^\mu \bar k^\nu\right) \right]e^{i\bar k\cdot \hat x(\tau) }\,.\label{1grav em}
\ee
If we regularize the worldline path integral using dimensional regularization, the only contribution from the counterterm potential is $V_{CT}=R/8$. It is worth mentioning that we could have discovered the need for the $R/8$ term in the dressed propagator \eqref{cspi}, had we required that the 3-point QFT vertex of a minimally coupled scalar field is reproduced by the worldline amplitude $\mathcal M_1$, which corresponds to the emission/absorption of one off-shell graviton. See \cite{Mogull:2020sak, Du:2023nzo}.   By the same token, we could consider a particular case of a non-minimally coupled scalar with an additional QFT interaction $R\phi^2$ and find the coefficient of this term such that the worldline description does not require any counterterm $V_{CT}$, and therefore eliminate the nonlinear terms of the worldline gravitational vertex operator, in which case \eqref{1grav em} becomes exact.
However, we will see that in the classical limit the contribution from the potential counterterm is higher order in $\hbar$ and therefore it can be neglected.

\subsection{ The Classical Limit of Worldline Formalism: WQFT Rules for Scalar Yukawa Theory} \label{subsect:worldline-to-wqft}

In general, we will be interested in the scattering of particles that are distinct from the interaction mediator particles, and we will let the latter become dynamic. The worldlines continue to be a useful description of the interactions of the scattering particles with mediators, but, the dynamics of the mediators will be described in terms of Feynman diagrams derived from quantum field theory. We can think of the worldlines as first quantized fields, and the mediators being described by second quantized fields. This is an appropriate distinction in the eikonal or classical limit. 
We will also address the radiative corrections (to the worldlines, when the mediator propagators attach to the same worldline, or through virtual loops to the mediators) in addition to  the so-called conservative contributions to the scattering amplitudes (which are described by ``ladder" diagrams, with the mediator propagators linking the two worldlines).
For the purpose of this section, which is meant as a toy model and used to demonstrate the use of the worldline formalism, we will ignore the radiative corrections. 
 
We will see that the worldline formalism offers a more efficient way of computing the scattering amplitudes in the classical limit.
Since we set up the eikonal expansion as an expansion in $\hbar$, we must also keep track of $\hbar$ factors in the worldline formalism.   As reviewed in the previous section, we work with a dimensionless QFT action and we set $d=4$. The bosonic fields have dimensions of inverse length
    $[\phi] = L^{-1}.$
Thus, the position-space Feynman propagator
\begin{equation}
    G_0(x', x) = \expval{\mathcal{T}(\hat \phi(x) \hat \phi(x'))}=\left\langle x' \left| \frac{1}{-\Box + \bar m^2-i\epsilon} \right| x \right\rangle\,,
\end{equation}
has dimensions of inverse length squared:
$
    [G_0] = L^{-2}\,.
$
The transition to worldline formalism is done by  introducing the Schwinger time and exponentiating, as in \eqref{freepropx}. 
To be dimensionally consistent, we need
 $   [T] = L^{2}.$ 
The rest checks out:
\begin{equation}
    [G_0] = L^{-2} = [T] [\ket{x}]^2 = L^{2} L^{-4} = L^{-2}\,,
\end{equation}
where we used that\footnote{ This follows from the normalization of the position eigenstates
$
    \braket{x'}{x} = \delta^4(x - x')$.}
 $   [\ket{x}] = L^{-2}.$ 
Lastly,  as shown earlier in\eqref{freepropx} and \eqref{freeL}, the free propagator can be cast in its worldline path integral form
\begin{equation}
G_0(x',x) = \int^{\infty}_0 \frac{dT}2 \int^{x(T)=x}_{x(0)=x'} \mathcal{D}x(\tau)\ e^{i\int^T_0 d\tau \mathcal{L}_0(x(\tau), \dot{x}(\tau))}\,.
\end{equation}
This is also dimensionally consistent since
$
    [\tau] = [T]= L^2,$ $[x]=L $ and 
$
    [\tau] [\dot x^2]=1.$
We argued earlier in \eqref{gnan}  that the scattering amplitude is computed in the worldline formalism as in \eqref{An}, where the asymptotic particle states are created by the action of the vertex operators $V_{\text{in}}$ and $V_{\text{out}}$, while the rest of the vertex operators account for the emission/absorption of background quanta.
After amputating the external legs, 
the finite interval $\tau\in [0,T]$ is extended to the infinite line $\tau\in(-\infty,\infty)$. We can  check that the infinite line worldline propagator is  dimensionally consistent:
\begin{equation}
    \expval {\mathcal T\bigg( \hat x^\mu(\tau) \hat x^\nu(\tau')\bigg)} = - \frac{i}{2} \abs{\tau- \tau'} \eta^{\mu\nu}\,.\label{xxprop}\
\end{equation}
In these units,  the canonical conjugate to $x$, 
$\bar p_\mu = \dot x_\mu$,  is the wave vector since 
$[\bar p] = [\dot x_\mu]=L^{-1}.
$

 In \cite{Mogull:2020sak} Mogull, Plefka and Steinhoff formulated and gave a set of "Feynman" rules for WQFT. The latter was viewed as an extension of the worldline effective field theory method, with the main distinction that the worldline was quantized around the classical trajectory, parametrized as a straight line whose slope is the averaged  wave vector $\bar v$:
\begin{equation}
    \bar v \equiv \frac{\bar p + \bar p'}{2}\,.
\end{equation} 
 We will show that the worldline formalism leads to   WQFT  \cite{Mogull:2020sak} by taking the classical limit of the worldline formulation of the scattering amplitude. This is not an effective field theory result. It is a repackaging of the QFT scattering amplitude.
 
 Let us start with the worldline scattering amplitude \eqref{An} and perform all the contractions\footnote{This is done using \begin{equation}
     \expval{\prod_{i}:e^{\hat A_i}:}  = \exp{\sum_{i < j}  \expval{\hat A_i \hat A_j} }\,.
\end{equation}} with the external legs using \eqref{xxprop}:
\begin{align}
    \mathcal{M}_N(\bar p', \bar p) &= (-i\bar g)^N \qty(\prod_{j=2}^{N} \int_{-\infty}^\infty\frac{d\tau_j}2) \Bigg[\exp{\sum_{i=1}^N\qty(- \frac{i}2 \bar p' \cdot \bar k_i (\tau_{N+1} - \tau_i)) + \sum_{i=1}^N\qty( \frac{i}2 \bar p \cdot \bar k_i (\tau_i - \tau_0))} \nn & \qquad \qquad \qquad \times \exp{-\frac{i}2 (\bar p' \cdot \bar p+\bar m^2)(\tau_{N+1}-\tau_0)}\expval{\mathcal{T}\qty(e^{i \bar k_N \cdot \hat x(\tau_N)}  e^{i \bar k_{N-1} \cdot \hat x(\tau_{N-1})} \dots e^{i \bar k_1 \cdot \hat x(0)})}\Bigg]\,,
\end{align}
where we recall that $\tau_0$ and $\tau_{N+1}$ are the asymptotic times, minus infinity and plus infinity,  respectively.
For simplicity of notation, in the previous expression, we rendered the convergence factors $e^{-\epsilon|\tau_i|}$ from \eqref{An} implicit. The terms that multiply $\tau_0$ and $\tau_{N+1}$ cancel 
\begin{equation}
\tau_0\left[\bar p\cdot \left(\sum_{i=1}^N\bar k_i\right)-\bar p\cdot \bar p'-m^2\right]
=-\tau_0(\bar p^2+m^2) =0  \end{equation}
using momentum conservation and the on-shell condition for the asymptotic states.

At this point, we notice that the vertex operators responsible for creating the asymptotic states have been eliminated from the scattering amplitude, their effect being accounted by a shift $\bar k\cdot x \to \bar k\cdot x+ \bar k\cdot \bar v \tau$ in the  vertex operators describing the interaction with the background particles:
\begin{equation}
   \label{newMn} \mathcal{M}_N(\bar p', \bar p) = (-i\bar g)^N \qty(\prod_{j=2}^{N} \int_{-\infty}^\infty\frac{d\tau_j}2) \expval{\mathcal{T}\left\{e^{i \bar k_N \cdot (\bar v \tau_N + \hat x(\tau_N))}  e^{i \bar k_{N-1} \cdot (\bar v \tau_{N-1}+\hat x(\tau_{N-1}))} \dots e^{i \bar k_1 \cdot \hat x(0)}\right\}}\,.
\end{equation}
Note that the new, shifted form of the background vertex operators simply follows from the expression of the scattering amplitude by performing the contractions with the asymptotic in-out vertex operators. We did not have to assert ab initio that the worldline coordinates $x$ are deviations from the classical trajectory of the scattered particle, as it was done in \cite{Mogull:2020sak}. 

We are now ready to derive the WQFT rules from an $\hbar$ eikonal expansion of the scattering amplitude in worldline formalism. Recall that in the classical limit, the external momenta are of the order $\hbar^0$ and the background particle momenta are of the order $\hbar^1$.

For simplicity, let us first consider the amplitude describing the interactions of a scalar particle  with   $N=2$ background scalars:
\begin{equation}
    \mathcal{M}_2(\bar p', \bar p) = (-i\bar g)^2 \int_0^\infty\frac{d\tau}2 \expval{\mathcal{T}\qty(e^{i \bar k_1 \cdot (\bar v \tau + \hat x(\tau))}  e^{i \bar k_2 \cdot \hat x(0)})}\label{a2wqft}
\end{equation}
We now restore $\hbar$ on the external wave numbers
\begin{equation}
    \bar v = \frac{v}{\hbar}\,,
\end{equation}
and find it convenient to perform the re-scaling
\begin{equation}
    \Tilde{\tau} = \frac{\tau}{\hbar}\,.
\end{equation}
Then, the worldline two-point correlation function becomes
\begin{equation}
    \expval{\mathcal{T}(\hat x^\mu(\Tilde{\tau}) \hat x^\nu(\Tilde{\tau}'))} = - \frac {i}2  \eta^{\mu \nu} \hbar \abs{\Tilde{\tau} - \Tilde{\tau}'}\label{tildecontract}\,.
\end{equation}
The worldline scattering amplitude \eqref{a2wqft} becomes
\begin{equation}
     \mathcal{M}_2(  p',  p) = (-i\bar g)^2 \hbar \int_{-\infty}^\infty\frac{d\tilde{\tau}}2 e^{\bar k_1 \cdot v \Tilde{\tau}}\expval{\mathcal{T}\qty(e^{i \bar k_1 \cdot\hat x(\Tilde{\tau})}  e^{i \bar k_2 \cdot \hat x(0)})}
     \end{equation}
and evaluates to
\begin{equation}
    \label{m2exp} \mathcal{M}_2( \bar p',  \bar p) = (-i\bar g)^2 \hbar \int_{-\infty}^\infty\frac{d\tilde{\tau}}2 e^{i\bar k_1 \cdot v \Tilde{\tau}} \exp{\frac{i\hbar }2 \bar k_1 \cdot \bar k_2  \abs{\Tilde{\tau}}}\,,
\end{equation}
after performing the worldine contractions according to \eqref{tildecontract}.
We have now arrived at the eikonal expansion of the scattering amplitude describing interactions with two background particles.
\begin{equation}
    \mathcal{M}_2(  p',   p) = (-i\bar g)^2 \hbar \int_{-\infty}^\infty\frac{d\tilde{\tau}}2 e^{i\bar k_1 \cdot v \Tilde{\tau}} \sum_{n=0}^\infty \frac{\hbar^n}{ n!} \qty( \frac i2\bar k_1 \cdot \bar k_2  \abs{\Tilde{\tau}})^n\,.\label{am2}
\end{equation}
In this expression, as in the preceding ones,  we have been suppressing an energy-momentum conservation delta-function $(2\pi)^4\delta(\bar p'-\bar p-\bar k_1-\bar k_2)$.

Diagrammatically, the $\hbar$ expansion can be represented as
\be \label{hbarwqftfig}
\includegraphics[valign=c]{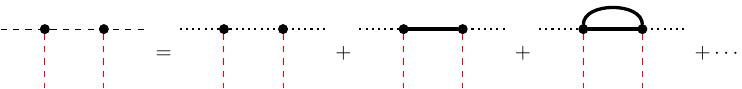}
\ee 
Here, we are following the conventions of \cite{Mogull:2020sak}, though we will not Fourier transform the worldline Schwinger time ($\tau$) to energy space, as done in \cite{Mogull:2020sak}. The worldline of the scalar particle is depicted by the dashed line. The black dots denote the worldline vertices, and the background field quanta, which are soft (i.e., their momenta are of the order $\hbar$), are represented by the dotted red lines. The vertices are free to slide on the worldline, corresponding to the integral(s) over the Schwinger time $\tau_i$   in the 
amplitude. There are $N-1$ integrals 
(one less than the number of background interactions), and so one vertex
is fixed, and the rest can slide. On the right-hand side of \eqref{hbarwqftfig}, we are depicting the $\hbar$ expansion of \eqref{m2exp}.
The worldline has become dotted because all $x$ contractions have been performed, and the dotted line is only for visual reference. Each contraction is represented by a solid line, and a factor of $\hbar$ accompanies each solid line.
We are equating the  $\hbar$ expansion of the worldline scattering amplitude to this {\it contraction expansion}.

 If $N$ background particles are emitted or absorbed from the worldline, the scattering amplitude for this process is given by a straightforward generalization of \eqref{am2}:
\begin{equation}
    \mathcal{M}_N( p',   p) = \left.(-i\bar g)^N \hbar^{N-1} \prod_{j=1}^{N-1}\int_{-\infty}^\infty\frac{d\tilde{\tau_j}}2e^{i\bar k_j \cdot v \Tilde{\tau_j}} \sum_{n=0}^\infty \frac{\hbar^n}{n!} \qty(\frac{i}2 \sum_{\stackrel{i,j=1}{i<j}}^N\bar k_i \cdot \bar k_j  \abs{\Tilde{\tau_i}-\tilde{\tau_j}})^n\right|_{\tau_{N}=0}\,.
    \label{am2-gen}
\end{equation}

Since we are ultimately interested in the scattering of two scalar particles which interact with each other through exchanges of massless mediators, we are now allowing the background fields (which for massless scalar Yukawa interactions are massless scalar fields) to become dynamical, and we integrate them out. As a result, every internal line associated with the massless mediators is accompanied by an integral $\int d^4\bar k/(2\pi)^4$ and an insertion of a propagator $-i/(\bar k^2-i\epsilon)$.

More concretely, consider two incoming particles with momenta $p_1$ and $p_2$ and two outgoing particles with momenta $p'_1$ and $p'_2$. The particles are each described by worldlines, and, as before, we denote the momentum transfer by $q$: $p'_1-p_1=q=p_2-p'_2$. 

To lowest order in perturbation theory,
the one mediator exchange amplitude is
\begin{equation}
 \includegraphics[valign=c]{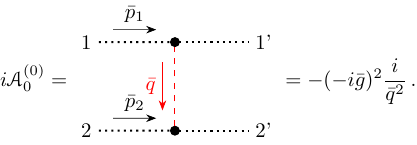}
\end{equation}
In the next order in perturbation theory, we are looking at two mediator exchange diagrams. The contribution which is the leading order in $\hbar$  is obtained from the diagram
\begin{equation}
    \includegraphics[valign=c]{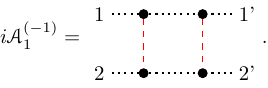}
\end{equation}
Each worldline (labeled $a=1,2$) yields a factor of $(-i\bar g)^2 2 \pi \hbar\delta(2\bar k_1\cdot v_a)(2\pi)^4\delta^4(2\bar q_a-\bar k_1-\bar k_2)$ where $\bar q_1=\bar p'_1-\bar p_1$ on the first worldline and $\bar q_2=\bar p_2-\bar p'_2$ on the second worldline. After factoring out the total energy-momentum conservation delta function, we set $\bar q_1=\bar q=-\bar q_2$.  Each mediator propagator labeled $i=1,2$ is $1/(i (\bar k_i^2-i\epsilon))$. If we put all the elements together, we find 
\be
i \mathcal  A_1^{(-1)}= \bar g^4 (2\pi \hbar)^2 \int_{\bar k_1,\bar k_2}\frac{(2\pi)^4\delta^4(\bar k_1+\bar k_2-\bar q)}{(\bar k_1^2-i\epsilon)( \bar k_2^2-i\epsilon)} \delta (2\bar k_1\cdot  v_1) \delta(2\bar k_1\cdot  v_2)
\ee
where we suppressed the overall energy-momentum conservation delta functions. 
This matches the leading order eikonal of the QFT amplitude derived earlier in \eqref{m200} and \eqref{m2001} once we replace the delta functions $\delta (2\bar k_1\cdot  v_1) \delta(2\bar k_1\cdot  v_2)$ by $\delta (2\bar k_1\cdot  p_1) \delta(2\bar k_1\cdot  p_2)
$, which is certainly correct to the leading order in $\hbar$. 

To further match the eikonal QFT expansion, we have two options. On one hand, we may go to the so-called eikonal frame $p_1=(E_1, \bm q/2, p_{1,z}), \ p'_1=(E_1, -\bm q/2, p_{1,z})$ in the context of the QFT calculation, in which case we can keep the WQFT $\hbar$-expansion as it is since both the QFT and WQFT $\hbar $-expansions will be expressed in terms of the average momenta $v_a=(E_a, \bm 0, p_{a,z})$ and momentum transferred $\bm q$. On the other hand, we may choose to remain in the frame where $p_1$ is longitudinal at the cost of having $q=(0,\bm q_\perp,q_z)$ with $\bm q_\perp$ of order $\hbar$ and $q_z$ of order $\hbar^2$ (see \eqref{qpishbar2}). Then we would have to reshuffle the WQFT $\hbar$-expansion  replacing the average momentum $v_a$ by the incoming momentum $p_a$, to match the QFT calculation\footnote{This is the root of the difference between KMOC and the eikonal-based method or WQFT method:  a change of momentum basis~\cite{Georgoudis:2023eke,Damgaard:2023vnx} from the scattering states asymptotic momenta to the averaged momenta that arise naturally in WQFT.}. 
In the latter case, we substitute
\begin{equation}
   2 v = p' + p = 2p + \hbar \bar q\,.
\end{equation}
Then the WQFT eikonal expansion for one worldline interacting with two  background field quanta becomes
\begin{equation}
    \mathcal{M}_2( p,  p') = (-i\bar g)^2 \hbar \int_{-\infty}^\infty\frac{d\tilde\tau }{2} e^{i\bar k_1 \cdot p \Tilde{\tau}} \sum_{n=0}^\infty \frac{\hbar^n}{n!} \qty(\frac{i}2( \bar k_1 \cdot \bar q \Tilde{\tau}+ \bar k_1 \cdot \bar k_2  \abs{\Tilde{\tau}}))^n.
\end{equation}

To find the next order term in the eikonal expansion,  we let one of the worldlines go up to the next order in $\hbar$. This is (after a change of integration variable)
\begin{align}
    \mathcal{M}_2^{(0)}(p', p) &= (-i\bar g)^2 \hbar^2 \intinf{\tau} e^{2i \bar k_1 \cdot p \tau} i (\bar k_1 \cdot \bar q \tau + \bar k_1 \cdot \bar k_2 \abs{\tau}) \nn
    &= (-i\bar g)^2 \hbar^2 \qty[\intoinf{\tau}e^{2i \bar k_1 \cdot p \tau} i (\bar k_1 \cdot \bar q \tau - \bar k_1 \cdot \bar k_2 \tau) - \intoinf{\tau}e^{-2i \bar k_1 \cdot p \tau} i (\bar k_1 \cdot \bar q \tau + \bar k_1 \cdot \bar k_2 \tau) ]\,.
\end{align}
Recalling that we made the $e^{-\epsilon|\tau|}$ implicit, we are now making use of it to evaluate the integrals. 
We obtain
\begin{align}
    \mathcal{M}_2^{(0)}(p', p) &= (-i\bar g)^2 \hbar^2 (-i) \frac{\bar k_1 \cdot \bar q - \bar k_1 \cdot \bar k_2 - \bar k_1 \cdot \bar q - \bar k_1 \cdot \bar k_2 }{(2 \bar k_1 \cdot p)^2} \nn
    &=  (-i\bar g)^2 \hbar^2 i \frac{2 \bar k_1 \cdot \bar k_2}{(2 \bar k_1 \cdot p)^2}\,.
\end{align}
If we now glue two worldlines, one with the lowest order expansion and the other with next to lowest order expansion, we get the classical contribution to the one loop QFT scattering amplitude. Diagrammatically, this is given by: 
\begin{equation}
    i\mathcal{A}_1^{(0)} =\includegraphics[valign=c]{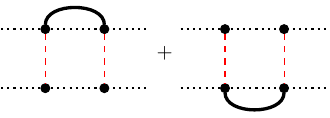}
\end{equation}
and translates to  
\begin{equation}
    i\mathcal{A}_1^{(0)} = \frac{(-i\bar g)^4 \hbar^3}{2!} \int_{\bar k_1} \int_{\bar k_2} (2\pi)^4 \delta^4(\bar k_1 + \bar k_2 -\bar q) \frac{1}{\bar k_1^2 \bar k_2^2} \qty[(2 \pi)\delta(2p_1 \cdot \bar k_1)  \frac{2 \bar k_1 \cdot \bar k_2}{(2 \bar k_1 \cdot p_2)^2} + (2 \pi)\delta(2p_2 \cdot \bar k_1)  \frac{2 \bar k_1 \cdot \bar k_2}{(2 \bar k_1 \cdot p_1)^2}].
\end{equation}
The factor $1/2!$ comes from the number of ways of rearranging the internal lines. As we showed earlier, this is precisely what one gets from the usual QFT treatment. 

In the next sections, we will discuss the eikonal expansion in scalar QED using QFT and worldline formalism techniques at length.



\section{WQFT and The Eikonal Method}\label{sect:wqft-eikonal}


In the previous sections, we have shown the equivalence between the worldline
formalism and QFT scattering amplitudes. 
We have seen how taking the classical limit of the worldline formalism leads to the WQFT
rules,  while taking the classical limit of the QFT scattering amplitudes leads to the eikonal
method. We can show these relations
with the following chart\\
\begin{figure}[h]
\includegraphics[scale=0.9]{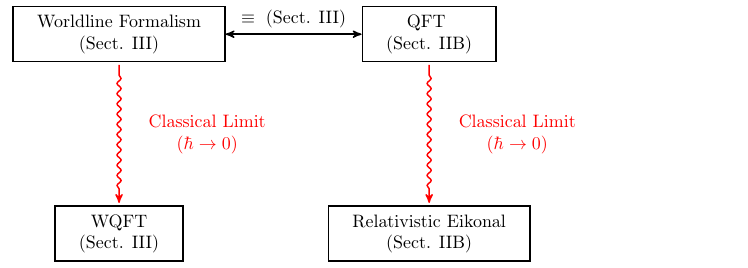}
\caption{A flowchart showing the relationships elaborated on throughout this work.}
\label{fig:eikonalW-QFT-part}
\end{figure}

This section will complete this chart by relating WQFT
and the eikonal method. More specifically, we will discuss how  the eikonal expansion, 
and the exponentiation of the
eikonal phase are realized and efficiently computed using WQFT.

\subsection{The Eikonal Expansion}\label{subsect:wqft-eikonal-expansion}

The eikonal expansion is essentially the $\hbar$ expansion with the assumption
that mediators are in the soft region, which is relevant for taking
classical limit. In Sections \ref{sect:eikonal-qm-qft} and \ref{sect:wordline} we have already shown how the
eikonal expansion works for the scattering amplitude in QFT  and in the worldline
formalism, respectively, for a simple case, that of scalar-mediated interactions. 

Here, we extend those results for electromagnetic and gravitational mediated interactions between the scalar scattering states.
On the one hand, in the QFT-based eikonal,  we sum up 
Feynman diagrams in the eikonal limit, which is equivalent to  an $\hbar$ expansion,  keeping the momenta $p_a$
for external particles and the wave numbers $\bar k_i$ for mediator particles to be
finite.

On the other hand, in  the  worldline formalism, the $\hbar$ expansion is equivalent to the
contraction expansion between background vertex operators. There are two ways
to do this contraction expansion depending on the order of the calculation. If
the contraction expansion is done before the worldline time integration, the
WQFT rules of Mogull, Plefka and Steinhoff are recovered. We are essentially
summing up expressions for different WQFT diagrams. If the worldline time
integration is done before the the contraction expansion, then the QFT expressions are recovered. 
However,  in worldline formalism we can
still preserve the information of the contractions by introducing several
coefficients $t_{i j}$, which we will discuss more  in Section \ref{ssec:amp-gen-WQFT}.
To see how this works out, let us start with the worldline amplitude:
\begin{eqnarray}
  \mathcal{M}_{N} & = &  2(-i\lambda)^N\left\langle  \mathcal{T}  \left\{ e^{- i \bar{p}_1' \cdot
  \hat x_(+ \infty)} \left( \prod_{i = 2}^N \int_{- \infty}^{ \infty} d \tau_i \hat V_i
  (\bar{k}_i, x_i) \right) \hat V_1 (\bar{k}_1, x_1) e^{i \bar{p}_1 \cdot \hat x(-
  \infty)}  \right\}\right\rangle\nonumber \\
  & = &2(-i\lambda)^N \left( \prod_{i = 2}^N \int_{- \infty}^{ \infty} d \tau_i  \right)
  \left\langle \mathcal{T} \left\{ \hat V_N \left( \bar{k}_N, \bar v_1\tau_N +
  x(\tau_N) \right) \ldots \hat V_1 \left( \bar{k}_1,  x(0)
  \right) \right\} \right\rangle\,,\label{mnnew1}
\end{eqnarray}
where
$\lambda$ is a placeholder for the relevant coupling constant. 
For example, for 
 scalar  QED or gravitational couplings we make the substitutions:\begin{equation}
\lambda =\bar e=  \frac{e}{\sqrt{\hbar}}\qquad {\text or}\qquad  \lambda= \sqrt{8 \pi \hbar G}\,
\end{equation}
where $G$ is Newton's constant.
In the second line of \eqref{mnnew1} we fully performed the contractions with the in and out vertex operators  as in \eqref{newMn}. 
Their effect is reflected in the shift $x\rightarrow x+\bar v \tau $ made in the background vertex operators.
Notice that the number of integrals in the expression of the amplitude $\mathcal M_N$ given in \eqref{mnnew1} is $N - 1$ instead of $N$. However, since we ultimately need to go to impact parameter
space with the Fourier transformation, we will combine the amplitude with the
corresponding $\delta$ functions from the integral measure\footnote{We recall that the reason behind the delta function is that $\bar{q}$ is not an arbitrary wave number; it is constrained such that the external momenta are on-shell. When Fourier-transforming to impact parameter space by integrating over  $\bar{q}$, the role of the delta function $\delta(2 \bar q\cdot \bar v)$ is to impose the mass shell constraint for the asymptotic scalar states.} and obtain a more symmetric expression for the worldline scattering amplitude:
\begin{eqnarray}
  (2\pi)\delta \left(2 \frac{v_1}{\hbar} \cdot \bar{q} \right) \mathcal{M}_{N}
  (v_1, \bar{q}) & = & \left( \prod_{i = 1}^N (-i\lambda)\int_{- \infty}^{ \infty} d
  \tau_i  \right) \left\langle \mathcal{T} \left\{ V_N \left( \bar{k}_N,
  \bar v_1 \tau_N + x_N \right) \ldots V_1 \left( \bar{k}_1,
  \bar v_1 \tau_1 + x_1 \right) \right\} \right\rangle. \nn
\end{eqnarray}
For simplicity, we will use $\widetilde{\mathcal{M}}_N$ to denote this combination,
\be
\widetilde{\mathcal{M}}_N \equiv (2\pi)\delta \left(2 \frac{v_1}{\hbar} \cdot \bar{q} \right) \mathcal{M}_N.
\ee
To highlight how the $\hbar$ expansion maps into the contraction expansion, we perform one more  rescaling, $\tau\rightarrow\tilde\tau=\tau/\hbar$, to arrive at
\begin{eqnarray}
\widetilde{\mathcal{M}}_N & =&(-i\lambda\hbar)^N  \left( \prod_{i = 1}^N \int_{- \infty}^{ \infty} d\tilde\tau_i  \right)
  \left\langle \mathcal{T} \left\{ \hat V_N \left( \bar k_N, v_1 \tilde\tau_N +
  x(\tilde \tau_N) \right) \ldots \hat V_1 \left( \bar k_1,  v_1 \tilde\tau_1 +x(\tilde\tau_1)
  \right) \right\} \right\rangle\label{mnnew3}.
\end{eqnarray}
In \eqref{mnnew3}, the vertex operators for the emission/absorption of the background quanta are shifted as a result of the contraction with the in and out scalar vertex operators and depend on the rescaled worldline time $\tau$.
For background photons, we insert 
\be\label{vph}
\hat V_j (\bar k_j,{v} \tilde\tau_j +
x(\tau_j))=\epsilon_\mu\bigg[\bigg(\frac{v^\mu}{\hbar} + \frac{\dot {\hat x}^\mu (\tilde\tau_j)} {\hbar}\bigg)e^{i\bar k_j \cdot v \tau_j }
e^{i\bar k_j\cdot \hat x(\tilde \tau_j)}\bigg]\,.
\ee
Similarly, the (linear) background graviton vertex operator is 
\be\label{vgrav}
\hat  V_j (\bar k_j,{v} \tilde \tau_j +
x(\tilde\tau_j))=-\frac 12 (\epsilon_j)_{\mu\nu}\bigg[\bigg(\frac{v^\mu}\hbar + \frac{\dot {\hat x}^\mu (\tilde \tau_j)}\hbar \bigg)
\bigg(\frac{v^\nu}\hbar  + \frac {\dot {\hat x}^\nu (\tilde \tau_j)} \hbar \bigg)e^{i\bar k_j \cdot v \tilde \tau_j }
e^{i\bar k_j\cdot \hat x(\tilde\tau_j)}\bigg].
\ee
The worldline amplitude is calculated by summing up all possible $\langle x_i x_j
\rangle$ contractions performed using \eqref{tildecontract}. The calculation is organized order by
order in terms of the number of contractions effectuated:
\begin{eqnarray*}
  \text{worldline amplitude} & = & \left( \text{term with 0 contraction}
  \right)\\
  &  & + \left( \text{terms with 1 contraction} \right)\\
  &  & + \left( \text{terms with 2 contractions} \right)\\
  &  & + \ldots
\end{eqnarray*}
where each contraction brings a factor of $\hbar$. Let us consider one term such that there
is a group of vertex operators which does not contract with any operator
outside the group,
\be \mathcal{T} \left\{ \left\langle \ldots \left( \prod_{i \in G} \hbar\int_{-
   \infty}^{ \infty} d \tilde\tau_i V_i \left( \bar{k}_i, {v} \tilde\tau_i +
   x_i \right) \right) \ldots \right\rangle \right\} . \ee
Then the correlation function  factorizes naturally
\ba
  \mathcal{T} \left\{ \left\langle \ldots \left( \prod_{i \in G}\hbar \int_{- \infty}^{
  \infty} d \tilde\tau_i V_i \left( \bar{k}_i, {v}\tilde \tau_i + x_i \right)
  \right) \ldots \right\rangle \right\} & = & \mathcal{T} \{ \left\langle \ldots \right\rangle
  \}  \mathcal{T} \left\{ \left\langle \prod_{i \in G} \hbar \int_{- \infty}^{ \infty} d
  \tilde \tau_i V_i \left( \bar{k}_i, {v}\tilde\tau_i + x_i \right) \right\rangle
  \right\} .\nonumber\\
\ea
Notice that we do not impose any condition on the contractions inside the
group $G$, which means that the group $G$ may be further factorized  into smaller
groups. Each factor  can be seen as  a lower order worldline
amplitude. For example,
\be  \mathcal{T} \left\{ \left\langle \hbar \prod_{i \in G} \int_{- \infty}^{+ \infty} d
   \tilde\tau_i V_i \left( \bar{k}_i, {v_1}\tilde \tau_i + x_i \right) \right\rangle
   \right\}, \ee
can be interpreted as the worldline amplitude $\delta \left(
\frac{v_1}{\hbar} \cdot \sum_{i \in G} k_i \right) \mathcal{A}_G \{ v_1, \{
\bar k_i \} \}$. Thus, in general, this factorization of
the expression can be written as
\be \label{fac-one-worldine}
\widetilde{\mathcal{M}}_{G\cup G^c}\left(
  v_1, \{ \bar{q} \} \right) =  \widetilde{\mathcal{M}}_G \left(
  v_1, \{ \bar{q}_1 \} \right)  \widetilde{\mathcal{M}}_{G^c} \left(v_1, \{ \bar{q}_2 \} \right).
\ee
Here, $G$ and $G^c$ do not denote the number of background mediators. Instead, they respectively represent a group of background mediators with such a configuration that there is no contraction between the two groups.

It is the information of the $x{-}x$ contractions that gives WQFT the advantage in taking
the classical limit. In the QFT eikonal method, each Feynman diagram contains terms of
different order in $\hbar$. 
In the WQFT method,  the Feynman diagrams are
reorganized and further expanded into WQFT diagrams. Each diagram is at one
specific order of $\hbar$. The $\hbar$ expansion, now in terms of the contraction
expansion, has a diagrammatic representation and shows a rich inner
structure. For example,
in Sections \ref{sect:eikonal-qm-qft}  and \ref{sect:wordline}, we saw how the one loop ladder diagrams in scalar Yukawa theory QFT are computed in WQFT:
\be
\includegraphics[width=\textwidth]{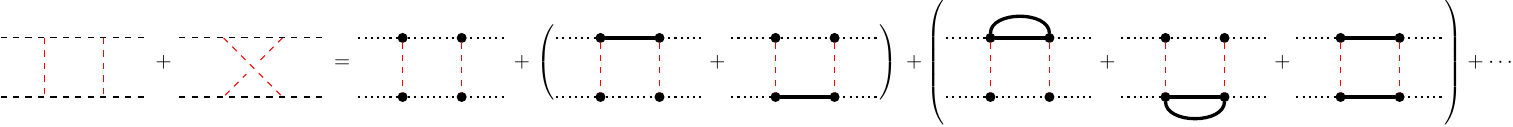}
\ee
where each set of terms within the round brackets is of a given order in $\hbar$. Notice in particular that the 
diagrams with the same number of contractions (which are depicted by the solid links) are at the same order of $\hbar$.
Of course, it is then important to be able to properly count the $\hbar$ to for a given diagram to figure out whether it is superclassical, classical or quantum. The $\hbar$ counting also tells us whether a quantum theory is ``good" or not for taking classical limit, which leads to a proper exponentiation. 

In practical calculations, based on the assumption of exponentiation, this
feature of WQFT allows to target classical contributions directly.   In fact, with WQFT, the idea of exponentiation
on longer needs to be treated as an assumption, but can be investigated and
analyzed directly. We  discuss the exponentiation of the eikonal to all loop orders  in a separate article~\cite{ADRV1}.

Before going directly into next subsection to discuss about the $\hbar$ counting in WQFT, we would like to make two comments about the eikonal expansion.

The first comment is about causality. The original propagator in QFT has both
a particle pole and an anti-particle pole. However, under the eikonal
expansion, which is also the $\hbar$ expansion, only the particle pole is
kept. Thus, the propagator now has a fixed energy flow and the Feynman
description becomes equivalent to the retarded description. In other words,
causality emerges from the $\hbar \rightarrow 0$ limit. This might be relevent
for understanding the relation between the in-in formalism in worldline
EFT/WQFT and the scattering amplitude-based methods, such as KMOC and eikonal
method, which in principle only use Feynman propagators. There has been some
discussion about the relation between KMOC and worldline formalism in
\cite{Damgaard:2023vnx}.

{The second one is about the freedom of choosing different momenta as basis. As discussed in Section \ref{subsect:worldline-to-wqft}, one can either use average momenta $v_a$ or longitudinal momenta $p_a$ to perform the $\hbar$ expansion. The result will be a reshuffling of the expansion, as we saw earlier in Section  \ref{sect:wordline},  since $v_a$ and $p_a$ differ by a term of order $\mathcal O(\hbar)$, namely ${\hbar \bar{q}}/{2}$. Obviously, this reshuffling should not affect anything physical. However, once Fourier transformed into the impact parameter space, the two different bases correspond to slightly different impact parameters. Thus, in matching the results from different methods, one needs to be careful with these different choices of impact parameters. For example, the KMOC formalism and the eikonal method are shown to match in \cite{Georgoudis:2023eke} by taking into account their different choices of impact parameters.}

\subsection{$\hbar$ counting in WQFT}\label{app:hbar-counting}
 In the previous sections we showed how the 2-body QFT scattering amplitude is computed in   worldline formalism by allowing the background fields to become dynamical and integrating out the mediators between two worldlines. A general WQFT diagram is built from vertex operators (either linear or non-linear) whose contractions are evaluated on each  worldline, and  n-point connected diagrams of the mediators which  attach to the worldlines. The notion of ``connected WQFT diagram" will be very useful in our analysis and further discussion. By ``connected WQFT diagram", we mean that all the n-point connected mediator diagram are  linked to one another by contractions on one of the worldlines. In Section \ref{subsect:wqft-eikonal-exponentiation}, we will see that the connected WQFT diagrams correspond to irreducible terms of the amplitude in impact parameter
space, while disconnected WQFT diagrams correspond to reducible terms that are factorizable.

We will start with diagrams with only linear vertex operators and connected
tree diagrams of the mediators. Then we  address the issue of the non-linear vertex
operators  (which arise for gravitational interactions from the worldline counterterm as discussed in Section \ref{sect:wordline} ), and the
issue of the induced virtual loops.

For a given diagram,
$\hbar$ can either come from the coupling constants or from the background vertex operators {along with their corresponding worldline time integrals}. Each
background vertex operator has two factors: an exponential part $e^{i\bar k_j \cdot (v_a\tilde \tau_j+x(\tilde\tau_j))}$, which contributes to the denominators of
the final loop integral after worldline integrals, and a spin prefactor, which
directly contributes to the numerators. For example, the prefactor is 1 for scalar emissions from the worldline; it is $(v_a^\mu+\dot x^\mu)/\hbar$ for photons as in \eqref{vph}; and it is $[(v_a^\mu+\dot x^\mu)/\hbar][(v_a^\nu+\dot x^\nu)/\hbar]$ for gravitons, as in \eqref{vgrav}. For simplicity, let us call the
exponential part the ``denominator part" and the rest the ``spin part". Thus, in
all, the $\hbar$ factors have three sources: the coupling constants,  spin parts (with or without contractions), and integrals of denominator
parts (again with or without contractions). We will explain them one by one and assume the spin of the mediators is
$s$ ($s = 0, 1, 2$). 
{As we already know, each contraction between background vertex operators
brings an additional $\hbar$. Thus, an easy way to do the $\hbar$ counting is
to first count $\hbar$s from a corresponding uncontracted diagram, and then
multiply the additional $\hbar$s coming from contractions.}

First, consider the coupling constants. Let us assume classical couplings are
$g$ for scalar-Yukawa, $e$ for scalar-QED or scalar-QCD, and $G$ for scalar
interacting with gravity. Through matching the QFT tree amplitude with the
classical potential, we can get the corresponding quantum coupling constants:
$\frac{g \sqrt{m_1 m_2}}{\hbar^{3 / 2}}$ (for $s = 0$ mediators), $\frac{e}{\sqrt{\hbar}}$ (for $s
= 1$ photons) and $\sqrt{8 \pi \hbar G}$ (for $s = 2$ gravitons). Thus, each coupling constant
produces $\hbar^{- \frac{3}{2} + s}$. Second, consider the contribution from spin parts. For
scalar-Yukawa, the
spin part is 1 and there are no $\hbar$ factors. For scalar-QED or
scalar-QCD, 
the {uncontracted} spin part (through $v/\hbar$)  contributes a factor of 
${\hbar}^{-1}$. For scalars interacting gravitationally, 
the {uncontracted} spin part yields a factor of $\hbar^{-2}$. In all, each spin part contributes a factor of $\hbar^{- s}$. 
Lastly, each vertex operator is accompanied by an integral over $\tilde \tau$ which comes with an $\hbar$ factor as in \eqref{mnnew3}.
All these $\hbar$ factors arise from the leading order term of the worldline contraction expansion, the one which does not include any additional $x{-}x$ contractions.

Let us assume that the WQFT diagram has $2 N$ coupling constants in total and $n$ tree diagrams of mediators. Assuming each tree diagram has $m_i$ legs attaching to the worldlines, we have the following constraint\footnote{A connected tree diagram with $m$ external legs gives $m-2$ coupling constants. The $m$ vertices on worldlines produce another $m$ coupling constants.}:
\begin{eqnarray*}
  N & = & \left( \sum_{i = 1}^n m_i \right) - n
\end{eqnarray*}
Based on our analysis above, the $\hbar$ counting for the diagram without
contractions among background vertex operators is
\[ \left( \hbar^{- \frac{3}{2} + s} \right)^{2 N}  (\hbar^{- s})^{\sum_{i =
   1}^n m_i} \hbar^{\sum_{i = 1}^n m_i} {= \hbar^{(s - 2) N + (1 - s) n}}\,.  \]

Allowing now for  $x{-}x$ contractions, for a diagram with $2 N$ coupling constants, $n$ tree diagrams of
mediators, and $n_c$ contractions among background vertex operators, the
$\hbar$ counting is
\[ \hbar^{(s - 2) N + (1 - s) n + n_c} \,.\]
Now, let us analyze each case according to the different spin $s$ of the
mediators.

For the case of gravity, $s = 2$, the $\hbar$ counting is
\[ \hbar^{- n + n_c} . \]
Since the classical terms are of order $O (\hbar^{- 1})$,  a diagram with $n_c = n
- 1$ contractions will be a classical term. Similarly, a diagram with $n_c < n
- 1$ contractions will be {called }a superclassical term, while a diagram with $n_c > n - 1$
contractions will be a quantum term. Also, notice that in order to get a diagram where
all tree diagrams of the mediators are connected through contractions, one
needs at least $n_c = n - 1$ contractions. Thus, a connected WQFT diagram with
only linear vertex operators and tree diagrams of gravitons is either
classical or quantum.

For the case of scalar-QED or scalar-QCD, the $\hbar$ counting is
\[ \hbar^{- N + n_c} . \]
This is a rather interesting result. Based on this counting, a diagram with
$n_c = N - 1$ contractions is a classical term. Diagrams with $n_c < N - 1$
are superclassical and diagrams with $n_c > N - 1$ are quantum. The difference
between scalar-QED and scalar-QCD is that photons do not self-interact, unlike  gluons. For scalar-QED, the n-point connected diagrams
of photons can only be trivial photon propagators, which means $N = n$. Thus, a
connected WQFT diagram in scalar-QED with only tree diagrams of photons, just
like in the gravity case, is either classical or quantum. However, in the case of scalar-QCD,
the gluons can self-interact, which means $N \geqslant n$.
Thus, a connected WQFT diagram in scalar-QCD with only  n-point gluon tree diagrams
can also be superclassical. This is an obstruction to the exponentiation of the eikonal.

For the case of scalar-Yukawa, the $\hbar$ counting is
\[ \hbar^{- 2 N + n + n_c} . \]
Similar to the earlier discussion on scalar-QED and scalar-QCD, if the mediator
scalars cannot self-interact, which means $N = n$, a connected WQFT diagram
can be either classical or quantum. However, if the mediator scalars can self-interact, a connected WQFT diagram can be more singular and become superclassical.

This essentially implies that not all quantum theories are ``good" for taking
the classical limit\footnote{We are leaving here aside the question of whether the theory has a perturbative description in the IR, which already points out that $\lambda \phi^3$ theories or Yang-Mills theories are distinct from ``good" theories like electromagnetism or gravity.}. Based on the idea of exponentiation, a good theory should have all its superclassical contributions be reducible, and 
its connected WQFT diagrams be either classical or quantum. Thus,
scalar-Yukawa with self-interaction of mediators and scalar-QCD are ``bad"
theories for taking classical limit. On the other hand, scalar-Yukawa without
self-interaction of mediators, scalar-QED, and scalar-Gravity are all
potentially ``good" theories since their connected WQFT diagrams cannot be
superclassical.

Next, let us address the issue of non-linear vertex operators. These vertex operators\footnote{The leading order term in the expansion in the metric fluctuations is actually a linear vertex operator. Nonetheless, we still use the word ``non-linear" since this counterterm yields, in general, non-linear vertices. } come from expanding the so-called worldline counterterm, which is proportional to the pull-back of the Ricci scalar $R$.   When expanded in 
$h_{\mu \nu}$ fluctuations, it always contains two derivatives.
 Thus, the spin
part of the non-linear vertex operator only consists of polarization tensors
and products of two wavenumbers of specific gravitons. The exponential part of the non-linear vertex
operator is still simply $e^{i\bar k \cdot (v_a\tilde \tau+x)}$, with $\bar k$ being the sum
of the wavenumbers of those gravitons. In all, a non-linear vertex operator
only contributes a single $\hbar$ (from the rescaled $\tau\to\hbar \tilde \tau$ integral) to the $\hbar$ counting. For a WQFT diagram
with only tree diagrams of gravitons, if there are $n_{n.l.}$ non-linear
vertex operators with $m_{n.l.}$ legs in total, the $\hbar$ counting is
\[ \hbar^{- n + n_c + m_{n.l.} + n_{n.l.}} . \]
In other words, a connected WQFT diagram with non-linear vertex operators is always
quantum. 

Lastly, let us address the issue of induced loops in the propagator of the
mediators. A simple treatment would be staying in the soft region. This
matches with the intuitive picture that physics at large length scales should
only {depend on} the soft modes. In the soft region, there can not be
massive scalar loops since they only give vanishing contributions due to
scaleless integrals. Since the photons do not self-interact, there are no induced scalar loop contributions in the soft region. Next let us consider  gravitational interactions and the effect of graviton
loops. A general n-point connected diagram of the gravitons with $l$ loops
and $m$ external legs gives $m + 2 l - 2$ coupling constants. Thus, the constraint  for a WQFT
diagram with $2 N$ coupling constants in total and $n$ n-point connected
diagrams of gravitons,  assuming each n-point diagram has $l_i$ loops
inside and $m_i$ legs attaching to the worldlines, is
\begin{eqnarray*}
  N & = & \left( \sum_{i = 1}^n m_i \right) + l - n,
\end{eqnarray*}
where \(l\) is the total number of loops. The $\hbar$ counting for a diagram with
only linear vertex operators then becomes
\[ \hbar^{- n + l + n_c}\,. \]
Turning a linear graviton vertex into a non-linear one will only increase the $\hbar$ factors. 
Thus, a connected WQFT diagram with  virtual graviton loops  will always be quantum.

\subsection{The Exponentiation}\label{subsect:wqft-eikonal-exponentiation}

The key idea of the eikonal method is the exponentiation of the scattering
amplitude. We already saw in Section \ref{sect:eikonal-qft} examples from non-relativistic QM and
QFT. The idea of exponentiation states that, in the classical limit, the full
amplitude in impact parameter space has the following exponentiation
form\footnote{This is (4.1) in  \cite{DiVecchia:2023frv} with the eikonal phase rescaled by a factor of two.},
\begin{eqnarray}\label{expConj}
  1 + i \widetilde{\mathcal{A}} (s, b) & = & [1 +  i \Delta (s, b)] e^{i
  \delta} .
\end{eqnarray}
Order by order in the loop expansion, this is\footnote{This is (4.4)-(4.6) in \cite{DiVecchia:2023frv} where the eikonal phase is again rescaled by a factor of two.}
\begin{eqnarray}
  i \widetilde{\mathcal{A}}_0(b) & = &  i \delta_0\nonumber\\
  i \widetilde{\mathcal{A}}_1(b) & = & \frac{1}{2!}  ( i \delta_0)^2 + i
  \delta_1 +  i \Delta_1\nonumber\\
  i \widetilde{\mathcal{A}}_2(b) & = & \frac{1}{3!}  ( i \delta_0)^3 +  i
  \delta_0  i \delta_1 + [ i \delta_2 +  i \delta_0  i \Delta_1] +  i
  \Delta_2\label{expConjExpand}
\end{eqnarray}
where $\delta$ is the classical eikonal phase, and $\delta_i$ are classical contributions at a given loop order,  while $\Delta$ is the quantum contribution to the scattering amplitude and
$\Delta_i$ are quantum contributions at a given loop order.
The lesson here is that  the classical contributions $\delta_i$ can be obtained from the scattering amplitude recursively. The reason is the amplitude contains both
reducible contributions, such as $\frac{1}{3!}  ( i \delta_0)^3$, $ i
\delta_0  i \delta_1$ and $ i \delta_0  i \Delta_1$, and irreducible
contributions, such as $ i \delta_2$. The reducible contributions do not
contain new information about the interaction and need to be subtracted. For
example, to compute $\delta_2$, we have to know $\delta_0$, $\delta_1$ and
$\Delta_1$ beforehand and subtract those reducible contributions from $i
\widetilde{\mathcal{A}}_2(b)$, which is not convenient.

The exponentiation of the leading order is very well-known.
However, even the leading-order exponentiation is non-trivial in QFT. The reason  is that the exponentiation is not apparent
 in momentum space, where we usually work with the amplitude, and
becomes explicit only in the impact parameter space,  after several algebraic manipulations, as we reviewed in Section \ref{sect:eikonal-qm-qft}.

WQFT provides us a way to better understand the idea of exponentiation. With WQFT, we can actually investigate
and prove the exponentiation structure from first principles rather than treating it as an
assumption. We cover the details in a separate article \cite{ADRV1}. Here, we will give a brief overview of the logic and some results.
First, we will explain how the contraction expansion in WQFT,
which was recast as an $\hbar$ expansion, reveals the factorization of the amplitude. Then, we will discuss how this factorization leads to the exponentiation of the amplitude in the eikonal limit.

 With this knowledge of the factorization of the calculation in WQFT, let us consider the $2 \rightarrow 2$ scattering. {Similar to the one worldline case, we will use $\widetilde{\mathcal{A}}_L$ to denote the following combination,}
\be
\widetilde{\mathcal{A}}_L \equiv (2\pi)^2\delta \left(2 \frac{v_1}{\hbar} \cdot \bar{q} \right) \delta \left(
  2\frac{v_2}{\hbar} \cdot \bar{q} \right) \mathcal{A}_L.
\ee
{Notice the implicit dependence on the momentum transfer $\bar{q}$ in $\widetilde{\mathcal{A}}_L$. We will use $\widetilde{\mathcal{A}}_L(b)$ to denote the amplitude in the impact parameter space, which is}
\be
\widetilde{\mathcal{A}}_L(b) \equiv \int \frac{d^4 \bar{q}}{(2\pi)^4} e^{i b\cdot \bar{q}} \widetilde{\mathcal{A}}_L
\ee
A general expression for $\widetilde{\mathcal{A}}_L$ in worldline
formalism can be schematically represented as the following:
\begin{eqnarray}\label{wl2to2}
  \widetilde{\mathcal A}_L&\equiv&
  (2\pi)^2\delta \left( 2\frac{v_1}{\hbar} \cdot \bar{q} \right) \delta \left(2
  \frac{v_2}{\hbar} \cdot \bar{q} \right) \mathcal{A}_L \nonumber\\& = & \# \left( \prod_{i
  = 1}^{N_1} \int_{- \infty}^{ \infty} d \tilde\tau_i  \right) \left\langle \mathcal{T} \left\{
  \ldots \left(V_n^{(2)} V_{n - 1}^{(2)} \ldots V_{i + 1}^{(2)}\right) \left(V_i^{(1)} V_{i -
  1}^{(1)} \ldots V_1^{(1)}\right)\right\} \right\rangle  \nonumber\\
  &  & \times (\ldots T^{(2)} T^{(1)})\times (-i\lambda \hbar)^{N_1+N_2}\nonumber\\
  &  & \times \left( \prod_{j = 1}^{N_2} \int_{- \infty}^{ \infty} d \tilde\tau_j 
  \right)\left\langle \mathcal{T} \left\{ \ldots \left( V_m^{(2)} V_{m - 1}^{(2)}
  \ldots V_{j + 1}^{(2)}\right) \left(V_j^{(1)} V_{j - 1}^{(1)} \ldots V_1^{(1)}\right)\right\}
  \right\rangle\,.
\end{eqnarray}
Here, for simplicity, we only consider the case where all background vertex operators are linear, and $T$ represents a connected tree diagram of the mediators, which we will call a tree diagram hereafter. The superscripts $(1), (2)$ mark the
corresponding vertex operators connecting to the mediator tree diagram, and \#
represents the symmetry factor due to indistinguishable tree diagrams of the mediators. We also make the loop momentum integrals implicit. For a more general analysis, taking into account the various radiative corrections (with $T$ standing in for a connected n-point diagram of the mediators) and the possible non-linearities of the graviton vertex operators (which stem from the worldline counterterm $-R/8)$, please see \cite{ADRV1}.

Suppose that there are $N$ different kinds of tree diagrams, and that each kind has $c_i$ copies with $u_i$ legs attached to the first worldline and $d_i$ legs attached to the second worldline. We have $N_1=\sum_{i=1}^{n}c_i u_i$ and $N_2=\sum_{i=1}^{n}c_i d_i$. Then, the  symmetry factor in \eqref{wl2to2} is given by
\begin{equation} \label{eq-sym-factor}
    \#=\prod_{i=1}^N \frac{1}{c_i!(u_i!)^{c_i}(d_i!)^{c_i}}.
\end{equation}
Each $\widetilde {\mathcal A}_L$ amplitude will be written as a sum over worldline diagrams, which all are at the same loop order $L$. We are further classifying these worldline diagrams based on the structure of the connected mediator diagrams, and how they are linked to one another on the worldline.  

To illustrate this point, consider one worldline diagrammatic example with three connected mediator tree diagrams:
\be 
\includegraphics[valign=c]{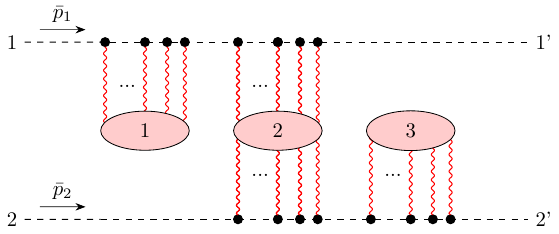}
\ee 
Highlighting only the connected mediator tree diagrams in the same WQFT diagram, we see: 
\be 
\includegraphics[valign=c]{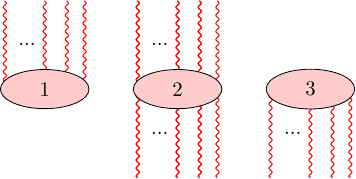}\label{photontreeparts}
\ee

Next,  we do the contraction expansion in \eqref{wl2to2}. That means performing the $x{-}x$ contraction of each background vertex operator associated with the external mediator lines in \eqref{photontreeparts}.  With each contraction being performed, we draw a solid line linking the external mediator lines in \eqref{photontreeparts}. In particular we need to  pay attention to the way
 the connected mediator tree diagrams are linked by these contractions.  
We will call these ``contractions between different trees" for simplicity.   Based on the factorization exhibited by a single worldline, one can intuitively see that a similar factorization will happen in the $2 \to 2$ scattering case. Through a careful combinatorial analysis in \cite{ADRV1}, we proved  the factorization
of a general WQFT diagram 
$\tilde {\mathcal A}_G$ 
\be \label{sep2to2}
\int \frac{d^4 \bar{q}}{(2 \pi)^4}  i\widetilde{\mathcal{A}}_G e^{i\bar q\cdot b}=
\qty[\prod_{j=1}^{n_G} \frac{1}{g_j!}]\prod_{k=1}^{N_G} 
\int \frac{d^4 \bar{q}_k}{(2 \pi)^4} i\widetilde{\mathcal{A}}_{G_k}e^{i\bar q_k\cdot b},
\ee
where $G=\bigcup_{k=1}^{N_G} G_k$. Here, the subscripts $G$ and $G_k$ do not represent the number of loops. Instead, $G_k$ represents a group of tree diagrams with no solid line links (contractions) between the various groups labeled $G_1, G_2,\dots$. We are allowing for the possibility that some of these groups are identical. 
$N_G$ is the number of groups of tree parts under the condition that there is no contraction among different groups and all tree parts within the same group are connected through contractions. $n_G$ is the number of distinct groups, and $g_j$ is the multiplicity for each distinct group. Obviously, $N_G=\sum_{j=1}^{n_G} g_j$. 
 For example, in \eqref{photontreeparts}, consider the three  tree diagrams labeled 1,2 and 3. If there are no links between these groups (i.e. the worldline is represented by a dotted line, without any solid lines, or the solid lines are only between the external legs of each tree), the worldline diagram factorizes into a product over the three worldline diagrams, each drawn with a dotted line and containing one mediator tree. In this case $N_G=3$.  If we draw a solid line between any of the external legs of tree 1 and tree 2, we factorize into a product of two worldline diagrams, so $N_G=2$. Since all three diagrams are distinct (which is clear by noticing how they attach to the worldlines), $g_j=1$ and $n_g=N_G$.
In~\cite{ADRV1}, we also show that ${\mathcal{A}}_{G_k}$ corresponding to WQFT diagrams with disconnected radiative corrections (i.e. radiative corrections with no momentum transferred between the worldlines, such as a self-energy diagram) vanish in dimensional regularization. Such diagrams are otherwise singular due to the $\delta(2 v_{1,2}\cdot \bar q)^2$ factors accompanying them. This is the case, for example,  if the tree labeled 3 in \eqref{photontreeparts} is not linked by a solid line to any of the other trees.
Based on ~\eqref{sep2to2}, we can classify the various terms in the  WQFT expansion in two classes: reducible
contributions and irreducible contributions. 
Diagrammatically, the irreducible contributions are connected $\widetilde {\mathcal A}_G$ WQFT diagrams, where all the
trees  are linked to one another through contractions. The reducible contributions are disconnected WQFT diagrams. We will see this classification matches beautifully with the one conjectured by the eikonal method of QFT.

According to the $\hbar$ counting done in Section \ref{app:hbar-counting},  in a ``good" theory for taking a classical limit, the connected WQFT diagrams are either classical or quantum, and the superclassical diagrams are factorizable (reducible). Let us use $\delta_L$ and $\Delta_L$ to denote the sum of the irreducible classical terms and the sum of the  irreducible quantum terms, respectively, where $L$ denotes the loop order. Using ~\eqref{sep2to2} we see that all the terms can be build from products of irreducible terms. In particular we can easily check that the lowest orders in the expansion of the scattering amplitude reproduce term-by-term the eikonal exponentiation conjecture \eqref{expConjExpand}:
\begin{align}
    i \widetilde{\mathcal{A}}_0(b) &= i\delta_0\nn
    i \widetilde{\mathcal{A}}_1(b) &= \frac{(i\delta_0)^2}{2!}+ i\delta_1 + i \Delta_1\nn
    i \widetilde{\mathcal{A}}_2(b) &= \frac{(i\delta_0)^3}{3!}+ (i \delta_0)(i \delta_1) + [i\delta_2 + (i \delta_0)(i \Delta^{(0)}_1)] + [(i \delta_0)(i \Delta_1 - i \Delta^{(0)}_1)+i \Delta_2],
\end{align} 
In the writing the two-loop amplitude $i\widetilde{\mathcal A}_2$, the first term on the right hand side is the maximally reducible WQFT diagram (with $G=G_1^3$) and of order $\hbar^{-3}$. The second term is also reducible (with $G=G_1 \times G_2$)  and of order $\hbar^{-2}$. The third group of terms, in between square brackets, contains the classical contribution $\delta_2$, which is the minimally connected WQFT diagram. At the same $\hbar^{-1}$ order,  we have the contribution of a reducible diagram, with $G=G_1\times G_2$, but with a contraction among the vertex operators that belong to the group $G_2$, which we represent by a solid line connecting two of the external legs of the tree group $G_2$.  We denoted this by $(i\delta_0)(i\Delta^{(0)}_1)$, where  $\Delta^{(0)}_1$ are terms of order $O(\hbar^0)$ in $\Delta_1$.  The next set of terms, again grouped within square brackets, are  the quantum contributions of order $\hbar^0$ and higher. Some of them are reducible, while others, such as $\Delta_2$ are irreducible. In particular $\Delta_2$ is a fully connected (by contractions on the worldline) diagram, but the number of contractions is not minimal.  
By summing up all the terms with the factorization relation \eqref{sep2to2}, one can prove the exponentiation \eqref{expConj} with the classical eikonal phase being identified with $\delta=\sum_L \delta_L$. Please see \cite{ADRV1} for details.

As a quick
example, let us show how the exponentiation of the leading eikonal, of which
the Feynman diagrams are generalized ladder diagrams, arises almost trivially
from the WQFT method. To show this clearly, let us start with the expression
involving only one worldline \eqref{mnnew3}.
The leading order corresponds to the term without any contraction: 
\begin{eqnarray}
  \widetilde{\mathcal{M}}_N (v_1, \{
  \bar{q} \}) & \xrightarrow{\text{leading order}} & \left( \prod_{i = 1}^N
  (-i\lambda\hbar )\int_{- \infty}^{ \infty} d \tilde\tau_i \left\langle \mathcal{} V_i \left( \bar{k}_i,
 {v_1}\tilde\tau_i + x_i \right) \right\rangle \right)\nonumber\\
  & = & \prod_{i = 1}^N \widetilde{\mathcal{M}}_1 (v_1, \bar{k}_i).
\end{eqnarray}
Thus, for the whole amplitude, we have
\begin{eqnarray}
  \int \frac{d^4 \bar{q}}{(2 \pi)^4} e^{i b \cdot \bar{q}} i\widetilde{\mathcal{A}}^{(-L)}_L & = & \frac{1}{(L+1)!} \left( \int \frac{d^4
  \bar{q}}{(2 \pi)^4} e^{i b \cdot \bar{q}}
  i\widetilde{\mathcal{A}}_0 \right)^{(L+1)}\,.
\end{eqnarray}Since the next step is the resummation of the leading eikonal terms at each loop order we have $i\widetilde{\mathcal  A}(b)\simeq i\sum_L \widetilde{\mathcal  A}^{(-L)}_L(b)=\exp(i\widetilde{\mathcal  A}_0(b))=\exp(i\delta_0)-1$,  
which concludes the proof of the exponentiation of the leading eikonal. 

\subsection{The Scattering Amplitude as the Generating Function of WQFT Diagrams}\label{ssec:amp-gen-WQFT}
In the previous discussion, we have shown that 
 there is a particular set of WQFT diagrams that yield the eikonal phase: diagrams that are irreducible,  minimally connected (by  $x{-}x$ contractions on the worldlines). The most straightforward way to calculate the eikonal phase would be to first handpick all the minimally connected WQFT diagrams, calculate them individually, and then sum all the contributions to get the final result. However,  the number of diagrams needed at higher loop order will increase quickly due to either a large number of possible contractions or complicated tree diagrams of mediators. Thus, finding a way to generate all the WQFT diagrams algorithmically could be helpful. It is not hard to see a good candidate is just the worldline diagrams  which yield the scattering amplitude at a specific loop order. On the one hand, the worldline diagrams  contain all the WQFT diagrams, including the irreducible diagrams. On the other hand, the number of worldline diagrams is much smaller than that of the WQFT diagrams which are needed at higher loop orders. The key then is to find a way to store the information on the contractions among the mediator trees within  the worldline diagram. This information will ultimately be used to select irreducible terms. 
 In other words,  the scattering amplitude in the worldline formalism at a specific loop order now serves as the generating function of  all the minimally connected WQFT diagrams at that loop order. Of course, the sum of the minimally connected WQFT diagrams is none other than the eikonal phase.
 
 Here, we will give an overview of this method, and we will use scalar-QED as an example to explain the details in the next section.

We already laid the foundation to implement this algorithm, in  \eqref{wl2to2}. Notice that we promoted the background vertex operators $V_i$ to $V_i^{(a)}$, where superscript $(a)$ marks the mediator tree to which that particular background quantum belongs. More specifically, we promote each $x^{\mu}(\tau_i)$ in $V_i$ to be $x^{\mu}_{(a)}(\tau_i)$, while $x^{\mu}(\tau_{\pm \infty})$ remain unchanged. We further promote the contractions
  $ \expval{x^{\mu}(\tau_i) x^{\nu}(\tau_j)} = -\frac{i}{2}\eta^{\mu\nu} |\tau_i-\tau_j|$
 to
\begin{align}
    \expval{x^{\mu}_{(a)}(\tau_i) x^{\nu}_{(b)}(\tau_j)} &= t_{ab}(-\frac{i}{2})\eta^{\mu\nu} |\tau_i-\tau_j|,
\end{align}
while any contraction involving $x^{\mu}(\tau_{\pm \infty})$ remain unchanged.  The coefficient $t_{ab}$ is a bean counter that contains the information of all the contractions among the mediator trees, and each WQFT diagram contains a polynomial of $t_{ab}$, which describes its contraction configuration. For example, $t_{ab}$ with $a$ and $b$ different indices, marks a solid line (an $x-x$ contraction) joining trees labeled $a$ and $b$. Likewise, $t_{aa}$ marks a contraction between vertex operators attached to the same tree labeled $a$. 
It is worth remembering that our goal is to extract the {\it minimally} connected irreducible diagrams. These diagrams will only contain $t_{ab}$ factors with $a\neq b$.
The scattering amplitude calculated from the worldline formalism  now includes these coefficients
\be
\widetilde{\mathcal{A}}(v_1, v_2, b) \rightarrow \widetilde{\mathcal{A}}(v_1, v_2, b, \{t_{ab}\}).
\ee
Notice the scattering amplitude itself can be recovered by setting all $t_{ab}=1$,
\be
\widetilde{\mathcal{A}}(v_1, v_2, b) =
\eval{\widetilde{\mathcal{A}}(v_1, v_2, b, \{t_{ab}\})}_{t_{ab}=1}.
\ee
The $\hbar$ expansion, or contraction expansion equivalently, is essentially expanding around $t_{ab}=0$. Let us now consider the polynomials of $t_{ab}$ corresponding to irreducible classical terms. The math problem is to find all the ways to form minimally connected diagrams from $N$ points, where $N$ is the number of mediator trees. The corresponding polynomials can be generated recursively,
\begin{align}
    P_1(a)=& 1 \nn
    P_2(a,b)=& t_{ab} \nn
    P_N(1,2,\dots N) =& (\text{1 connected subdiagram}) \text{ connected to } (\text{point $N$}) \nn
    & +(\text{2 connected subdiagrams}) \text{ separately connected to } (\text{point $N$}) \nn
    & +\dots \nn
    & +(\text{$N-1$ points}) \text{ separately connected to } (\text{point $N$}).
\end{align}
For example,
\begin{align}
    P_3(1,2,3) =& P(1,2) (t_{13}+t_{23})\nn
    & +P(1)P(2) t_{13} t_{23}\nn
    =& t_{12}t_{13}+t_{12}t_{23}+t_{13}t_{23}\nn
    P_4(1,2,3,4) =& P_3(1,2,3) (t_{14}+t_{24}+t_{34})\nn
    & +P_2(1,2)P_1(3) (t_{14}+t_{24}) t_{34}\nn
    & +P_2(1,3)P_1(2) (t_{14}+t_{34}) t_{24}\nn
    & +P_2(2,3)P_1(1) (t_{24}+t_{34}) t_{14}\nn
    & +P_1(1)P_1(2)P_1(3) t_{14}t_{24}t_{34}\,.
\end{align}
This also indicates that there will be a large number of terms in the  WQFT diagrams as we go to higher loop order. The last step is to sum up the coefficients of all the terms in the polynomial, while each coefficient can be calculated by taking derivatives. We represent the calculation schematically by
\be
\eval{\frac{\partial}{\partial P(\{t_{ab}\})} i \widetilde{\mathcal{A}}(v_1, v_2, b, \{t_{ab}\})}_{t_{ab}=0},
\ee
where the operation $\frac{\partial}{\partial P(\{t_{ab}\})}$ means the sum of partial derivatives of each term in the polynomial. For example,
\begin{align}
    \frac{\partial}{\partial P_2(1,2)} &= \frac{\partial}{\partial t_{12}}\nn
    \frac{\partial}{\partial P_3(1,2,3)} &= \qty(\frac{\partial^2}{\partial t_{12} \partial t_{13}}+\frac{\partial^2}{\partial t_{12} \partial t_{23}}+\frac{\partial^2}{\partial t_{13} \partial t_{23}}).
\end{align}
In conclusion, the eikonal phase, which is also the sum of the minimally connected WQFT diagrams, is generated through 
\be
i \delta_L = \eval{\frac{\partial}{\partial P(\{t_{ab}\})} i\widetilde{\mathcal{A}}(v_1, v_2, b, \{t_{ab}\})}_{t_{ab}=0}.
\ee
This method preserves the relation between the scattering amplitude and WQFT diagrams throughout the calculation. It shows how each piece in the scattering amplitude is mapped into WQFT diagrams at each loop order. Thus, calculations in the eikonal method and the WQFT method can be translated into each other seamlessly. We will see how this works out in more detail in the next section, where we will use scalar QED as an example.

\section{WQFT for Scalar QED}\label{sect:wqft-sQED}

In this section, we calculate the scattering amplitude for two body processes in scalar QED using the eikonal approximation WQFT method.  
Throughout, we shall attempt to make connections between the WQFT and conventional QFT calculations.
We relegate the details of the QFT calculation in detail in Appendix~\ref{app:sQED}, and we shall explain why WQFT can be used to effectively navigate the perturbation series if one is solely interested in classical/ soft contributions to the potential.

\subsection{Leading eikonal}\label{subsect:wqft-sQED-leading}
First, let us discuss the eikonal expansion for the tree level diagram.
The whole amplitude can be calculated as 
\begin{eqnarray}
i  \tilde {\mathcal{A}}_0 & = & \widetilde{\mathcal{M}}_1^{\mu}(p_1, p_1')  \frac{-i \eta_{\mu \nu}}{\bar{q}^2} 
  \widetilde{\mathcal{M}}_1^{\nu} (p_2, p_2')\nonumber\\
 & = & (-i\bar e)^2\frac{-i \bar v_1\cdot \bar v_2}{\bar q^2}\int_{-\infty}^\infty d\tau\,  e^{i \bar q\cdot \bar v_1\tau}
 \int_{-\infty}^\infty d\tau'  \,e^{-i \bar q\cdot \bar v_2\tau'}\nonumber\\
 &=&i \frac{e^2}{\hbar}  \frac{1}{ \bar q^2} \left(2 v_1 \cdot 2 v_2\right) \,(2\pi)^2 \,\delta(2 \bar q\cdot v_1)\delta(2\bar q\cdot v_2),
\end{eqnarray}
which gives the classical contribution. Note that the $1/\hbar$ dependence is predicated on the presence of the two delta functions. Stripping them off and writing the amplitude $\mathcal A_0 $ as opposed to $\tilde{\mathcal A}_0$ brings back the $1/\hbar^3$ scaling of the classical  scattering amplitude which we saw earlier in section \ref{sect:eikonal-qm-qft}.

\subsection{Subleading eikonal}\label{subsect:wqft-sQED-subleading}
We now move to the next order of perturbation, the subleading order of the eikonal expansion.
Since photons do not  self-interact, the inner tree structure is very
simple, consisting of simply two photon propagators. The QFT Feynman diagrams
include  ladder and cross-ladder diagrams, triangle diagrams, bubble diagrams,
self-energy diagrams, and vertex correction diagrams:
\ba 
&\includegraphics[width=\textwidth]{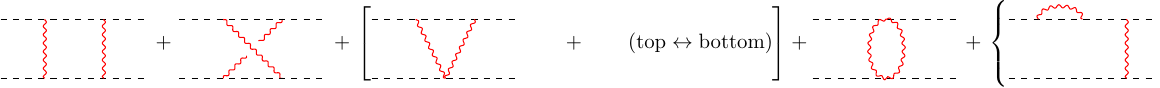}
\nonumber\\
&\includegraphics[width=\textwidth]{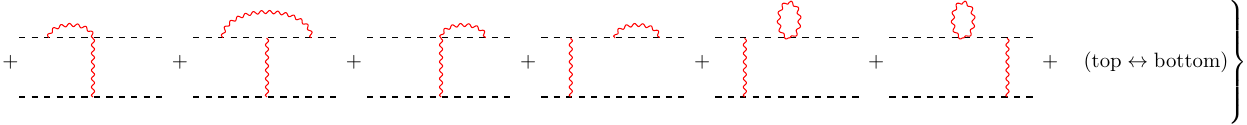}\nonumber\\
\ea 
In the worldline formalism, these are combined into three diagrams, since the
vertices are allowed to slide freely on the worldlines\\
\be
\includegraphics[]{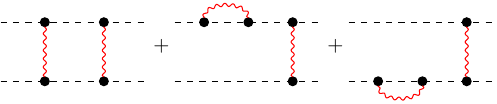}.
\ee
The dashed lines on the worldlines symbolize that we need to consider all worldline contractions (as opposed to the contraction expansion we perform when employing WQFT diagrams).

Let us first consider the conservative contribution to the 2-body scattering. The relevant diagrams are those where each photon propagator connects to both worldlines. The five corresponding Feynman diagrams include generalized ladder diagrams, triangle diagrams, and bubble diagrams, as analyzed in Appendix~\ref{app:sQED}. This is one of the advantages of the worldline formalism:
all the generalized ladder diagrams and triangle diagrams are contained within one
expression and represented by one worldline diagram given by
\begin{equation}
\begin{tikzpicture}[baseline=0cm,scale=0.5]
    \begin{feynhand}
       \vertex (b) at (-2.5,-1){2}; 
       \vertex (d) at (2.5,-1){2'}; 
       \vertex (a) at (-2.5,1){1}; 
       \vertex (c) at (2.5,1){1'};
       \vertex [dot](tr) at (1,1){};
       \vertex (tm) at (0,1);
       \vertex [dot](tl) at (-1,1){};
       \vertex [dot](br) at (1,-1){};
       \vertex (bm) at (0,-1);
       \vertex [dot](bl) at (-1,-1){};
       \propag[scalar] (a) to (c);
       \propag[scalar] (b) to (d);
       \propag[photon, red] (tr) to (br);
       \propag[photon, red] (tl) to (bl);
    \end{feynhand}
 \end{tikzpicture},
\end{equation}
which translates to an amplitude of the form
\begin{eqnarray}
i  \tilde{\mathcal{A}}_1 & = & \frac{1}{2!} \int \frac{d^4 \bar{k}_1}{(2 \pi)^4}  \widetilde{\mathcal{M}}_2^{\mu_1
  \mu_2} (p_1, p_1')  \frac{- i \eta_{\mu_1 \nu_1}}{\bar{k}_1^2}  \frac{-
  i \eta_{\mu_2 \nu_2}}{\bar{k}_2^2}  \widetilde{\mathcal{M}}_2^{\nu_1 \nu_2} (p_2, p_2')
 \,,
\end{eqnarray}
where we recall that the worldline amplitude
for a specific  worldline is given by \eqref{mnnew3} and the vertex operators are given in \eqref{vph}. Evaluating the correlation function of the two photon vertex  operators yields 
\begin{eqnarray}
  \widetilde{\mathcal{M}}_2^{\mu_1 \mu_2} (p_1, p_1') & = & - 2 e^2\hbar \int^{\infty}_{-
  \infty} d \tilde{\tau}_1 \int^{ \infty}_{-
  \infty} d \tilde{\tau}_2 \exp \left\{i v_1 \cdot \bar{k}_1  \tilde{\tau}_1 + i v_1 \cdot \bar{k}_2  \tilde{\tau}_2 +
  t_{12} \hbar (\bar{k}_2 \cdot \bar{k}_1) \left( \frac{i}{2} \right) |
  \tilde{\tau}_2 -\tilde\tau_1| \right\}\nonumber\\
  &  & \times \left( \frac{1}{\hbar} \right)^2 \left\{ \left( v_1 + \frac{t_{12}}{2}
  \hbar \bar{k}_1 \text{sgn} (\tilde{\tau}_2-\tilde\tau_1) \right)^{\mu_2} \left( v_1 -
  \frac{t_{12}}{2} \hbar \bar{k}_2 \text{sgn} (\tilde{\tau}_2-\tilde\tau_1) \right)^{\mu_1}
  \nonumber\right.\\
  & +&
  i t_{12} \hbar \delta (\tilde{\tau}_2-\tilde\tau_1) \eta^{\mu_2 \mu_1}  \big\}(2\pi)\delta(2\bar q\cdot  v_1) .
\end{eqnarray}
It is clear that the expansion of $t_{12}$ captures the expansion in
$\hbar$ as well. After we glue together both wordlines, we can also verify the terms that enter at linear in $t_{12}$ in the whole amplitude  
are at order $\frac{1}{\hbar}$, which is a classical contribution.

Let us analyze the worldline amplitude $\widetilde{\mathcal{M}}_2^{\mu_1 \mu_2} (p_1, p_1')$ in more detail to better clarify the relation between QFT eikonal calculation and WQFT method. On one hand, keeping contractions to all orders, we have
\begin{align}
  \widetilde{\mathcal{M}}_2^{\mu_1 \mu_2} (p_1, p_1') =& 2\left( \frac{-
  e^2}{\hbar} \right) \bigg[ \int^{+ \infty}_0 d \tilde{\tau}_2 \exp \left\{ i v_1
  \cdot \bar{k}_2  \tilde{\tau}_2 + t_{12} \hbar (\bar{k}_2 \cdot \bar{k}_1)
  \left( \frac{i}{2} \right) \tilde{\tau}_2 \right\} \left( v_1 -
  \frac{t_{12}}{2} \hbar \bar{k}_1 \right)^{\mu_2} \left( v_1 +
  \frac{t_{12}}{2} \hbar \bar{k}_2 \right)^{\mu_1}\nonumber\\
  & + 
  \int^0_{- \infty} d
  \tilde{\tau}_2 \exp \left\{ i v_1 \cdot \bar{k}_2  \tilde{\tau}_2 - t_{12} \hbar
  (\bar{k}_2 \cdot \bar{k}_1) \left( \frac{i}{2} \right) \tilde{\tau}_2 \right\}
  \left( v_1 + \frac{t_{12}}{2} \hbar \bar{k}_1 \right)^{\mu_2} \left( v_1 -
  \frac{t_{12}}{2} \hbar \bar{k}_2 \right)^{\mu_1}\nonumber\\
  & + 
  (i t_{12} \hbar) 
  \eta^{\mu_2
  \mu_1}\bigg](2\pi)\delta(2\bar q\cdot  v_1)\nonumber\\
  =& 2\left( \frac{- e^2}{\hbar} \right)\bigg[\ \left( v_1 + \frac{t_{12}}{2}
  \hbar \bar{k}_1 \right)^{\mu_2} \left( v_1 - \frac{t_{12}}{2} \hbar
  \bar{k}_2 \right)^{\mu_1}  \frac{(- 1)}{i v_1 \cdot \bar{k}_2 + t_{12} \hbar
  (\bar{k}_2 \cdot \bar{k}_1) \left( \frac{i}{2} \right) - \epsilon}\nonumber\\
  & + 
  \left( v_1 - \frac{t_{12}}{2}
  \hbar \bar{k}_1 \right)^{\mu_2} \left( v_1 + \frac{t_{12}}{2} \hbar
  \bar{k}_2 \right)^{\mu_1}  \frac{1}{i v_1 \cdot \bar{k}_2 - t_{12} \hbar
  (\bar{k}_2 \cdot \bar{k}_1) \left( \frac{i}{2} \right) + \epsilon}\nonumber\\
  & + 
  (t_{12} \hbar) \eta^{\mu_2
  \mu_1} \bigg](2\pi)\delta(2\bar q\cdot \bar v_1).
\end{align}
where we made use of the implicit $\epsilon$ regulators\footnote{Recall that this is just the $i \epsilon$
Feynman propagator prescription in QFT.}. We can see that this is indeed the ``Compton scattering"
for scalar QED. The first two terms which come from
$\int^{+ \infty}_0 d \tilde{\tau}$ and $\int_{- \infty}^0 d \tilde{\tau}$
correspond to two Feynman diagrams with different pole structures, while the
last term 
corresponds to the diagram with a quartic vertex.

On the other hand, let us calculate $\widetilde{\mathcal{M}}_2^{\mu_1 \mu_2} (p_1,
p_1')$ order by order in $t_{12}$(or $\hbar$, equivalently),
\ba
    \widetilde{\mathcal{M}}_2^{\mu_1 \mu_2} (p_1, p_1') &=& \bigg\{4\left( \frac{- i e^2}{\hbar^2} \right) v_1^{\mu_1} v_1^{\mu_2} (2\pi) \delta
  (2 v_1 \cdot \bar{k}_2)+ 2\left( \frac{-
  e^2}{\hbar^2} \right) \left[v_1^{\mu_2}  (t_{12} \hbar)  \bar{k}_2^{\mu_1} 
  \frac{1}{i v_1 \cdot \bar{k}_2}- v_1^{\mu_1}  (t_{12} \hbar) 
  \bar{k}_1^{\mu_2}  \frac{1}{i v_1 \cdot \bar{k}_2}\right.\nonumber
  \\ &&\left.+  v_1^{\mu_1} v_1^{\mu_2} 
  \frac{i t_{12} \hbar (\bar{k}_2 \cdot \bar{k}_1)}{(i v_1 \cdot \bar{k}_2)^2}+   (i t_{12} \hbar) \eta^{\mu_2
  \mu_1}\right] + \dots\bigg\}(2\pi)\delta(2  v_1\cdot \bar q)
\ea
This order-by-order calculation is essentially the WQFT method, which can be
represented diagrammatically as
\begin{equation}
 \includegraphics[valign=c]{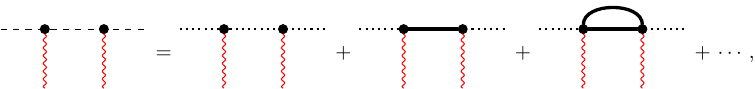}
\end{equation}
where 
\begin{equation}\label{eqn-definitionofwldiagramnocontraction}
  \begin{tikzpicture}[baseline=0cm,scale=0.5]
    \begin{feynhand}
       \vertex (b) at (-2.5,-1); 
       \vertex (d) at (2.5,-1); 
       \vertex (a) at (-2.5,1); 
       \vertex (c) at (2.5,1);
       \vertex [dot](tr) at (1,1){};
       \vertex (tm) at (0,1);
       \vertex [dot](tl) at (-1,1){};
       \vertex (br) at (1,-1);
       \vertex (bm) at (0,-1);
       \vertex (bl) at (-1,-1);
       \propag[ghost] (a) to (c);
       \propag[photon, red] (tr) to (br);
       \propag[photon, red] (tl) to (bl);
    \end{feynhand}
 \end{tikzpicture} =4[2\pi\delta(2v_2\cdot \bar q)] \left( \frac{- i e^2}{\hbar^2} \right) v_1^{\mu_1} v_1^{\mu_2} (2\pi) \delta
  (v_1 \cdot \bar{k}_2)
\end{equation}
and
\ba
\begin{tikzpicture}[baseline=0cm,scale=0.5]
    \begin{feynhand}
       \vertex (b) at (-2.5,-1); 
       \vertex (d) at (2.5,-1); 
       \vertex (a) at (-2.5,1); 
       \vertex (c) at (2.5,1);
       \vertex [dot](tr) at (1,1){};
       \vertex (tm) at (0,1);
       \vertex [dot](tl) at (-1,1){};
       \vertex (br) at (1,-1);
       \vertex (bm) at (0,-1);
       \vertex (bl) at (-1,-1);
       \propag[ghost] (a) to (c);
       \propag[photon, red] (tr) to (br);
       \propag[photon, red] (tl) to (bl);
        \propag[plain, ultra thick, black] (tl) to (tr);
    \end{feynhand}
 \end{tikzpicture} &=& 
    2[2\pi\delta(2v_2\cdot \bar q)] \left( \frac{-
  e^2}{\hbar^2} \right) \left[v_1^{\mu_2}  (t_{12} \hbar)  \bar{k}_2^{\mu_1} 
  \frac{1}{i v_1 \cdot \bar{k}_2}- v_1^{\mu_1}  (t_{12} \hbar) 
  \bar{k}_1^{\mu_2}  \frac{1}{i v_1 \cdot \bar{k}_2}\right.
   \nonumber\\&&\left.+  v_1^{\mu_1} v_1^{\mu_2} 
  \frac{i t_{12} \hbar (\bar{k}_2 \cdot \bar{k}_1)}{(i v_1 \cdot \bar{k}_2)^2}+   (i t_{12} \hbar) \eta^{\mu_2
  \mu_1}\right].
\ea
Notice that the delta function in the leading order term in \eqref{eqn-definitionofwldiagramnocontraction} is a combination of the two pole structures coming from the integration domains 
$\int_{- \infty}^0 d \bar{\tau}$ and $\int_0^{+ \infty} d \bar{\tau}$, respectively. In QFT,
 this delta function structure, which is crucial for the exponentiation
of the amplitude, is hidden. To see it, one needs to expand the propagator, symmetrize the integrand
and utilize the relation
 $ \frac{1}{x - i \epsilon} - \frac{1}{x + i \epsilon} 
 \to
 (2 \pi
  i) \delta (x),$
as we reviewed in \ref{sect:eikonal-qft}.
However, in WQFT, the delta function is so obvious that the exponentiation
looks almost trivial. When a background  vertex operator only contracts with
the asymptotic in-out vertex operators, it immediately yields the delta function, as a result of the overall integral factor $\int_{-
\infty}^{+ \infty} d \tilde{\tau} e^{i v \cdot \bar{k}  \tilde{\tau}} = (2\pi) \delta (v
\cdot \bar{k})$.

With each contraction between the  vertex operators, we are expanding 
 one order higher in $\hbar$, corresponding to the eikonal expansion
in QFT. More specifically, there are two $x (\tau)$s in each interaction
vertex operator $\epsilon \cdot \dot{x} e^{i \bar{k} \cdot x}$, which leads to
three types of contractions that correspond to three different sources of
$\hbar$ expansion in QFT. The $\langle x (\tau) x (\tau') \rangle$ coming from
contracting exponentials $e^{i \bar{k} \cdot x}$ corresponds to expanding the
propagator in QFT. The $\langle \dot{x} (\tau) x (\tau') \rangle$ coming from
contracting $\epsilon \cdot \dot{x}$ and $e^{i \bar{k} \cdot x}$ corresponds
to expanding the numerator in QFT. Lastly, the $\langle \dot{x} (\tau) \dot{x}
(\tau') \rangle$ coming from contracting two $\epsilon \cdot \dot{x}$
corresponds to the quartic vertex in QFT.

The calculation is trivial at one-loop order, and there is no need to use $t_{12}$  since all classical contributions
are irreducible. In other words, we have
\begin{eqnarray}
  \eval{\left( \frac{\partial}{\partial t_{12}}  \widetilde{\mathcal{A}}_1 \right)}_{t_{12} = 0} & = & \eval{\widetilde{\mathcal{A}}_1}_{\text{terms of order
  $1 / \hbar$}}.
\end{eqnarray}
However, when we get to higher order, e.g. two-loop order, the classical contributions correspond to terms consisting of two
$\langle x (\tau) x (\tau') \rangle$ contractions, where the use of $t_{ab}$ will be crucial for us to correctly select all irreducible terms. In other words, we have
\begin{eqnarray}
  \eval{\left( \left( \frac{\partial^2}{\partial t_{12} \partial t_{13}} +
  \frac{\partial^2}{\partial t_{12} \partial t_{23}} +
  \frac{\partial^2}{\partial t_{13} \partial t_{23}} \right) \widetilde{\mathcal{A}}_2
  \right)}_{t_{i j} = 0}  & \neq & \eval{\widetilde{\mathcal{A}}_2}_{\text{terms of order $1 / \hbar$}}.
\end{eqnarray}
We will show this in detail in the next subsection.

Next, let us consider other diagrams that involve radiative
corrections. These correspond to fourteen Feynman diagrams that have photon
propagators  connecting to one worldline only. In worldline formalism, they
are represented by two worldline diagrams
\begin{equation}
    \begin{tikzpicture}[baseline=0cm,scale=0.5]
    \begin{feynhand}
       \vertex (b) at (-2.5,-1); 
       \vertex (d) at (2.5,-1); 
       \vertex (a) at (-2.5,1); 
       \vertex (c) at (2.5,1);
       \vertex (tr) at (1,1);
       \vertex [dot](tm) at (0,1){};
       \vertex (tl) at (-1,1);
       \vertex (br) at (1,-1);
       \vertex (bm) at (0,-1);
       \vertex (bl) at (-1,-1);
       \vertex [dot](tsr) at (1.5, 1){};
       \vertex [dot](tsl) at (-1.5, 1){};
       \vertex [dot](bsr) at (1.5, -1){};
       \vertex (bsl) at (-1.5, -1);
       \propag[scalar] (a) to (c);
       \propag[scalar] (b) to (d);
       \propag[photon, red] (tsr) to (bsr);
       \propag[photon, red] (tsl) to [in=90, out=90] (tm);
    \end{feynhand}
 \end{tikzpicture} + \begin{tikzpicture}[baseline=0cm,scale=0.5]
    \begin{feynhand}
       \vertex (b) at (-2.5,-1); 
       \vertex (d) at (2.5,-1); 
       \vertex (a) at (-2.5,1); 
       \vertex (c) at (2.5,1);
       \vertex (tr) at (1,1);
       \vertex (tm) at (0,1);
       \vertex (tl) at (-1,1);
       \vertex (br) at (1,-1);
       \vertex [dot] (bm) at (0,-1){};
       \vertex (bl) at (-1,-1);
       \vertex [dot] (tsr) at (1.5, 1){};
       \vertex [dot] (tsl) at (-1.5, 1);
       \vertex [dot] (bsr) at (1.5, -1){};
       \vertex [dot] (bsl) at (-1.5, -1){};
       \propag[scalar] (a) to (c);
       \propag[scalar] (b) to (d);
       \propag[photon, red] (tsr) to (bsr);
       \propag[photon, red] (bsl) to [in=270, out=270] (bm);
    \end{feynhand}
 \end{tikzpicture}.
\end{equation}
We will show that these diagrams (which contribute to self-energy diagrams and vertex corrections) vanish in the soft region. 
 In \cite{ADRV1} we gave a  general argument that the pure radiative corrections obtained after factorization vanish since they reduce to scaleless integrals. Here, we give an example to show the details explicitly.

Due to symmetry, it is sufficient to only consider the first diagram. Let us focus on the upper worldline. Its contribution to the worldline diagram, $\widetilde{\mathcal{M}}_3^{\mu} (v_1, q)$, has a contraction expansion that diagrammatically looks like
\begin{equation}
    \begin{tikzpicture}[baseline=0cm,scale=0.5]
    \begin{feynhand}
       \vertex (b) at (-2.5,-1); 
       \vertex (d) at (2.5,-1); 
       \vertex (a) at (-2.5,1); 
       \vertex (c) at (2.5,1);
       \vertex (tr) at (1,1);
       \vertex [dot](tm) at (0,1){};
       \vertex (tl) at (-1,1);
       \vertex (br) at (1,-1);
       \vertex (bm) at (0,-1);
       \vertex (bl) at (-1,-1);
       \vertex [dot](tsr) at (1.5, 1){};
       \vertex [dot](tsl) at (-1.5, 1){};
       \vertex (bsr) at (1.5, -1);
       \vertex (bsl) at (-1.5, -1);
       \propag[scalar] (a) to (c);
       \propag[photon, red] (tsr) to (bsr);
       \propag[photon, red] (tsl) to [in=90, out=90] (tm);
    \end{feynhand}
 \end{tikzpicture} = \begin{tikzpicture}[baseline=0cm,scale=0.5]
    \begin{feynhand}
       \vertex (b) at (-2.5,-1); 
       \vertex (d) at (2.5,-1); 
       \vertex (a) at (-2.5,1); 
       \vertex (c) at (2.5,1);
       \vertex (tr) at (1,1);
       \vertex [dot](tm) at (0,1){};
       \vertex (tl) at (-1,1);
       \vertex (br) at (1,-1);
       \vertex (bm) at (0,-1);
       \vertex (bl) at (-1,-1);
       \vertex [dot](tsr) at (1.5, 1){};
       \vertex [dot](tsl) at (-1.5, 1){};
       \vertex (bsr) at (1.5, -1);
       \vertex (bsl) at (-1.5, -1);
       \propag[ghost] (a) to (c);
       \propag[photon, red] (tsr) to (bsr);
       \propag[photon, red] (tsl) to [in=90, out=90] (tm);
    \end{feynhand}
 \end{tikzpicture} + \qty[\begin{tikzpicture}[baseline=0cm,scale=0.5]
    \begin{feynhand}
       \vertex (b) at (-2.5,-1); 
       \vertex (d) at (2.5,-1); 
       \vertex (a) at (-2.5,1); 
       \vertex (c) at (2.5,1);
       \vertex (tr) at (1,1);
       \vertex [dot](tm) at (0,1){};
       \vertex (tl) at (-1,1);
       \vertex (br) at (1,-1);
       \vertex (bm) at (0,-1);
       \vertex (bl) at (-1,-1);
       \vertex [dot](tsr) at (1.5, 1){};
       \vertex [dot](tsl) at (-1.5, 1){};
       \vertex (bsr) at (1.5, -1);
       \vertex (bsl) at (-1.5, -1);
       \propag[ghost] (a) to (c);
       \propag[photon, red] (tsr) to (bsr);
       \propag[photon, red] (tsl) to [in=90, out=90] (tm);
       \propag[plain, ultra thick] (tm) to (tsr);
    \end{feynhand}
 \end{tikzpicture} + \begin{tikzpicture}[baseline=0cm,scale=0.5]
    \begin{feynhand}
       \vertex (b) at (-2.5,-1); 
       \vertex (d) at (2.5,-1); 
       \vertex (a) at (-2.5,1); 
       \vertex (c) at (2.5,1);
       \vertex (tr) at (1,1);
       \vertex [dot](tm) at (0,1){};
       \vertex (tl) at (-1,1);
       \vertex (br) at (1,-1);
       \vertex (bm) at (0,-1);
       \vertex (bl) at (-1,-1);
       \vertex [dot](tsr) at (1.5, 1){};
       \vertex [dot](tsl) at (-1.5, 1){};
       \vertex (bsr) at (1.5, -1);
       \vertex (bsl) at (-1.5, -1);
       \propag[ghost] (a) to (c);
       \propag[photon, red] (tsr) to (bsr);
       \propag[photon, red] (tsl) to [in=90, out=90] (tm);
       \propag[plain, ultra thick] (tm) to (tsl);
    \end{feynhand}
 \end{tikzpicture}] + \cdots
\label{V15}
\end{equation}
One can easily see that, once combined with $(2\pi)\delta(2 \frac{v_1}{\hbar}\cdot \bar{q})$, the first and third terms with no contraction between the ladder rung and the radiative photon loop admit factorization as in \eqref{fac-one-worldine}. Let us write down the mathematical expression these   diagrams. The leading order in $\hbar$ diagram/term is
\begin{equation}
    \begin{aligned}
        \begin{tikzpicture}[baseline=0cm,scale=0.5]
    \begin{feynhand}
       \vertex (b) at (-2.5,-1); 
       \vertex (d) at (2.5,-1); 
       \vertex (a) at (-2.5,1); 
       \vertex (c) at (2.5,1);
       \vertex (tr) at (1,1);
       \vertex [dot](tm) at (0,1){};
       \vertex (tl) at (-1,1);
       \vertex (br) at (1,-1);
       \vertex (bm) at (0,-1);
       \vertex (bl) at (-1,-1);
       \vertex [dot](tsr) at (1.5, 1){};
       \vertex [dot](tsl) at (-1.5, 1){};
       \vertex (bsr) at (1.5, -1);
       \vertex (bsl) at (-1.5, -1);
       \propag[ghost] (a) to (c);
       \propag[photon, red] (tsr) to (bsr);
       \propag[photon, red] (tsl) to [in=90, out=90] (tm);
    \end{feynhand}
 \end{tikzpicture} &= \int \frac{d^4 \bar{l}}{(2\pi)^4} \frac{-\bar e^3 \hbar^3}{\bar{l}^2-i \epsilon} v_1^2 v_1^{\mu} \int_{-\infty}^{\infty} d\tilde\tau_1\int_{-\infty}^{+\infty}  d\tilde\tau_2 \int_{-\infty}^{+\infty}d\tilde\tau_3
    \exp{i v_1 \cdot \bar{l} \tau_3 -i v_1 \cdot \bar{l}\tau_2 + i v_1 \cdot \bar{q}\tau_1}\nonumber\\
    & =  \bigg[\delta(0) (2\pi i \bar{e})^2 v_1^2 \int \frac{d^4 \bar{l}}{(2\pi)^4} \frac{-i}{\bar{l}^2-i \epsilon}
    \delta(v_1 \cdot l)\bigg] \times \bigg[ (2\pi i \bar{e}) v_1^{\mu} \delta( v_1\cdot \bar{q})\bigg]
    \end{aligned}
\end{equation}
and the subleading reducible diagram/term  is 
\ba
 \begin{tikzpicture}[baseline=0cm,scale=0.5]
    \begin{feynhand}
       \vertex (b) at (-2.5,-1); 
       \vertex (d) at (2.5,-1); 
       \vertex (a) at (-2.5,1); 
       \vertex (c) at (2.5,1);
       \vertex (tr) at (1,1);
       \vertex [dot](tm) at (0,1){};
       \vertex (tl) at (-1,1);
       \vertex (br) at (1,-1);
       \vertex (bm) at (0,-1);
       \vertex (bl) at (-1,-1);
       \vertex [dot](tsr) at (1.5, 1){};
       \vertex [dot](tsl) at (-1.5, 1){};
       \vertex (bsr) at (1.5, -1);
       \vertex (bsl) at (-1.5, -1);
       \propag[ghost] (a) to (c);
       \propag[photon, red] (tsr) to (bsr);
       \propag[photon, red] (tsl) to [in=90, out=90] (tm);
       \propag[plain, ultra thick] (tm) to (tsl);
    \end{feynhand}
 \end{tikzpicture} &=&
    \left[(2\pi i \bar{e}) v_1^{\mu} \delta(v_1\cdot \bar{q})\right] \left[\delta(0) (2\pi) (i \bar{e})^2\right]  \nonumber\\ && 
    \times\quad \bigg\{\left[\int \frac{d^4 \bar{l}}{(2\pi)^4} \frac{-i}{\bar{l}^2-i \epsilon} \frac{2 i \hbar v_1 \cdot \bar{l}}{(v_1\cdot\bar{l}+i\epsilon)(v_1\cdot\bar{l}-i\epsilon)}
   \right.\nonumber\\
   &&+ \left.
   \int \frac{d^4 \bar{l}}{(2\pi)^4} \frac{-i}{\bar{l}^2-i \epsilon} \frac{i \hbar (v_1 \cdot \bar{l})^2 \bar{l}^2}{(v_1\cdot\bar{l}+i\epsilon)^2(v_1\cdot\bar{l}-i\epsilon)^2}
    \right] 
    \nonumber\\
    & & 
    \quad +  \int \frac{d^4 \bar{l}}{(2\pi)^4} \frac{-i}{\bar{l}^2-i \epsilon} (4i\hbar)
    \bigg\} \label{1loop-1}
\ea
We see that, as advertised, in the first and third terms \eqref{V15} the integrals over the loop momentum $\bar{l}$ integrals are scaleless integrals and vanish in dimensional regularization.

Now, we are only left with the second term in \eqref{V15}, the connected WQFT diagram.
  Of course, we can still follow the same procedure by writing down the expression explicitly and noticing that the $\bar{l}$ integral is scaleless. However, we will take a shortcut here because it leads to a conclusion that is valid for any higher order eikonal. The shortcut is to think about possible forms of the denominator coming from worldline time integrals. Alternatively, one can think about possible denominators coming from the Feynman diagrams after the eikonal expansion. Under the eikonal expansion, if we ignore all the $i \epsilon$, the denominator could only be some products of $v_1 \cdot (\pm\bar{l})$ and $v_1 \cdot (\pm\bar{l}\pm\bar{q})$. However, the two are the same due to the on-shell condition $v_1 \cdot \bar{q}=0$. Thus, the denominator scales homogeneously under a $\bar{l}$ rescaling, which then leads to a scaleless integral.

This can be easily generalized to any higher order eikonal. For any higher order eikonal, the most general building block for the denominator is $v_1 \cdot (\sum_i (\pm \bar{l}_i)+\sum_j (\pm \bar{q}_j))$, where $l_i$ represents some loop momenta for radiative corrections and $q_j$ represents some exchange momenta between the two worldlines. In general, the existence of multiple exchange momenta $q_j$ prevents the homogeneous scaling of the denominator  under  $\bar{l}_i$ rescaling. However, if only one momentum exchange $q$ exists between the two worldlines, then the on-shell condition $v_1 \cdot \bar{q}=0$ essentially eliminates the $\bar{q}$ dependence in the denominator, which becomes a homogeneous function of the loop momentum, and the integral is zero in dimensional regularization. Thus, we get a simple conclusion. One needs diagrams with at least two exchange momenta in order for the radiative corrections to contribute to the  classical scattering amplitude.
This argument also holds for scalars interacting gravitationally.

\subsection{Subsubleading eikonal}\label{subsect:wqft-sQED-subsubleading}
Just as in the subleading eikonal calculation, let us first
consider the case  the conservative part of the scattering amplitude. The relevant diagrams are those where each photon propagator connects to
both worldlines. The single worldline diagram which sums up all the two loop ladders and cross ladders is
\begin{equation}
    \begin{tikzpicture}[baseline=0cm]
    \begin{feynhand}
       \vertex (b) at (-2.5,-1){2}; 
       \vertex (d) at (2.5,-1){2'}; 
       \vertex (a) at (-2.5,1){1}; 
       \vertex (c) at (2.5,1){1'};
       \vertex [dot](tr) at (1.3,1){};
       \vertex [dot](tm) at (0,1){};
       \vertex [dot](tl) at (-1.3,1){};
       \vertex [dot](br) at (1.3,-1){};
       \vertex [dot](bm) at (0,-1){};
       \vertex [dot](bl) at (-1.3,-1){};
       \propag[scalar] (a) to (c);
       \propag[scalar] (b) to (d);
       \propag[photon, red] (tr) to (br);
       \propag[photon, red] (tl) to (bl);
       \propag[photon, red] (tm) to (bm);
    \end{feynhand}
 \end{tikzpicture}\,,
\end{equation}
and has the mathematical expression
\begin{align} \label{amp-qed-2loop}
i  \widetilde{\mathcal{A}}_2 =& \frac{1}{3!} \int \frac{d^4 \bar{k}_1}{(2 \pi)^4}  \int \frac{d^4
  \bar{k}_2}{(2 \pi)^4}  \widetilde{\mathcal{M}}_3^{\mu_1 \mu_2 \mu_3} (p_1, p_1') 
  \frac{- i \eta_{\mu_1 \nu_1}}{\bar{k}_1^2}  \frac{- i \eta_{\mu_2
  \nu_2}}{\bar{k}_2^2}  \frac{- i \eta_{\mu_3 \nu_3}}{\bar{k}_3^2}
  \widetilde{\mathcal{M}}_3^{\nu_1 \nu_2 \nu_3} (p_2, p_2')
\end{align}
where we have implicitly used $\bar{k}_3 = \bar{q} - \bar{k}_1 - \bar{k}_2$.
In evaluating this expression, we will mark the photon trees  with indices $(1), (2)$ and $(3)$ and use $t_{ab}$ to mark the worldline contractions of the vertex operators which connect to a particular tree. 
Then, after we obtained the amplitude 
\[ \widetilde{\mathcal{A}}_2 = \widetilde{\mathcal{A}}_2 (v_1, v_2, \bar{q}, t_{12}, t_{13}, t_{23})  \]
we proceed to expand in terms of $\hbar$, as one would do in the QFT eikonal method
\begin{eqnarray}
  \widetilde{\mathcal{A}}_2 & = & \widetilde{\mathcal{A}}_2^{(-2)} + \widetilde{\mathcal{A}}_2^{(-1)} + \widetilde{\mathcal{A}}_2^{(0)}
  + \ldots,\\
  \widetilde{\mathcal{A}}_2^{(-2)} & = & \eval{\widetilde{\mathcal{A}}_2}_{\text{terms of order $1/ \hbar^3$}}= \widetilde{\mathcal{A}}_2 (v_1, v_2, \bar{q}, 0, 0, 0),\\
  \widetilde{\mathcal{A}}_2^{(-1)} & = & \eval{\widetilde{\mathcal{A}}_2}_{\text{terms of order $1/ \hbar^2$}}= \sum_{3 \geqslant b > a \geqslant 1} \frac{\partial
  \widetilde{\mathcal{A}}_2}{\partial t_{ab}} (v_1, v_2, \bar{q}, 0, 0, 0),\\
  \widetilde{\mathcal{A}}_2^{(0)} & = & \eval{\widetilde{\mathcal{A}}_2}_{\text{terms of order $1/
  \hbar$}}= \sum_{3 \geqslant b > a \geqslant 1} \sum_{3 \geqslant d > c
  \geqslant 1} \frac{\partial^2 \widetilde{\mathcal{A}}_2}{\partial t_{ab} \partial t_{cd}} (v_1, v_2, \bar{q}, 0, 0, 0).
\end{eqnarray}
While $\mathcal{A}^{(0)}_2$  contains the classical contributions,
it also contains reducible contributions that ought to be
dropped. The most straightforward way to get rid of all the irreducible
contributions is to compute the amplitude $\mathcal{A}$ recusively in QFT, as detailed in \cite{DiVecchia:2023frv}.
However, using the worldline formalism, the  reducible and irreducible terms are captured by the
structure of $\langle x_a x_b \rangle$ contractions, through the 
the polynomials of $t_{ab}$, and we can extract those irreducible contributions directly. In
this subsubleading order, the polynomials representing irreducible
contributions are
\begin{eqnarray*}
  P_3 (1, 2, 3) & = & t_{12} t_{13} + t_{12} t_{23} + t_{13} t_{23} .
\end{eqnarray*}
Thus, the classical contribution is
\begin{eqnarray*}
  &  & i \eval{\left( \frac{\partial^2 \widetilde{\mathcal{A}}_2}{\partial t_{12} \partial t_{13}}
  + \frac{\partial^2 \widetilde{\mathcal{A}}_2}{\partial t_{12} \partial t_{23}} +
  \frac{\partial^2 \widetilde{\mathcal{A}}_2}{\partial t_{13} \partial t_{23}} \right)}_{t_{i j} = 0}
\end{eqnarray*}
Let us first explain how this expression directly relates to the diagrammatic method of WQFT in \cite{Mogull:2020sak}. For example, the term
$\eval{\frac{\partial^2 (i \widetilde{\mathcal{A}}_2)}{\partial t_{12} \partial t_{13}}}_{t_{i j} = 0}$ corresponds to the sum of following
diagrams of WQFT
\begin{equation}
    \begin{tikzpicture}[baseline=0cm,scale=0.5]
    \begin{feynhand}
       \vertex (T1) at (-1.6, 0){1};
       \vertex (T2) at (-0.3, 0){2};
       \vertex (T3) at (1.0, 0){3};
    
       \vertex (b) at (-2.5,-1); 
       \vertex (d) at (2.5,-1); 
       \vertex (a) at (-2.5,1); 
       \vertex (c) at (2.5,1);
       \vertex [dot](tr) at (1.3,1){};
       \vertex [dot](tm) at (0,1){};
       \vertex [dot](tl) at (-1.3,1){};
       \vertex [dot](br) at (1.3,-1){};
       \vertex [dot](bm) at (0,-1){};
       \vertex [dot](bl) at (-1.3,-1){};
       \propag[ghost] (a) to (c);
       \propag[ghost] (b) to (d);
       \propag[photon, red] (tr) to (br);
       \propag[photon, red] (tl) to (bl);
       \propag[photon, red] (tm) to (bm);
       \propag[plain, ultra thick] (tl) to (tm);
       \propag[plain, ultra thick] (tl) to [in = 90, out = 90](tr);
    \end{feynhand}
 \end{tikzpicture} + \begin{tikzpicture}[baseline=0cm,scale=0.5]
    \begin{feynhand}
       \vertex (T1) at (-1.6, 0){1};
       \vertex (T2) at (-0.3, 0){2};
       \vertex (T3) at (1.0, 0){3};
       \vertex (b) at (-2.5,-1); 
       \vertex (d) at (2.5,-1); 
       \vertex (a) at (-2.5,1); 
       \vertex (c) at (2.5,1);
       \vertex [dot](tr) at (1.3,1){};
       \vertex [dot](tm) at (0,1){};
       \vertex [dot](tl) at (-1.3,1){};
       \vertex [dot](br) at (1.3,-1){};
       \vertex [dot](bm) at (0,-1){};
       \vertex [dot](bl) at (-1.3,-1){};
       \propag[ghost] (a) to (c);
       \propag[ghost] (b) to (d);
       \propag[photon, red] (tr) to (br);
       \propag[photon, red] (tl) to (bl);
       \propag[photon, red] (tm) to (bm);
       \propag[plain, ultra thick] (tl) to (tm);
       \propag[plain, ultra thick] (bl) to [in = 270, out = 270](br);
    \end{feynhand}
 \end{tikzpicture} + \begin{tikzpicture}[baseline=0cm,scale=0.5]
    \begin{feynhand}
       \vertex (T1) at (-1.6, 0){1};
       \vertex (T2) at (-0.3, 0){2};
       \vertex (T3) at (1.0, 0){3};
       \vertex (b) at (-2.5,-1); 
       \vertex (d) at (2.5,-1); 
       \vertex (a) at (-2.5,1); 
       \vertex (c) at (2.5,1);
       \vertex [dot](tr) at (1.3,1){};
       \vertex [dot](tm) at (0,1){};
       \vertex [dot](tl) at (-1.3,1){};
       \vertex [dot](br) at (1.3,-1){};
       \vertex [dot](bm) at (0,-1){};
       \vertex [dot](bl) at (-1.3,-1){};
       \propag[ghost] (a) to (c);
       \propag[ghost] (b) to (d);
       \propag[photon, red] (tr) to (br);
       \propag[photon, red] (tl) to (bl);
       \propag[photon, red] (tm) to (bm);
       \propag[plain, ultra thick] (bl) to (bm);
       \propag[plain, ultra thick] (bl) to [in = 270, out = 270](br);
    \end{feynhand}
 \end{tikzpicture}
\end{equation}
By marking the tree diagrams with indices 1,2 and 3, we pretend that those trees are distinguishable, which is exactly how we calculate the worldline expression \eqref{amp-qed-2loop}. However, the tree parts are actually indistinguishable, and the actual WQFT diagram is the sum of different labeled diagrams. For  
example,
%
\begin{equation}
    \begin{tikzpicture}[baseline=0cm,scale=0.5]
    \begin{feynhand}
       \vertex (b) at (-2.5,-1); 
       \vertex (d) at (2.5,-1); 
       \vertex (a) at (-2.5,1); 
       \vertex (c) at (2.5,1);
       \vertex [dot](tr) at (1.3,1){};
       \vertex [dot](tm) at (0,1){};
       \vertex [dot](tl) at (-1.3,1){};
       \vertex [dot](br) at (1.3,-1){};
       \vertex [dot](bm) at (0,-1){};
       \vertex [dot](bl) at (-1.3,-1){};
       \propag[ghost] (a) to (c);
       \propag[ghost] (b) to (d);
       \propag[photon, red] (tr) to (br);
       \propag[photon, red] (tl) to (bl);
       \propag[photon, red] (tm) to (bm);
       \propag[plain, ultra thick] (tl) to (tm);
       \propag[plain, ultra thick] (tl) to [in = 90, out = 90](tr);
    \end{feynhand}
 \end{tikzpicture}
    =\begin{tikzpicture}[baseline=0cm,scale=0.5]
    \begin{feynhand}
       \vertex (T1) at (-1.6, 0){1};
       \vertex (T2) at (-0.3, 0){2};
       \vertex (T3) at (1.0, 0){3};
    
       \vertex (b) at (-2.5,-1); 
       \vertex (d) at (2.5,-1); 
       \vertex (a) at (-2.5,1); 
       \vertex (c) at (2.5,1);
       \vertex [dot](tr) at (1.3,1){};
       \vertex [dot](tm) at (0,1){};
       \vertex [dot](tl) at (-1.3,1){};
       \vertex [dot](br) at (1.3,-1){};
       \vertex [dot](bm) at (0,-1){};
       \vertex [dot](bl) at (-1.3,-1){};
       \propag[ghost] (a) to (c);
       \propag[ghost] (b) to (d);
       \propag[photon, red] (tr) to (br);
       \propag[photon, red] (tl) to (bl);
       \propag[photon, red] (tm) to (bm);
       \propag[plain, ultra thick] (tl) to (tm);
       \propag[plain, ultra thick] (tl) to [in = 90, out = 90](tr);
    \end{feynhand}
 \end{tikzpicture} + \begin{tikzpicture}[baseline=0cm,scale=0.5]
    \begin{feynhand}
       \vertex (T1) at (-1.6, 0){1};
       \vertex (T2) at (-0.3, 0){2};
       \vertex (T3) at (1.0, 0){3};
       \vertex (b) at (-2.5,-1); 
       \vertex (d) at (2.5,-1); 
       \vertex (a) at (-2.5,1); 
       \vertex (c) at (2.5,1);
       \vertex [dot](tr) at (1.3,1){};
       \vertex [dot](tm) at (0,1){};
       \vertex [dot](tl) at (-1.3,1){};
       \vertex [dot](br) at (1.3,-1){};
       \vertex [dot](bm) at (0,-1){};
       \vertex [dot](bl) at (-1.3,-1){};
       \propag[ghost] (a) to (c);
       \propag[ghost] (b) to (d);
       \propag[photon, red] (tr) to (br);
       \propag[photon, red] (tl) to (bl);
       \propag[photon, red] (tm) to (bm);
       \propag[plain, ultra thick] (tl) to (tm);
       \propag[plain, ultra thick] (tm) to (tr);
    \end{feynhand}
 \end{tikzpicture} + \begin{tikzpicture}[baseline=0cm,scale=0.5]
    \begin{feynhand}
       \vertex (T1) at (-1.6, 0){1};
       \vertex (T2) at (-0.3, 0){2};
       \vertex (T3) at (1.0, 0){3};
       \vertex (b) at (-2.5,-1); 
       \vertex (d) at (2.5,-1); 
       \vertex (a) at (-2.5,1); 
       \vertex (c) at (2.5,1);
       \vertex [dot](tr) at (1.3,1){};
       \vertex [dot](tm) at (0,1){};
       \vertex [dot](tl) at (-1.3,1){};
       \vertex [dot](br) at (1.3,-1){};
       \vertex [dot](bm) at (0,-1){};
       \vertex [dot](bl) at (-1.3,-1){};
       \propag[ghost] (a) to (c);
       \propag[ghost] (b) to (d);
       \propag[photon, red] (tr) to (br);
       \propag[photon, red] (tl) to (bl);
       \propag[photon, red] (tm) to (bm);
       \propag[plain, ultra thick] (tm) to (tr);
       \propag[plain, ultra thick] (tl) to [in = 90, out = 90](tr);
    \end{feynhand}
 \end{tikzpicture}
\end{equation}

Notice this is already reflected in the permuting structure of $\frac{\partial}{\partial P_3(1,2,3)}$.
The diagrammatic method is equivalent to selecting all the relevant expressions
at an earlier stage before doing all $\tau$ integrals. More specifically, one
single term in $\eval{\frac{\partial^2 \mathcal{A}_2}{\partial t_{12} \partial t_{13}}}_{t_{i j} = 0}$,
\be
\int \frac{d^4 \bar{k}_1}{(2 \pi)^4}  \int \frac{d^4 \bar{k}_2}{(2 \pi)^4} 
   \left( \eval{\frac{\partial^2 \mathcal{M}_3^{\mu_1 \mu_2 \mu_3}}{\partial
   t_{12} \partial t_{13}}}_{t_{i j} = 0} \right)  \frac{- i
   \eta_{\mu_1 \nu_1}}{\bar{k}_1^2}  \frac{- i \eta_{\mu_2
   \nu_2}}{\bar{k}_2^2}  \frac{- i \eta_{\mu_3 \nu_3}}{\bar{k}_3^2}
   \left( \eval{\mathcal{M}_3^{\nu_1 \nu_2 \nu_3}}_{t_{i j} = 0} \right)
\ee
corresponds to one term in the labeled WQFT diagram
\begin{equation}
    \begin{tikzpicture}[baseline=0cm,scale=0.5]
    \begin{feynhand}
       \vertex (T1) at (-1.6, 0){1};
       \vertex (T2) at (-0.3, 0){2};
       \vertex (T3) at (1.0, 0){3};
       \vertex (b) at (-2.5,-1); 
       \vertex (d) at (2.5,-1); 
       \vertex (a) at (-2.5,1); 
       \vertex (c) at (2.5,1);
       \vertex [dot](tr) at (1.3,1){};
       \vertex [dot](tm) at (0,1){};
       \vertex [dot](tl) at (-1.3,1){};
       \vertex [dot](br) at (1.3,-1){};
       \vertex [dot](bm) at (0,-1){};
       \vertex [dot](bl) at (-1.3,-1){};
       \propag[ghost] (a) to (c);
       \propag[ghost] (b) to (d);
       \propag[photon, red] (tr) to (br);
       \propag[photon, red] (tl) to (bl);
       \propag[photon, red] (tm) to (bm);
       \propag[plain, ultra thick] (tl) to (tm);
       \propag[plain, ultra thick] (tl) to [in = 90, out = 90](tr);
    \end{feynhand}
 \end{tikzpicture}\,.
\end{equation}
Let us also show explicitly how the other terms we dropped correspond to
those reducible contributions predicted by the exponentiation form in eikonal method. For example, the term $\eval{\frac{\partial^2 \mathcal{A}_2}{\partial t_{12}^2}}_{t_{i j} = 0}$ corresponds to the sum of following labeled
diagrams of WQFT
\begin{equation}
    \begin{tikzpicture}[baseline=0cm,scale=0.5]
    \begin{feynhand}
       \vertex (T1) at (-1.6, 0){1};
       \vertex (T2) at (-0.3, 0){2};
       \vertex (T3) at (1.0, 0){3};
       \vertex (b) at (-2.5,-1); 
       \vertex (d) at (2.5,-1); 
       \vertex (a) at (-2.5,1); 
       \vertex (c) at (2.5,1);
       \vertex [dot](tr) at (1.3,1){};
       \vertex [dot](tm) at (0,1){};
       \vertex [dot](tl) at (-1.3,1){};
       \vertex [dot](br) at (1.3,-1){};
       \vertex [dot](bm) at (0,-1){};
       \vertex [dot](bl) at (-1.3,-1){};
       \propag[ghost] (a) to (c);
       \propag[ghost] (b) to (d);
       \propag[photon, red] (tr) to (br);
       \propag[photon, red] (tl) to (bl);
       \propag[photon, red] (tm) to (bm);
       \propag[plain, ultra thick] (tl) to (tm);
       \propag[plain, ultra thick] (tl) to [in = 90, out = 90](tm);
    \end{feynhand}
 \end{tikzpicture} + \begin{tikzpicture}[baseline=0cm,scale=0.5]
    \begin{feynhand}
       \vertex (T1) at (-1.6, 0){1};
       \vertex (T2) at (-0.3, 0){2};
       \vertex (T3) at (1.0, 0){3};
       \vertex (b) at (-2.5,-1); 
       \vertex (d) at (2.5,-1); 
       \vertex (a) at (-2.5,1); 
       \vertex (c) at (2.5,1);
       \vertex [dot](tr) at (1.3,1){};
       \vertex [dot](tm) at (0,1){};
       \vertex [dot](tl) at (-1.3,1){};
       \vertex [dot](br) at (1.3,-1){};
       \vertex [dot](bm) at (0,-1){};
       \vertex [dot](bl) at (-1.3,-1){};
       \propag[ghost] (a) to (c);
       \propag[ghost] (b) to (d);
       \propag[photon, red] (tr) to (br);
       \propag[photon, red] (tl) to (bl);
       \propag[photon, red] (tm) to (bm);
       \propag[plain, ultra thick] (tl) to (tm);
       \propag[plain, ultra thick] (bl) to (bm);
    \end{feynhand}
 \end{tikzpicture} + \begin{tikzpicture}[baseline=0cm,scale=0.5]
    \begin{feynhand}
       \vertex (T1) at (-1.6, 0){1};
       \vertex (T2) at (-0.3, 0){2};
       \vertex (T3) at (1.0, 0){3};
       \vertex (b) at (-2.5,-1); 
       \vertex (d) at (2.5,-1); 
       \vertex (a) at (-2.5,1); 
       \vertex (c) at (2.5,1);
       \vertex [dot](tr) at (1.3,1){};
       \vertex [dot](tm) at (0,1){};
       \vertex [dot](tl) at (-1.3,1){};
       \vertex [dot](br) at (1.3,-1){};
       \vertex [dot](bm) at (0,-1){};
       \vertex [dot](bl) at (-1.3,-1){};
       \propag[ghost] (a) to (c);
       \propag[ghost] (b) to (d);
       \propag[photon, red] (tr) to (br);
       \propag[photon, red] (tl) to (bl);
       \propag[photon, red] (tm) to (bm);
       \propag[plain, ultra thick] (bl) to (bm);
       \propag[plain, ultra thick] (bl) to [in = 270, out = 270](bm);
    \end{feynhand}
 \end{tikzpicture}
\end{equation}
Considering one specific term in it,
\be \label{reducible-term-example}
\int \frac{d^4 \bar{k}_1}{(2 \pi)^4}  \int \frac{d^4 \bar{k}_2}{(2 \pi)^4} 
   \left( \eval{\frac{\partial^2 \mathcal{M}_3^{\mu_1 \mu_2 \mu_3}}{\partial t_{12}^2}}_{t_{i j} = 0} \right)  \frac{- i \eta_{\mu_1
   \nu_1}}{\bar{k}_1^2}  \frac{- i \eta_{\mu_2 \nu_2}}{\bar{k}_2^2}  \frac{- i
   \eta_{\mu_3 \nu_3}}{\bar{k}_3^2} \left( \eval{\mathcal{M}_3^{\nu_1 \nu_2 \nu_3}}_{t_{i j} = 0} \right)
\ee
which is represented diagrammatically by the labeled diagram:
\begin{equation}
    \begin{tikzpicture}[baseline=0cm,scale=0.5]
    \begin{feynhand}
       \vertex (T1) at (-1.6, 0){1};
       \vertex (T2) at (-0.3, 0){2};
       \vertex (T3) at (1.0, 0){3};
       \vertex (b) at (-2.5,-1); 
       \vertex (d) at (2.5,-1); 
       \vertex (a) at (-2.5,1); 
       \vertex (c) at (2.5,1);
       \vertex [dot](tr) at (1.3,1){};
       \vertex [dot](tm) at (0,1){};
       \vertex [dot](tl) at (-1.3,1){};
       \vertex [dot](br) at (1.3,-1){};
       \vertex [dot](bm) at (0,-1){};
       \vertex [dot](bl) at (-1.3,-1){};
       \propag[ghost] (a) to (c);
       \propag[ghost] (b) to (d);
       \propag[photon, red] (tr) to (br);
       \propag[photon, red] (tl) to (bl);
       \propag[photon, red] (tm) to (bm);
       \propag[plain, ultra thick] (tl) to (tm);
       \propag[plain, ultra thick] (tl) to [in = 90, out = 90](tm);
    \end{feynhand}
 \end{tikzpicture}\,.
\end{equation}
Based on our discussion about the factorization in worldline calculation in \eqref{fac-one-worldine}, we see
\begin{eqnarray*}
  &  & (2\pi)^2 \int \frac{d^4 \bar{q}}{(2 \pi)^4} \delta \left( \bar{q} \cdot
  \frac{v_1}{\hbar} \right) \delta \left( \bar{q} \cdot \frac{v_2}{\hbar}
  \right) \int \frac{d^4 \bar{k}_1}{(2 \pi)^4}  \int \frac{d^4 \bar{k}_2}{(2
  \pi)^4} \\
  &  & \times\left( \eval{\frac{\partial^2 \mathcal{M}_3^{\mu_1 \mu_2 \mu_3}}{\partial t_{12}^2}}_{t_{i j} = 0} \right) 
  \frac{- i \eta_{\mu_1 \nu_1}}{\bar{k}_1^2}  \frac{- i \eta_{\mu_2
  \nu_2}}{\bar{k}_2^2}  \frac{- i \eta_{\mu_3 \nu_3}}{(\bar{q} - \bar{k}_1 -
  \bar{k}_2)^2}  \left( \eval{\mathcal{M}_3^{\nu_1 \nu_2 \nu_3}}_{t_{i j} = 0} \right)\\
  & = & (2\pi)^2 \int \frac{d^4 \bar{q}_1}{(2 \pi)^4} \delta \left( \bar{q}_1 \cdot
  \frac{v_1}{\hbar} \right) \delta \left( \bar{q}_1 \cdot \frac{v_2}{\hbar}
  \right) \int \frac{d^4 \bar{k}_1}{(2 \pi)^4}\\
  &  & \times\left( \eval{\frac{\partial^2 \mathcal{M}_2^{\mu_1 \mu_2}}{\partial t_{12}^2}}_{t_{i j} = 0} \right) \frac{- i \eta_{\mu_1
  \nu_1}}{\bar{k}_1^2}  \frac{- i \eta_{\mu_2 \nu_2}}{(\bar{q}_1 -
  \bar{k}_1)^2}  \left( \eval{\mathcal{M}_2^{\nu_1 \nu_2}}_{t_{i j} = 0} \right)\\
  &  & \times (2\pi)^2 \int \frac{d^4 \bar{q}_2}{(2 \pi)^4} \delta \left( \bar{q}_2
  \cdot \frac{v_1}{\hbar} \right) \delta \left( \bar{q}_2 \cdot
  \frac{v_2}{\hbar} \right) \left( \eval{\mathcal{M}_1^{\mu_3}}_{t_{i j} = 0} \right)  \frac{- i \eta_{\mu_3 \nu_3}}{\bar{q}_2^2}  \left( \eval{\mathcal{M}_1^{\nu_3}}_{t_{i j} = 0} \right)
\end{eqnarray*}
Notice that this structure is just the factorization in \eqref{sep2to2}, which matches perfectly with the exponentiation form
\eqref{expConjExpand} with
\begin{eqnarray*}
   i \delta_0 & = & (2\pi)^2 \int \frac{d^4 \bar{q}_2}{(2 \pi)^4} \delta \left(
  \bar{q}_2 \cdot \frac{v_1}{\hbar} \right) \delta \left( \bar{q}_2 \cdot
  \frac{v_2}{\hbar} \right) \left( \eval{\mathcal{M}_1^{\mu_3}}_{t_{i j} =
  0} \right)  \frac{- i \eta_{\mu_3 \nu_3}}{\bar{q}_2^2}  \left( \eval{\mathcal{M}_1^{\nu_3}}_{t_{i j} = 0} \right)
\end{eqnarray*}
while
\begin{eqnarray*}
  &  & (2\pi)^2 \int \frac{d^4 \bar{q}_1}{(2 \pi)^4} \delta \left( \bar{q}_1 \cdot
  \frac{v_1}{\hbar} \right) \delta \left( \bar{q}_1 \cdot \frac{v_2}{\hbar}
  \right) \int \frac{d^4 \bar{k}_1}{(2 \pi)^4} \left( \eval{\frac{\partial^2
  \mathcal{M}_2^{\mu_1 \mu_2 }}{\partial t_{12}^2}}_{t_{i j}
  = 0} \right) \frac{- i \eta_{\mu_1 \nu_1}}{\bar{k}_1^2}  \frac{- i
  \eta_{\mu_2 \nu_2}}{(\bar{q}_1 - \bar{k}_1)^2}  \left( \eval{\mathcal{M}_2^{\nu_1 \nu_2
  }}_{t_{i j} = 0} \right)
\end{eqnarray*}
is one term in $2 i \Delta_1$. Thus, the term \eqref{reducible-term-example}
is just the corresponding term in $2 i \delta_0 2 i \Delta_1$.
Diagramatically, this can be represented as,
\begin{equation}
    \begin{tikzpicture}[baseline=0cm,scale=0.5]
    \begin{feynhand}
       \vertex (b) at (-2.5,-1); 
       \vertex (d) at (2.5,-1); 
       \vertex (a) at (-2.5,1); 
       \vertex (c) at (2.5,1);
       \vertex [dot](tr) at (1.3,1){};
       \vertex [dot](tm) at (0,1){};
       \vertex [dot](tl) at (-1.3,1){};
       \vertex [dot](br) at (1.3,-1){};
       \vertex [dot](bm) at (0,-1){};
       \vertex [dot](bl) at (-1.3,-1){};
       \propag[ghost] (a) to (c);
       \propag[ghost] (b) to (d);
       \propag[photon, red] (tr) to (br);
       \propag[photon, red] (tl) to (bl);
       \propag[photon, red] (tm) to (bm);
       \propag[plain, ultra thick] (tl) to (tm);
       \propag[plain, ultra thick] (tl) to [in = 90, out = 90](tm);
    \end{feynhand}
 \end{tikzpicture} = \begin{tikzpicture}[baseline=0cm,scale=0.5]
    \begin{feynhand}
       \vertex (b) at (-2.5,-1); 
       \vertex (d) at (2.5,-1); 
       \vertex (a) at (-2.5,1); 
       \vertex (c) at (2.5,1);
       \vertex (tr) at (1.3,1);
       \vertex [dot](tm) at (0,1){};
       \vertex [dot](tl) at (-1.3,1){};
       \vertex (br) at (1.3,-1);
       \vertex [dot](bm) at (0,-1){};
       \vertex [dot](bl) at (-1.3,-1){};
       \propag[ghost] (a) to (tr);
       \propag[ghost] (b) to (br);
       \propag[photon, red] (tl) to (bl);
       \propag[photon, red] (tm) to (bm);
       \propag[plain, ultra thick] (tl) to (tm);
       \propag[plain, ultra thick] (tl) to [in = 90, out = 90](tm);
    \end{feynhand}  \end{tikzpicture} \times \begin{tikzpicture}[baseline=0cm,scale=0.5]
    \begin{feynhand}
       \vertex (b) at (-2.5,-1); 
       \vertex (d) at (2.5,-1); 
       \vertex (a) at (-2.5,1); 
       \vertex (c) at (2.5,1);
       \vertex (tr) at (1.3,1);
       \vertex [dot](tm) at (0,1){};
       \vertex (tl) at (-1.3,1);
       \vertex (br) at (1.3,-1);
       \vertex [dot](bm) at (0,-1){};
       \vertex (bl) at (-1.3,-1);
       \propag[ghost] (tl) to (tr);
       \propag[ghost] (bl) to (br);
       \propag[photon, red] (tm) to (bm);
    \end{feynhand}
 \end{tikzpicture}
\end{equation}
This illustrates how the factorization in WQFT corresponds to the exponentiation in the
eikonal method of QFT, which is the reason that we can select those classical
contributions directly by considering specific structure of contractions in
WQFT. For the complete proof please see \cite{ADRV1}.

Next, let us consider other diagrams that involve radiative
corrections. These correspond to Feynman diagrams that have
photon propagators only connecting to one worldline. Based on our conclusion from the last subsection, we need at least two mediator exchanges between the  worldlines to get a non-vanishing contribution from the worldline radiative corrections. In terms of diagrams of worldline formalism, these are\\
\be
\begin{tikzpicture}[baseline=-0cm]
   \begin{feynhand}
      \vertex (b) at (-2,0){2}; 
      \vertex (d) at (2,0){2'};
      \vertex (a) at (-2,1){1}; 
      \vertex (c) at (2,1){1'};
      \vertex [dot](tr) at (1,1){};
      \vertex [dot](trm) at (0.5,1){};
      \vertex [dot](tm) at (0,1){};
      \vertex [dot](tlm) at (-0.5,1);
      \vertex [dot](tl) at (-1,1){};
      \vertex [dot](br) at (1,0){};
      \vertex [dot](brm) at (0.5,0){};
      \vertex [dot](bm) at (0,0);
      \vertex [dot](blm) at (-0.5,0);
      \vertex [dot](bl) at (-1,0);
      \propag [scalar] (a) to (c); 
      \propag [scalar] (b) to (d);
      \propag [photon,red] (tr) to (br);
      \propag [photon,red] (tm) [in=90, out=90, looseness=1] to (tl);
      \propag [photon,red] (trm) to (brm);
   \end{feynhand}
\end{tikzpicture}
\begin{tikzpicture}[baseline=-0cm]
   \begin{feynhand}
      \vertex (b) at (-2,0){2}; 
      \vertex (d) at (2,0){2'};
      \vertex (a) at (-2,1){1}; 
      \vertex (c) at (2,1){1'};
      \vertex [dot](tr) at (1,1){};
      \vertex [dot](trm) at (0.5,1){};
      \vertex [dot](tm) at (0,1);
      \vertex [dot](tlm) at (-0.5,1);
      \vertex [dot](tl) at (-1,1);
      \vertex [dot](br) at (1,0){};
      \vertex [dot](brm) at (0.5,0){};
      \vertex [dot](bm) at (0,0){};
      \vertex [dot](blm) at (-0.5,0);
      \vertex [dot](bl) at (-1,0){};
      \propag [scalar] (a) to (c); 
      \propag [scalar] (b) to (d);
      \propag [photon,red] (tr) to (br);
      \propag [photon,red] (bm) [in=270, out=270, looseness=1] to (bl);
      \propag [photon,red] (trm) to (brm);
   \end{feynhand}
\end{tikzpicture}
\ee
To get the classical contribution, we want all irreducible WQFT diagrams with two contractions. These are\\
\be
\begin{tikzpicture}[baseline=-0cm]
   \begin{feynhand}
      \vertex (b) at (-2,0){2}; 
      \vertex (d) at (2,0){2'};
      \vertex (a) at (-2,1){1}; 
      \vertex (c) at (2,1){1'};
      \vertex [dot](tr) at (1,1){};
      \vertex [dot](trm) at (0.5,1){};
      \vertex [dot](tm) at (0,1);
      \vertex [dot](tlm) at (-0.5,1);
      \vertex [dot](tl) at (-1,1);
      \vertex [dot](br) at (1,0){};
      \vertex [dot](brm) at (0.5,0){};
      \vertex [dot](bm) at (0,0){};
      \vertex [dot](blm) at (-0.5,0);
      \vertex [dot](bl) at (-1,0){};
      \propag [gho] (a) to (c); 
      \propag [gho] (b) to (d);
      \propag [photon,red] (tr) to (br);
      \propag [photon,red] (bm) [in=270, out=270, looseness=1] to (bl);
      \propag [photon,red] (trm) to (brm);
      \propag [plain,ultra thick, black] (br)  to (bm);
   \end{feynhand}
\end{tikzpicture}
\begin{tikzpicture}[baseline=-0cm]
   \begin{feynhand}
      \vertex (b) at (-2,0){2}; 
      \vertex (d) at (2,0){2'};
      \vertex (a) at (-2,1){1}; 
      \vertex (c) at (2,1){1'};
      \vertex [dot](tr) at (1,1){};
      \vertex [dot](trm) at (0.5,1){};
      \vertex [dot](tm) at (0,1);
      \vertex [dot](tlm) at (-0.5,1);
      \vertex [dot](tl) at (-1,1);
      \vertex [dot](br) at (1,0){};
      \vertex [dot](brm) at (0.5,0){};
      \vertex [dot](bm) at (0,0){};
      \vertex [dot](blm) at (-0.5,0);
      \vertex [dot](bl) at (-1,0){};
      \propag [gho] (a) to (c); 
      \propag [gho] (b) to (d);
      \propag [photon,red] (tr) to (br);
      \propag [photon,red] (bm) [in=270, out=270, looseness=1] to (bl);
      \propag [photon,red] (trm) to (brm);
      \propag [plain,ultra thick, black] (brm)  to (bm);
      \propag [plain,ultra thick, black] (bm)  to [in=270, out=270, looseness=1] (br);
   \end{feynhand}
\end{tikzpicture}
\begin{tikzpicture}[baseline=-0cm]
   \begin{feynhand}
      \vertex (b) at (-2,0){2}; 
      \vertex (d) at (2,0){2'};
      \vertex (a) at (-2,1){1}; 
      \vertex (c) at (2,1){1'};
      \vertex [dot](tr) at (1,1){};
      \vertex [dot](trm) at (0.5,1){};
      \vertex [dot](tm) at (0,1);
      \vertex [dot](tlm) at (-0.5,1);
      \vertex [dot](tl) at (-1,1);
      \vertex [dot](br) at (1,0){};
      \vertex [dot](brm) at (0.5,0){};
      \vertex [dot](bm) at (0,0){};
      \vertex [dot](blm) at (-0.5,0);
      \vertex [dot](bl) at (-1,0){};
      \propag [gho] (a) to (c); 
      \propag [gho] (b) to (d);
      \propag [photon,red] (tr) to (br);
      \propag [photon,red] (bm) [in=270, out=270, looseness=1] to (bl);
      \propag [photon,red] (trm) to (brm);
      \propag [plain,ultra thick, black] (bl)  to [in=90, out=90, looseness=1] (brm);
      \propag [plain,ultra thick, black] (bm)  to [in=270, out=270, looseness=1] (br);
   \end{feynhand}
\end{tikzpicture}
+(1 \leftrightarrow 2)
\ee
However, only two of these diagrams are non-vanishing, which are
\be
\begin{tikzpicture}[baseline=-0cm]
   \begin{feynhand}
      \vertex (b) at (-2,0){2}; 
      \vertex (d) at (2,0){2'};
      \vertex (a) at (-2,1){1}; 
      \vertex (c) at (2,1){1'};
      \vertex [dot](tr) at (1,1){};
      \vertex [dot](trm) at (0.5,1){};
      \vertex [dot](tm) at (0,1);
      \vertex [dot](tlm) at (-0.5,1);
      \vertex [dot](tl) at (-1,1);
      \vertex [dot](br) at (1,0){};
      \vertex [dot](brm) at (0.5,0){};
      \vertex [dot](bm) at (0,0){};
      \vertex [dot](blm) at (-0.5,0);
      \vertex [dot](bl) at (-1,0){};
      \propag [gho] (a) to (c); 
      \propag [gho] (b) to (d);
      \propag [photon,red] (tr) to (br);
      \propag [photon,red] (bm) [in=270, out=270, looseness=1] to (bl);
      \propag [photon,red] (trm) to (brm);
      \propag [plain,ultra thick, black] (bl)  to [in=90, out=90, looseness=1] (brm);
      \propag [plain,ultra thick, black] (bm)  to [in=270, out=270, looseness=1] (br);
   \end{feynhand}
\end{tikzpicture}
+(1 \leftrightarrow 2)
\ee
The other diagrams vanish due to scaleless integrals. Notice that they all have an uncontracted vertex in the radiative loop, whose $\tau$ integral yields $(2\pi)\delta(v_i \cdot l)$, assuming that the loop momentum of the radiative loop is $l$. The delta function eliminates all the $l$-dependence in the denominator, which leads to scaleless integrals once more.

Comparing with the diagrams in \cite{Wang:2022ntx}, we see we have recovered all the diagrams used there. The difference is that we systematically generate these diagrams from the perspective of scattering amplitude and we are able to show the WQFT calculation is equivalent to the eikonal method, which is a QFT calculation. One can even make this correspondence explicit in a term-by-term matching. Thus, the discrepancy found in \cite{Wang:2022ntx} when comparing WQFT and QFT calculations should not originate from the WQFT framework of calculation. Instead, we suspect, it {is the intrinsic mismatch between the eikonal phase and the radial action, which is discussed in \cite{DiVecchia:2023frv}}.

\section{Comments on WQFT for scalar coupled to Gravity}\label{sect:wqft-sGravity}

The generalization from scalar-QED to scalar-Gravity is overall
straightforward. The main differences come from the fact that the mediators in
the scalar-Gravity case allow self-interaction. In this section, we will not
go into details with the calculation, but focus more on the general
diagrammatic structure, which is enough to highlight the similarities and differences between
scalar-Gravity and scalar-QED. 

\subsection{Leading eikonal}\label{sect:wqft-sGravity-leading}

The leading eikonal contribution, which is of the order $\kappa^2$  
is given by the tree exchange diagram, similar to the scalar-QED case except that the mediator is a graviton:
\be
\begin{tikzpicture}[baseline=0.4cm]
   \begin{feynhand}
      \vertex (b) at (-1,0){2}; 
      \vertex (d) at (1,0){2'};
      \vertex (a) at (-1,1){1}; 
      \vertex (c) at (1,1){1'};
      \vertex [dot](tr) at (1,1);
      \vertex [dot](trm) at (0.5,1);
      \vertex [dot](tm) at (0,1){};
      \vertex [dot](tlm) at (-0.5,1);
      \vertex [dot](tl) at (-1,1);
      \vertex [dot](br) at (1,0);
      \vertex [dot](brm) at (0.5,0);
      \vertex [dot](bm) at (0,0){};
      \vertex [dot](blm) at (-0.5,0);
      \vertex [dot](bl) at (-1,0);
      \propag [gho] (a) to (c); 
      \propag [gho] (b) to (d);
      \propag [graviton,red] (tm) to (bm);
   \end{feynhand}
\end{tikzpicture}
\ee

\subsection{Subleading eikonal}\label{sect:wqft-sGravity-subleading}

The subleading eikonal is of the order $\kappa^4$. The terms at this order which can come from
one tree (a ``Y" tree) or two graviton trees (each ``tree" being a graviton propagator) exchanges. 

The diagrams with two graviton trees are similar to the scalar-QED case,
\be
\begin{tikzpicture}[baseline=0.4cm]
   \begin{feynhand}
      \vertex (b) at (-1.5,0){2}; 
      \vertex (d) at (1.5,0){2'};
      \vertex (a) at (-1.5,1){1}; 
      \vertex (c) at (1.5,1){1'};
      \vertex [dot](tr) at (1,1);
      \vertex [dot](middle) at (1,1);
      \vertex [dot](trm) at (0.5,1){};
      \vertex [dot](tm) at (0,1);
      \vertex [dot](tlm) at (-0.5,1){};
      \vertex [dot](tl) at (-1,1);
      \vertex [dot](br) at (1,0);
      \vertex [dot](brm) at (0.5,0){};
      \vertex [dot](bm) at (0,0);
      \vertex [dot](blm) at (-0.5,0){};
      \vertex [dot](bl) at (-1,0);
      \propag [gho] (a) to (c); 
      \propag [gho] (b) to (d);
      \propag [graviton,red] (tlm) to (blm);
      \propag [graviton,red] (trm) to (brm);
      \propag [plain, ultra thick] (tlm) [in=90, out=90, looseness=1] to (trm);
   \end{feynhand}
\end{tikzpicture}
+
\begin{tikzpicture}[baseline=0.4cm]
   \begin{feynhand}
      \vertex (b) at (-1.5,0){2}; 
      \vertex (d) at (1.5,0){2'};
      \vertex (a) at (-1.5,1){1}; 
      \vertex (c) at (1.5,1){1'};
      \vertex [dot](tr) at (1,1);
      \vertex [dot](trm) at (0.5,1){};
      \vertex [dot](tm) at (0,1);
      \vertex [dot](tlm) at (-0.5,1){};
      \vertex [dot](tl) at (-1,1);
      \vertex [dot](br) at (1,0);
      \vertex [dot](brm) at (0.5,0){};
      \vertex [dot](bm) at (0,0);
      \vertex [dot](blm) at (-0.5,0){};
      \vertex [dot](bl) at (-1,0);
      \propag [gho] (a) to (c); 
      \propag [gho] (b) to (d);
      \propag [graviton,red] (tlm) to (blm);
      \propag [graviton,red] (trm) to (brm);
      \propag [plain, ultra thick] (blm) [in=270, out=270, looseness=1] to (brm);
   \end{feynhand}
\end{tikzpicture}
\ee
while the ``Y" graviton tree 
 diagram
is new (since gravitons self-interact, unlike photons):
\be
\begin{tikzpicture}[baseline=0.4cm]
   \begin{feynhand}
      \vertex (b) at (-1.5,0){2}; 
      \vertex (d) at (1.5,0){2'};
      \vertex (a) at (-1.5,1){1}; 
      \vertex (c) at (1.5,1){1'};
      \vertex [dot](tr) at (1,1);
      \vertex [dot](trm) at (0.5,1){};
      \vertex [dot](tm) at (0,1);
      \vertex [dot](tlm) at (-0.5,1){};
      \vertex [dot](tl) at (-1,1);
      \vertex [dot](br) at (1,0);
      \vertex [dot](brm) at (0.5,0);
      \vertex [dot](bm) at (0,0){};
      \vertex [dot](blm) at (-0.5,0);
      \vertex [dot](bl) at (-1,0);
      \vertex [dot](mm) at (0,0.5);
      \propag [gho] (a) to (c); 
      \propag [gho] (b) to (d);
      \propag [graviton,red] (tlm) to (mm);
      \propag [graviton,red] (trm) to (mm);
      \propag [graviton,red] (mm) to (bm);
   \end{feynhand}
\end{tikzpicture}
+
\begin{tikzpicture}[baseline=0.4cm]
   \begin{feynhand}
      \vertex (b) at (-1.5,0){2}; 
      \vertex (d) at (1.5,0){2'};
      \vertex (a) at (-1.5,1){1}; 
      \vertex (c) at (1.5,1){1'};
      \vertex [dot](tr) at (1,1);
      \vertex [dot](trm) at (0.5,1);
      \vertex [dot](tm) at (0,1){};
      \vertex [dot](tlm) at (-0.5,1);
      \vertex [dot](tl) at (-1,1);
      \vertex [dot](br) at (1,0);
      \vertex [dot](brm) at (0.5,0){};
      \vertex [dot](bm) at (0,0);
      \vertex [dot](blm) at (-0.5,0){};
      \vertex [dot](bl) at (-1,0);
      \vertex [dot](mm) at (0,0.5);
      \propag [gho] (a) to (c); 
      \propag [gho] (b) to (d);
      \propag [graviton,red] (blm) to (mm);
      \propag [graviton,red] (brm) to (mm);
      \propag [graviton,red] (mm) to (tm);
   \end{feynhand}
\end{tikzpicture}
\ee

Note that there are no radiative corrections at this order.
This follows our previous analysis for scalar-QED in Section \ref{subsect:wqft-sQED-subleading} and which extends to gravitational interactions.  We found that in order for the radiative corrections to be non-zero, we need at least two exchanges between the worldlines (two graviton trees). 

\subsection{Subsubleading eikonal}\label{sect:wqft-sGravity-subsubleading}

The subsubleading eikonal has coupling constants $\kappa^6$, which can come
from one or two or three tree parts.

\subsubsection{Conservative part of the dynamics}

Diagrams that contain three tree parts are the same as the scalar-QED case,
\be
\begin{tikzpicture}[baseline=0.4cm]
   \begin{feynhand}
      \vertex (b) at (-1.5,0){2}; 
      \vertex (d) at (2.5,0){2'};
      \vertex (a) at (-1.5,1){1}; 
      \vertex (c) at (2.5,1){1'};
      \vertex [dot](trr) at (1.5,1){};
      \vertex [dot](brr) at (1.5,0){};
      \vertex [dot](tr) at (1,1);
      \vertex [dot](trm) at (0.5,1){};
      \vertex [dot](tm) at (0,1);
      \vertex [dot](tlm) at (-0.5,1){};
      \vertex [dot](tl) at (-1,1);
      \vertex [dot](br) at (1,0);
      \vertex [dot](brm) at (0.5,0){};
      \vertex [dot](bm) at (0,0);
      \vertex [dot](blm) at (-0.5,0){};
      \vertex [dot](bl) at (-1,0);
      \vertex [dot](mm) at (0,0.5);
      \propag [gho] (a) to (c); 
      \propag [gho] (b) to (d);
      \propag [graviton,red] (trr) to (brr);
      \propag [graviton,red] (tlm) to (blm);
      \propag [graviton,red] (trm) to (brm);
      \propag [plain, ultra thick] (trr) [in=90, out=90, looseness=1] to (tlm);
      \propag [plain, ultra thick] (trm) [in=90, out=90, looseness=1] to (tlm);
   \end{feynhand}
\end{tikzpicture}
+
\begin{tikzpicture}[baseline=0.4cm]
   \begin{feynhand}
      \vertex (b) at (-1.5,0){2}; 
      \vertex (d) at (2.5,0){2'};
      \vertex (a) at (-1.5,1){1}; 
      \vertex (c) at (2.5,1){1'};
      \vertex [dot](trr) at (1.5,1){};
      \vertex [dot](brr) at (1.5,0){};
      \vertex [dot](tr) at (1,1);
      \vertex [dot](trm) at (0.5,1){};
      \vertex [dot](tm) at (0,1);
      \vertex [dot](tlm) at (-0.5,1){};
      \vertex [dot](tl) at (-1,1);
      \vertex [dot](br) at (1,0);
      \vertex [dot](brm) at (0.5,0){};
      \vertex [dot](bm) at (0,0);
      \vertex [dot](blm) at (-0.5,0){};
      \vertex [dot](bl) at (-1,0);
      \vertex [dot](mm) at (0,0.5);
      \propag [gho] (a) to (c); 
      \propag [gho] (b) to (d);
      \propag [graviton,red] (trr) to (brr);
      \propag [graviton,red] (tlm) to (blm);
      \propag [graviton,red] (trm) to (brm);
      \propag [plain, ultra thick] (brr) [in=270, out=270, looseness=1] to (brm);
      \propag [plain, ultra thick] (trm) [in=90, out=90, looseness=1] to (tlm);
   \end{feynhand}
\end{tikzpicture}
+
\begin{tikzpicture}[baseline=0.4cm]
   \begin{feynhand}
      \vertex (b) at (-1.5,0){2}; 
      \vertex (d) at (2.5,0){2'};
      \vertex (a) at (-1.5,1){1}; 
      \vertex (c) at (2.5,1){1'};
      \vertex [dot](trr) at (1.5,1){};
      \vertex [dot](brr) at (1.5,0){};
      \vertex [dot](tr) at (1,1);
      \vertex [dot](trm) at (0.5,1){};
      \vertex [dot](tm) at (0,1);
      \vertex [dot](tlm) at (-0.5,1){};
      \vertex [dot](tl) at (-1,1);
      \vertex [dot](br) at (1,0);
      \vertex [dot](brm) at (0.5,0){};
      \vertex [dot](bm) at (0,0);
      \vertex [dot](blm) at (-0.5,0){};
      \vertex [dot](bl) at (-1,0);
      \vertex [dot](mm) at (0,0.5);
      \propag [gho] (a) to (c); 
      \propag [gho] (b) to (d);
      \propag [graviton,red] (trr) to (brr);
      \propag [graviton,red] (tlm) to (blm);
      \propag [graviton,red] (trm) to (brm);
      \propag [plain, ultra thick] (brr) [in=270, out=270, looseness=1] to (blm);
      \propag [plain, ultra thick] (brm) [in=270, out=270, looseness=1] to (blm);
   \end{feynhand}
\end{tikzpicture}
\ee

Diagrams that contain two tree parts are new,
\be
\begin{tikzpicture}[baseline=0.4cm]
   \begin{feynhand}
      \vertex (b) at (-1.5,0){2}; 
      \vertex (d) at (2.5,0){2'};
      \vertex (a) at (-1.5,1){1}; 
      \vertex (c) at (2.5,1){1'};
      \vertex [dot](trr) at (1.5,1){};
      \vertex [dot](brr) at (1.5,0){};
      \vertex [dot](tr) at (1,1);
      \vertex [dot](trm) at (0.5,1){};
      \vertex [dot](tm) at (0,1);
      \vertex [dot](tlm) at (-0.5,1){};
      \vertex [dot](tl) at (-1,1);
      \vertex [dot](br) at (1,0);
      \vertex [dot](brm) at (0.5,0);
      \vertex [dot](bm) at (0,0){};
      \vertex [dot](blm) at (-0.5,0);
      \vertex [dot](bl) at (-1,0);
      \vertex [dot](mm) at (0,0.5);
      \propag [gho] (a) to (c); 
      \propag [gho] (b) to (d);
      \propag [graviton,red] (trr) to (brr);
      \propag [graviton,red] (tlm) to (mm);
      \propag [graviton,red] (trm) to (mm);
      \propag [graviton,red] (mm) to (bm);
      \propag [plain, ultra thick] (trr) [in=90, out=90, looseness=1] to (trm);
   \end{feynhand}
\end{tikzpicture}
+
\begin{tikzpicture}[baseline=0.4cm]
   \begin{feynhand}
      \vertex (b) at (-1.5,0){2}; 
      \vertex (d) at (2.5,0){2'};
      \vertex (a) at (-1.5,1){1}; 
      \vertex (c) at (2.5,1){1'};
      \vertex [dot](trr) at (1.5,1){};
      \vertex [dot](brr) at (1.5,0){};
      \vertex [dot](tr) at (1,1);
      \vertex [dot](trm) at (0.5,1){};
      \vertex [dot](tm) at (0,1);
      \vertex [dot](tlm) at (-0.5,1){};
      \vertex [dot](tl) at (-1,1);
      \vertex [dot](br) at (1,0);
      \vertex [dot](brm) at (0.5,0);
      \vertex [dot](bm) at (0,0){};
      \vertex [dot](blm) at (-0.5,0);
      \vertex [dot](bl) at (-1,0);
      \vertex [dot](mm) at (0,0.5);
      \propag [gho] (a) to (c); 
      \propag [gho] (b) to (d);
      \propag [graviton,red] (trr) to (brr);
      \propag [graviton,red] (tlm) to (mm);
      \propag [graviton,red] (trm) to (mm);
      \propag [graviton,red] (mm) to (bm);
      \propag [plain, ultra thick] (brr) [in=270, out=270, looseness=1] to (bm);
   \end{feynhand}
\end{tikzpicture}
+(1 \leftrightarrow 2)
\ee

Diagrams that contain one tree part are also new,
\begin{align} \label{Diagram1tree}
&\begin{tikzpicture}[baseline=0.4cm]
   \begin{feynhand}
      \vertex (b) at (-1.5,0){2}; 
      \vertex (d) at (1.5,0){2'};
      \vertex (a) at (-1.5,1){1}; 
      \vertex (c) at (1.5,1){1'};
      \vertex [dot](tr) at (1,1);
      \vertex [dot](middle) at (1,1);
      \vertex [dot](trm) at (0.5,1){};
      \vertex [dot](tm) at (0,1);
      \vertex [dot](tlm) at (-0.5,1){};
      \vertex [dot](tl) at (-1,1);
      \vertex [dot](br) at (1,0);
      \vertex [dot](brm) at (0.5,0){};
      \vertex [dot](bm) at (0,0);
      \vertex [dot](blm) at (-0.5,0){};
      \vertex [dot](bl) at (-1,0);
      \vertex [dot](mlm) at (-0.5,.5);
      \vertex [dot](mrm) at (0.5,.5);
      \propag [gho] (a) to (c); 
      \propag [gho] (b) to (d);
      \propag [graviton,red] (tlm) to (blm);
      \propag [graviton,red] (trm) to (brm);
      \propag [graviton,red] (mrm) to (mlm);
   \end{feynhand}
\end{tikzpicture}
+
\begin{tikzpicture}[baseline=0.4cm]
   \begin{feynhand}
      \vertex (b) at (-1.5,0){2}; 
      \vertex (d) at (1.5,0){2'};
      \vertex (a) at (-1.5,1){1}; 
      \vertex (c) at (1.5,1){1'};
      \vertex [dot](tr) at (1,1);
      \vertex [dot](middle) at (1,1);
      \vertex [dot](trm) at (0.5,1){};
      \vertex [dot](tm) at (0,1);
      \vertex [dot](tlm) at (-0.5,1){};
      \vertex [dot](tl) at (-1,1);
      \vertex [dot](br) at (1,0);
      \vertex [dot](brm) at (0.5,0){};
      \vertex [dot](bm) at (0,0);
      \vertex [dot](blm) at (-0.5,0){};
      \vertex [dot](bl) at (-1,0);
      \vertex [dot](mlm) at (-0.5,.5);
      \vertex [dot](mrm) at (0.5,.5);
      \vertex [dot](mb) at (0,.25);
      \vertex [dot](mt) at (0,.75);
      \propag [gho] (a) to (c); 
      \propag [gho] (b) to (d);
      \propag [graviton,red] (tlm) to (mt);
      \propag [graviton,red] (trm) to (mt);
      \propag [graviton,red] (blm) to (mb);
      \propag [graviton,red] (brm) to (mb);
      \propag [graviton,red] (mt) to (mb);
   \end{feynhand}
\end{tikzpicture}
+
\begin{tikzpicture}[baseline=0.4cm]
   \begin{feynhand}
      \vertex (b) at (-1.5,0){2}; 
      \vertex (d) at (1.5,0){2'};
      \vertex (a) at (-1.5,1){1}; 
      \vertex (c) at (1.5,1){1'};
      \vertex [dot](tr) at (1,1);
      \vertex [dot](middle) at (1,1);
      \vertex [dot](trm) at (0.5,1){};
      \vertex [dot](tm) at (0,1);
      \vertex [dot](tlm) at (-0.5,1){};
      \vertex [dot](tl) at (-1,1);
      \vertex [dot](br) at (1,0);
      \vertex [dot](brm) at (0.5,0){};
      \vertex [dot](bm) at (0,0);
      \vertex [dot](blm) at (-0.5,0){};
      \vertex [dot](bl) at (-1,0);
      \vertex [dot](mlm) at (-0.5,.5);
      \vertex [dot](mrm) at (0.5,.5);
      \vertex [dot](mb) at (0,.25);
      \vertex [dot](mt) at (0,.75);
      \vertex [dot](mm) at (0, 0.5);
      \propag [gho] (a) to (c); 
      \propag [gho] (b) to (d);
      \propag [graviton,red] (tlm) to (mm);
      \propag [graviton,red] (trm) to (mm);
      \propag [graviton,red] (blm) to (mm);
      \propag [graviton,red] (brm) to (mm);
   \end{feynhand}
\end{tikzpicture}\nn
& +\qty[
\begin{tikzpicture}[baseline=0.4cm]
   \begin{feynhand}
      \vertex (b) at (-1.5,0){2}; 
      \vertex (d) at (1.5,0){2'};
      \vertex (a) at (-1.5,1){1}; 
      \vertex (c) at (1.5,1){1'};
      \vertex [dot](tr) at (1,1){};
      \vertex [dot](trm) at (0.5,1);
      \vertex [dot](tm) at (0,1){};
      \vertex [dot](tlm) at (-0.5,1);
      \vertex [dot](tl) at (-1,1){};
      \vertex [dot](br) at (1,0);
      \vertex [dot](brm) at (0.5,0);
      \vertex [dot](bm) at (0,0){};
      \vertex [dot](blm) at (-0.5,0);
      \vertex [dot](bl) at (-1,0);
      \vertex [dot] (mlt) at (-0.5, 0.75);
      \vertex [dot](mm) at (0,0.5);
      \propag [gho] (a) to (c); 
      \propag [gho] (b) to (d);
      \propag [graviton,red] (tl) to (mlt);
      \propag [graviton, red] (mlt) to (mm);
      \propag [graviton,red] (tm) to (mlt);
      \propag [graviton,red] (tr) to (mm);
      \propag [graviton,red] (mm) to (bm);
   \end{feynhand}
\end{tikzpicture}
+
\begin{tikzpicture}[baseline=0.4cm]
   \begin{feynhand}
      \vertex (b) at (-1.5,0){2}; 
      \vertex (d) at (1.5,0){2'};
      \vertex (a) at (-1.5,1){1}; 
      \vertex (c) at (1.5,1){1'};
      \vertex [dot](tr) at (1,1){};
      \vertex [dot](trm) at (0.5,1);
      \vertex [dot](tm) at (0,1){};
      \vertex [dot](tlm) at (-0.5,1);
      \vertex [dot](tl) at (-1,1){};
      \vertex [dot](br) at (1,0);
      \vertex [dot](brm) at (0.5,0);
      \vertex [dot](bm) at (0,0){};
      \vertex [dot](blm) at (-0.5,0);
      \vertex [dot](bl) at (-1,0);
      \vertex [dot](mm) at (0,0.5);
      \propag [gho] (a) to (c); 
      \propag [gho] (b) to (d);
      \propag [graviton,red] (tl) to (mm);
      \propag [graviton,red] (tm) to (mm);
      \propag [graviton,red] (tr) to (mm);
      \propag [graviton,red] (mm) to (bm);
   \end{feynhand}
\end{tikzpicture}
+ (1 \leftrightarrow 2)]
\end{align}

\subsubsection{Radiative corrections}

We need at least two exchanges between the worldlines to get non-vanishing contributions of radiative corrections. By further discarding all diagrams that lead to scaleless integrals, the non-vanishing diagrams are the the same as the scalar-QED case,
\be
\begin{tikzpicture}[baseline=0.4cm]
   \begin{feynhand}
      \vertex (b) at (-2,0){2}; 
      \vertex (d) at (2,0){2'};
      \vertex (a) at (-2,1){1}; 
      \vertex (c) at (2,1){1'};
      \vertex [dot](tr) at (1,1){};
      \vertex [dot](trm) at (0.5,1){};
      \vertex [dot](tm) at (0,1);
      \vertex [dot](tlm) at (-0.5,1);
      \vertex [dot](tl) at (-1,1);
      \vertex [dot](br) at (1,0){};
      \vertex [dot](brm) at (0.5,0){};
      \vertex [dot](bm) at (0,0){};
      \vertex [dot](blm) at (-0.5,0);
      \vertex [dot](bl) at (-1,0){};
      \propag [gho] (a) to (c); 
      \propag [gho] (b) to (d);
      \propag [graviton,red] (tr) to (br);
      \propag [graviton,red] (bm) [in=270, out=270, looseness=1] to (bl);
      \propag [graviton,red] (trm) to (brm);
      \propag [plain,ultra thick, black] (bl)  to [in=90, out=90, looseness=1] (brm);
      \propag [plain,ultra thick, black] (bm)  to [in=270, out=270, looseness=1] (br);
   \end{feynhand}
\end{tikzpicture}
+(1 \leftrightarrow 2)
\ee

Thus, we have systematically generated all the diagrams used in past WQFT calculations and proved the equivalence between the WQFT method and the eikonal method.

Our understanding of the equivalence even allows us to combine the two methods. It is interesting to notice that in the eikonal method calculation \cite{Bjerrum-Bohr:2021din}, the second and the third diagrams in \eqref{Diagram1tree} are not included from the beginning since they are not 3-cut constructable. In worldline EFT, these two diagrams are calculated and indeed turn out to vanish \cite{Kalin:2020fhe}. Now with our understanding of the the equivalence, we are free to use all the tools from both sides. Not only we can target classical contributions directly with WQFT method, but also reduce the number of WQFT diagrams needed with the analysis of scattering amplitude. This could further improve the efficiency of the calculation.

\section{Conclusions}\label{sect:conclusion}
To summarize this work, for pedagogical reasons we began by reviewing two separate methods  relevant for extracting classical observables from scattering amplitudes. First we reviewed how the classical contribution to the effective potential in a two-body scattering can be extracted by taking the  eikonal limit of the QFT amplitude. A non-relativistic limit can be taken on top of the eikonal limit to match it with the quantum mechanics eikonal. We demonstrated this matching explicitly using the scalar Yukawa theory as an example. The leading order eikonal terms (i.e, the terms most singular in $\hbar$) at each  loop order add up and exponentiate into the leading eikonal phase. 
Next, we reviewed the worldline formalism, which is a first quantized version of quantum field theory. This formalism is especially suitable and efficient for studying scattering processes under the classical limit. We reviewed the path integral formulation  and operator approach of the worldline formalism for a scalar particle interacting with a scalar, electromagnetic or gravitational background field. We discussed the connection between the dressed propagator and the scattering amplitude. We mention in passing that, with some hints from Levy and Sucher \cite{Levy:1969cr}, we arrived at a symmetric form of the dressed propagator in the worldline formalism. This symmetric form allows taking the on-shell limit in the same fashion for both external legs of the scalar worldline. We would like to point out though that the same on-shell limit can be implemented much more straightforwardly in the operator formalism, through the action of vertex operators creating the asymptotic states.
\\

\par {We noted that the eikonal expansion of the scattering QFT amplitude can be cast as an $\hbar$ expansion under the classical limit. On the other hand, the $\hbar$ expansion is mapped into the contraction expansion in the worldline formalism. This contraction expansion is the link between the eikonal method, which is QFT-based, and the WQFT method, which was seen as an  EFT method. We showed how the expansion in fluctuations about the classical trajectory, parametrized in terms of the averaged in-out momentum, which is the starting point of the worldline EFT treatment, emerges naturally from the operator approach through contractions with the asymptotic states vertex operators. We also discussed how the various terms  of the scattering amplitude are mapped into the corresponding WQFT diagrams.}\\

\par {With this side-by-side comparison, we demonstrated that the  WQFT diagrams 
circumvent the work done by expanding the QFT amplitude order by order in $\hbar$, and the cumbersome algebraic manipulations that were needed to show the exponentiation of the eikonal phase, to the known lowest orders. With the understanding of how the WQFT diagrams map into the full scattering amplitude, 
 we were able to show the factorization structure of the scattering amplitude at each loop level and prove the complete exponentiation of the eikonal phase, which we detailed in a separate article \cite{ADRV1}. As anticipated in \cite{Mogull:2020sak}, the eikonal phase corresponds to the leading order of connected WQFT diagrams. We referred to such diagrams as minimally connected, irreducible diagrams.}\\

\par {In practice, the worldline formalism also provides an alternative way for generating and calculating WQFT diagrams, without the need  of going into the energy space (the canonical conjugate variable to the Schwinger worldline time)   as in \cite{Mogull:2020sak}. We showed how in the worldline formalism, the scattering amplitude can be modified into  a generating function of the WQFT diagrams. In particular, one can  generate and calculate the minimally connected irreducible WQFT diagrams corresponding to the eikonal phase in a systematic  way. We expect that this algorithmic method is relevant for the higher loop orders, when keeping track of the relevant WQFT diagrams  by hand becomes less and less practical.
\\

\par {We used scalar QED as an example to demonstrate in detail the equivalence between the eikonal method and the WQFT method. We showed how the calculations in the two methods can be translated to each other freely. Though it is guaranteed that they produce the same result in calculating eikonal phase, the WQFT method is certainly more advantageous since it  targets each term in the eikonal phase  directly, without any additional subtractions, recursively. We also showed this equivalence extends naturally to the case where scalar particles interacts gravitationally.}\\

\par {The equivalence between the eikonal method and the WQFT method establishes a bridge between the two methods, which then in principle allows one to use tools from both sides to further improve the efficiency in calculating classical observables. In the simplest case, one may simply use the analysis in QFT to narrow down the WQFT diagrams that are needed. 
Or we could first calculate the full amplitude with the worldline formalism, which just gives the generating function. One drawback is that we have to separately consider many different terns due to different time orderings of the vertex operators. Some progress has been made to do a worldline calculation without fixing the order of the vertex operators \cite{Edwards:2021elz,Ahmadiniaz:2022yam}, which is worth further investigation.}

 \appendix
 
 \section{Scalar QED eikonal with QFT}\label{app:sQED}
 
In this appendix, we will perform the eikonal expansion in the case of scalar QED within the framework of conventional QFT. This will provide an avenue for comparison with the discussion in Section \ref{sect:wqft-sQED}. The goal here is to find the \textit{classical} contribution of each set of diagrams, thereby giving classical potential information (see Section \ref{sect:eikonal-qm-qft} for a detailed discussion on extracting potentials from scattering amplitudes). 
We follow the  discussion in Section \ref{sect:eikonal-qm-qft} and work with  the barred quantities (i.e. $\bar k$ being interpreted as a wave-vector) first,  and then restore the $\hbar$ factors to reconstruct external particle momenta and perform the eikonal expansion. We will extract the terms at $\order{1/\hbar^3}$, which contribute to  the classical limit. 

We begin with the Feynman rules for scalar QED:
\begin{eqnarray*}
    \begin{tikzpicture}[baseline=-0.3cm]
       \begin{feynhand}
          \vertex (l) at (-1,0); \vertex (m) at (0,0);\vertex (r) at (1,0);\vertex (bm) at (0,-1){$\mu$};\vertex (bl) at (-1,-1);\vertex (br) at (-1,1);
        \propag[scalar,mom={$\bar k$}] (l) to (m);
        \propag[scalar,mom={$\bar k'$}] (m) to (r);
        \propag[photon,red] (m) to (bm);
       \end{feynhand}
    \end{tikzpicture}
    &=& i\bar e(\bar k + \bar k')_\mu,\\
    \begin{tikzpicture}[baseline=-0.3cm]
       \begin{feynhand}
          \vertex (l) at (-1,0); \vertex (m) at (0,0);\vertex (r) at (1,0);\vertex (bm) at (0,-1);\vertex (bl) at (-1,-1){$\mu$};\vertex (br) at (1,-1){$\nu$};
        \propag[scalar,mom={$\bar k$}] (l) to (m);
        \propag[scalar,mom={$\bar k'$}] (m) to (r);
        \propag[photon,red] (m) to (bl);
        \propag[photon,red] (m) to (br);
       \end{feynhand}
    \end{tikzpicture}
        &=& -2i\bar e^2\eta_{\mu \nu}.
   \end{eqnarray*}
For simplicity, we do not show the charge flow, though it is understood there is a conserved charge associated with the complex scalar field. In the following calculation, we shall suppress the $i\epsilon$ prescription on propagators related to soft mediators (notated with $\bar k_i$). The calculation will altogether be similar to Section \ref{sect:eikonal-qm-qft}, the only differences being a momentum-dependent factor from the difference in mediator fields and the presence of a new set of diagrams originating from the quartic vertex. Additionally, due to the quartic vertex, we shall get an infinite set of irreducible diagrams obtained by combining alternating triangle diagrams. Other examples of irreducible diagrams are the radiative loop diagrams discussed in Section \ref{sect:wqft-sQED}.

To begin, we outline the procedure to finding the  classical contributions to the effective potential. The steps for scalar QED are: 
\begin{itemize}
    \item {Write down the amplitude using usual Feynman rules; note we exclusively use ``barred'' quanitites at this initial stage.}
    \item {Make the replacements
    \be
    \left\{\bar e\to\frac e {\sqrt \hbar}, \bar m\to\frac m\hbar, \bar p^\mu\to\frac{p^\mu}{\hbar}\right\},
    \ee
    which recovers the $\hbar$ factors explicitly. }
    \item{We can now take the eikonal expansion, thereby imposing that the mediator particles, photons in this case, are soft and also that the transfer momentum is low. This is essentially satisfied by the limits $\hbar\bar k_i\ll p_j$ and $\hbar\bar q\ll p_j$, which are equivalently a series in ${\hbar}$ taken to be a small parameter with the replacements given above. }
    \item 
    To extract the classical piece that enters at $L^{\text{th}}$ loop order, we must expand the soft terms (taking $\hbar\ll1$) to $L$ orders. There will be redundant information packed into the amplitude from lower order terms, however, and at higher loop orders one must disentangle quantum pieces from the classical terms of interest, as in Eq.~\eqref{expConjExpand}. 
    \item{We extract the potential from the scattering amplitude by inverting Eq.~\eqref{qmamp} to solve for $V$, allowing us to extract the classical contribution once we find the classical eikonal terms.}
\end{itemize}
We with the tree diagram, whereafter we compute two types of 1-loop terms (triangle and box diagrams); the tree and box diagrams contribute to the  leading eikonal, while the triangle diagram contributes to the subleading eikonal.
By the discussion in Section \ref{sect:eikonal-qm-qft}, we see that the classical contribution to the scattering amplitude, denoted by $\mathcal A_{L}^{(0)}$, is an $\order{\hbar^{-3}}$ quantity. This turns out to be the leading order term in the $\hbar\ll1$ expansion for the tree and triangle, while the boxes enter \textit{superclassically} at $\order{\hbar^{-4}}$. 
At each order, we must go further in the eikonal expansion to extract the classical contribution due to such superclassical (denotes by $\mathcal {A}_{n}^{(-1)}$) and higher order terms entering the expansion. These terms that are $\order{1/\hbar}$ or higher than the classical term are thought to either contain IR divergent terms that will not contribute \cite{Weinberg:1965nx} or contain vanishing contributions. They are coined to be so since they are dominant over the classical pieces in the $\hbar\to 0$ limit. We shall label such superclassical terms by negative superscripts, and this label indicates how many orders of the $\hbar$ expansion we must keep in order to reach the classical limit. Terms that are $\order{\hbar^{-2}}$ or higher can be ignored since these are quantum corrections that enter at higher orders in perturbation than the classical bits of potential. When approaching the calculation na\"ively, the quantum terms must be considered in order to recover the classical pieces, but with the aid of factorization provided by WQFT, we can directly extract such classical contributions (see Section \ref{sect:wqft-eikonal}).

We begin with  the tree-level calculation:
\be
   \includegraphics[valign=c]{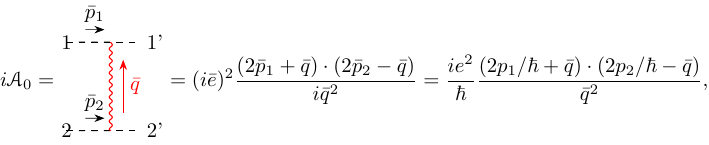}
\ee
and then, by expanding in orders of $\hbar$, we get the classical contribution from the leading eikonal expansion term, as seen by the $\order{\hbar^{-3}}$ dependence\footnote{Recall that the classical contribution to the amplitude in momentum space is of order $\hbar^{-3}$. In impact parameter space, or for what we denoted $\widetilde{
\mathcal A}$, the classical contribution is of order $\hbar^{-1}$.}. Thus we get
\be
i \mathcal A_0=\frac{ie^2}{\hbar^3}\frac{(2 p_1+\hbar \bar q)\cdot(2 p_2-\hbar \bar q)}{\bar q^2}=\frac{i4e^2}{\hbar^3}\frac{p_1\cdot p_2}{\bar q^2}+\order{\frac 1\hbar^2}=i \mathcal A^{(0)}_0+\order{\frac 1\hbar^2},
\ee
which completes the tree-level analysis.

The one loop amplitude is of order $\order{e^4}$. The diagrams that yield either superclassical or classical contributions are the following: 
\ba
&&\includegraphics[width=\textwidth]{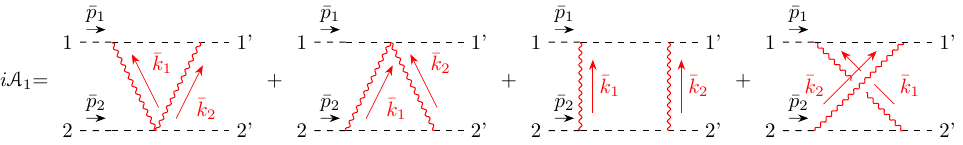}\nonumber\\
&&=i\mathcal A_{1,\triangle}+i\mathcal A_{1,\Box}.
\ea
Other one loop diagrams such as 
$\includegraphics[valign=c]{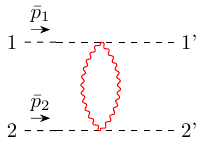}$
are suppressed since they yield quantum contributions, at $\order{\hbar^{-2}}$. This can be understood by noting that all $\hbar$ factors are coming from the coupling constant in this case.

Let us first consider the first two diagrams. These ``seagull'' or ``triangle'' diagrams have the value,
\begin{eqnarray}
    i\mathcal{A}_{1,\triangle} &=& -(2i\bar e^2)(i\bar e)^2\int \dfourk{\bar k_1}\int \dfourk{\bar k_2} \frac{\eta_{\mu \nu}}{i\,\bar k_1^2}\frac{1}{i\,\bar k_2^2}\frac{1}{i}\left[ \frac{(2 \bar p_1 + \bar k_1)^
    \mu(\bar p_1' + \bar p_1 + \bar k)_1^{\nu}}{(\bar p_1 + \bar k_1)^2 + \bar m_1^2 - i\epsilon}+\frac{(2\bar p_2-\bar k_1)^{\mu}(\bar p_2' + \bar p_2 - \bar k_1)^{\nu}}{(\bar p_2 - \bar k_1)^2 + \bar m_2^2 - i \epsilon}\right]\nonumber\\
    &&\times{(2\pi)^4} \delta^4(\bar k_1+\bar k_2-\bar q) \nonumber\\
    &=& -{\bar e^4}\int \dfourk{\bar k} \frac{\eta_{\mu \nu}}{\bar k^2}\frac{1}{(\bar q-\bar k)^2}\left[ \frac{(2 \bar p_1 + \bar k)^
    \mu[ 2\bar p_1 + (\bar k+\bar q)]^{\nu}}{ \bar k^2 + 2\bar p_1\cdot\bar k - i\epsilon}-\frac{(2\bar p_2-\bar k)^{\mu}[2\bar p_2 - (\bar k+\bar q)]^{\nu}}{-\bar k^2 + 2\bar p_2\cdot\bar k_1 +i\epsilon}\right.\nonumber\\
    &&\left.-\frac{[2 \bar p_1 + (\bar q-\bar k)]^
    \mu[2\bar p_1 + (2\bar q-\bar k)]^{\nu}}{\bar q ^2-(\bar q-\bar k)^2 + 2\bar p_1\cdot\bar k+ i\epsilon}+\frac{[2\bar p_2-(\bar q-\bar k)]^{\mu}[2\bar p_2 -(2\bar q- \bar k)]^{\nu}}{-\bar q^2+(\bar q-\bar k)^2 + 2\bar p_2\cdot\bar k- i\epsilon}\right] ,
\end{eqnarray}
where we have made use of the definition of momentum transfer $\bar q:= \bar p_1'-\bar p_1=\bar p_2-\bar p_2'$, relabelled $k:=k_1$, and used the on-shell conditions $\bar p_i^2=-\bar m_i^2$ and $\bar p'_i{}^2+\bar m_i^2=(\bar p_i+\eta_i\bar q)^2+\bar m_i^2=\bar q^2+2\eta_i\bar p_1\cdot\bar q\to 2\bar p_i\cdot \bar q=-\eta_i \bar q^2$ where $\eta_1=1,\,\eta_2=-1$.
We now perform the eikonal expansion by following the procedure outlined above starting with the second bullet point. 
We recover the $\hbar$ factors by taking $\left\{\bar e\to\frac e {\sqrt \hbar}, \bar p^\mu\to\frac{p^\mu}{\hbar}\right\}$ and expand in powers of $\hbar$:
\begin{eqnarray}
    i\mathcal{A}_{1,\triangle} 
    &=& -\frac{4e^4}{\hbar^{3}}\int \dfourk{\bar k} \frac{1}{\bar k^2}\frac{1}{(\bar q-\bar k)^2}\left\{\left[ \frac{p_1^2}{  2p_1\cdot\bar k - i\epsilon}-\frac{p_1^2}{ 2 p_1\cdot\bar k+ i\epsilon}\right]+\left[\frac{p_2^2}{ 2 p_2\cdot\bar k- i\epsilon}-\frac{p_2^2}{2 p_2\cdot\bar k_1 + i\epsilon}+\order{\hbar}\right]\right\}\nonumber\\
    &=&i\mathcal{A}_{1,\triangle}^{(0)}+\order{\frac 1 {\hbar^2}}\,.
\end{eqnarray}
Since the above expression enters at $\mathcal O(\hbar^{-3})$, we can see that we only need to consider the leading order term to recover the classical piece of the potential.
Practically speaking, we simply discard terms where $\hbar$ explicitly appears in the numerator and denominator of the propagator terms. 
In both square brackets above, we can see that the principal part will cancel out, leaving the imaginary delta function contribution of the propagator:
\begin{equation}
   i\mathcal{A}_{1,\triangle }^{(0)} =  \frac{ 4e^4}{\hbar^3}\int \dfourk{\bar k}\frac{1}{\bar k^2(\bar q-\bar k)^2}\qty[m_1^2 (2 \pi i) \delta(2 \fourdot{p_1}{\bar k}) + m_2^2 (2 \pi i) \delta(2 \fourdot{p_2}{\bar k})],
\end{equation}
where we again use the on-shell conditions given by $p_i^2=-m_i^2$.
Both of the above terms can be computed in essentially the same manner, so consider the integral 
\begin{equation}
    i\mathcal{A}^{(0)}_{1,\triangle ,j}
    =  \frac{4ie^4}{\hbar^3}\int \dfourk{\bar k}\frac{m_j^2 (2 \pi ) \delta( 2\fourdot{p_j}{\bar k})}{\bar k^2(\bar q-\bar k)^2},
\end{equation}
where $j=1$ or $2$ depending on which term is considered. We can choose the rest frame of $p_j^\mu$ where the only nonzero component is $p_j^0=m_j$, and we assume $q^\mu$ to be purely spatial (it is also transverse to $p_i$, but this argument is not needed here). This yields the integral 
\begin{eqnarray}\label{eq:ATriangleLO}
    i\mathcal{A}^{(0)}_{1,\triangle ,j}
    =  \frac{4ie^4}{\hbar^3}\int \dfourk{\bar k}\frac{m_j^2 (2 \pi ) \delta(2 m_j\bar k^0)}{\bar k^2[-(\bar k^0)^2+(\mathbf{\bar q}-\mathbf{\bar k})^2]}=\frac{4ie^4}{2\hbar^3}\int \frac{d^3\mathbf{\bar k}}{(2\pi)^3}\frac{m_j }{\mathbf{\bar k}^2(\mathbf{\bar q}-\mathbf{\bar k})^2},
\end{eqnarray}
which can be further manipulated into
\begin{eqnarray}\label{eq:ATriangleLOFP}
    i\mathcal{A}^{(0)}_{1,\triangle ,j}
    =\frac{2ie^4}{\hbar^3}\int_0^1 dx\int \frac{d^3\mathbf{\bar k}}{(2\pi)^3}\frac{m_j }{(\mathbf{\bar k}^2-2x\mathbf{\bar k}\cdot\mathbf{\bar q}+x\mathbf{\bar q}^2)^2}.
\end{eqnarray}
Then, we shift the integration variable $\mathbf k\to \mathbf k-x \mathbf q$ and  use the Schwinger trick to write
\begin{eqnarray}
    i\mathcal{A}^{(0)}_{1,\triangle ,j}
    &=&\frac{2im_je^4}{\hbar^3}\int_0^1 dx\int \frac{d^3\mathbf{\bar k}}{(2\pi)^3}\int^\infty_0dT\,Te^{-(\mathbf{\bar k}^2+x(1-x)\mathbf{\bar q}^2)T}\nonumber\\
    &=&\frac{im_je^4}{4\hbar^3\pi^{3/2}}\int_0^1 dx\int^\infty_0dT\frac{e^{-x(1-x)\mathbf{\bar q}^2T}}{\sqrt T}=\frac{im_je^4\pi^{3/2}}{4\hbar^3\pi^{3/2}|\mathbf {\bar q}|}=\frac{i m_j e^4}{4\hbar^3|\mathbf {\bar q}|}.
\end{eqnarray}
Thus, the amplitude obtained from the triangle diagrams is given by 
\begin{equation}
    i\mathcal{A}^{(0)}_{1,\triangle }=i\mathcal{A}^{(0)}_{1,\triangle ,j=1}+i\mathcal{A}^{(0)}_{1,\triangle ,j=2}=\frac{i e^4 (m_1+m_2)}{4\hbar^3|\mathbf {\bar q}|}.
\end{equation}
This matches exactly with the finite terms that are real-valued in Eq.~(3.9) of Ref.~\cite{Bern:2021xze}, which means this amplitude is the main contribution to the soft potential (we shall see that the box diagrams only give divergent pieces which do not contribute to the {subleading} eikonal phase). Additionally, we can see that the triangle diagram contribution is captured by the second line of Eq.~(3.4) of Ref.~\cite{Bern:2021xze} when $\epsilon\to 0$, confirming our calculation matches known results. 

Now let us look at the 1-loop ``ladder'' or ``box'' diagrams:
\be
    \includegraphics[valign=c]{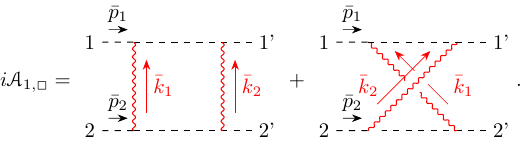}
\ee
The ladder diagrams are given by
\begin{eqnarray}
    i\mathcal{A}_{1,\Box} 
     &=& i\frac{(i\bar e)^4}{i^4}\int_{\bar {k}_1} \int_{\bar{k}_2}\, \left\{\left[\frac{{(2\bar p_2-\bar k_1)_\mu}{(2\bar p_2+[2\bar q+\bar k_2])}_\alpha}{\bar k_1^2-2\bar p_2\cdot \bar k_1-i\epsilon}
     + \frac{{(2\bar p_2+[2\bar q+\bar k_1])_\mu}{(2\bar p_2-\bar k_2)_\alpha}}{\bar k_2^2-2\bar p_2\cdot \bar k_2-i\epsilon}\right]\right.\nonumber\\
     &&\times
     \left. \frac{[2\bar p_1+\bar k_1]^\mu[2\bar p_1-(2\bar q+\bar k_2)]^\alpha}{\bar k_1^2+2\bar p_1\cdot \bar k_1-i\epsilon}\right\}\frac{1}{\bar k_1^2-i\epsilon}
     \frac{1}{\bar k_2^2-i\epsilon}
     (2\pi)^4 \delta^4\left(\bar k_1 +\bar  k_2 -\bar q\right),
\end{eqnarray}
where we have used the on-shell conditions to simplify the denominators and used $\{\bar p_1'=\bar p_1-\bar q,\bar p_2'=\bar p_2+\bar q\}$ to write everything in terms of the incoming momenta. 
We now recover the $\hbar$-dependence by taking $\left\{\bar e\to\frac e {\sqrt \hbar}, \bar p^\mu\to\frac{p^\mu}{\hbar}\right\}$. When taking the eikonal limit, to get the correct $\hbar$-counting, we must expand until subleading order. 
Recovering $\hbar$ factors explicitly and doing the $k_2$ integration (and taking $k_1:=k$) yields
\begin{eqnarray}
    i\mathcal{A}_{1,\Box} 
    &=& \frac{i^5}{2i^4}\frac{e^4}{ \hbar ^{4}}\int_{\bar {k}} \frac{1}{\bar k^2-i\epsilon}\frac{1}{(\bar q-\bar k)^2-i\epsilon}\, \left[\frac{{(2 p_2-\hbar\bar k)_\mu}{(2 p_2-\hbar \bar k)}_\alpha}{\hbar\bar k^2-2 p_2\cdot \bar k-i\epsilon}
    -\frac{{(2 p_2+\hbar\bar k)_\mu}{(2 p_2+\hbar \bar k)_\alpha}}{\hbar \bar q^2-\hbar(\bar q-\bar k)^2-2 p_2\cdot \bar k+i\epsilon}\right]\nonumber\\
     &&\times\left[
    \frac{(2 p_1+\hbar\bar k)^\mu(2 p_1-\hbar \bar k)^\alpha}{\hbar\bar k^2+2 p_1\cdot \bar k-i\epsilon}-\frac{(2 p_1-\hbar\bar k)^\mu(2 p_1+\hbar\bar k)^\alpha}{\hbar \bar q^2-\hbar(\bar q-\bar k)^2+2 p_1\cdot \bar k+i\epsilon}\right]+\order{\frac 1 \hbar^2},\nonumber\\
\end{eqnarray}
where we symmetrize over the loop momenta (basically splitting above into half of itself plus the same term with $k\to q-k$) and drop $\bar q$~-dependent term in the numerator using $2 p_i\cdot \bar q=-\hbar\eta_i \bar q^2$ with $\eta_1=1,\,\eta_2=-1$. 
We must still expand the denominators, which can be done as 
\be 
\frac{1}{\hbar \bar q^2-\hbar(\bar q-\bar k)^2+2\eta_i p_i\cdot \bar k+i\epsilon}=\frac{1}{2\eta_i p_i\cdot \bar k+i\epsilon}\left[1-\frac{\hbar \bar q^2-\hbar(\bar q-\bar k)^2}{2 \eta_ip_i\cdot \bar k+i\epsilon}\right]+\order{\frac 1 \hbar^2},
\ee 
and 
\be 
\frac{1}{\hbar\bar k^2+2 \eta_ip_i\cdot \bar k-i\epsilon}=\frac{1}{2 \eta_ip_i\cdot \bar k-i\epsilon}\left[1-\frac{\hbar\bar k^2}{2 \eta_ip_i\cdot \bar k-i\epsilon}\right]+\order{\frac 1 \hbar^2}.
\ee 
The expansion of the ladder diagrams becomes:
\begin{eqnarray}
    i\mathcal{A}_{1,\Box}&=& \frac{i^5}{2i^4}\frac{e^4}{ \hbar ^{4}}\int_{\bar {k}} \, \left\{
    \frac{4[ p_2{}_\mu p_2{}_\alpha\textcolor{orange}{-\hbar\bar k_{(\mu|} p_2{}_{|\alpha)}}]}{-2 p_2\cdot \bar k-i\epsilon}\left[1\textcolor{blue}{-\frac{\hbar\bar k^2}{-2 p_2\cdot \bar k-i\epsilon}}\right]-\frac{4[ p_2{}_\mu p_2{}_\alpha\textcolor{orange}{+\hbar\bar k_{(\mu|} p_2{}_{|\alpha)}}]}{-2 p_2\cdot \bar k+i\epsilon}\left[1\textcolor{blue}{-\frac{\hbar \bar q^2-\hbar(\bar q-\bar k)^2}{-2 p_2\cdot \bar k+i\epsilon}}\right]\right\}\nonumber\\
    &&\times\left\{
    \frac{4[ p_1{}^\mu p_1{}^\alpha\textcolor{orange}{+\hbar\bar k^{[\mu} p_1{}^{\alpha]}}]}{2 p_1\cdot \bar k-i\epsilon}\left[1\textcolor{blue}{-\frac{\hbar\bar k^2}{2 p_1\cdot \bar k-i\epsilon}}\right]-\frac{4[ p_1{}^\mu p_1{}^\alpha\textcolor{orange}{-\hbar\bar k^{[\mu} p_1{}^{\alpha]}}]}{2 p_1\cdot \bar k+i\epsilon}\left[1\textcolor{blue}{-\frac{\hbar \bar q^2-\hbar(\bar q-\bar k)^2}{2 p_1\cdot \bar k+i\epsilon}}\right]\right\}\nonumber\\
    &&\times\frac{1}{\bar k^2-i\epsilon}\frac{1}{(\bar q-\bar k)^2-i\epsilon}+\order{\frac 1 \hbar^2}\nonumber\\
    &=& \frac{i^5}{2i^4}\frac{e^4}{ \hbar ^{4}}\int_{\bar {k}} \, \left\{\frac{4[ p_2{}_\mu p_2{}_\alpha\textcolor{orange}{+\hbar\bar k_{(\mu|} p_2{}_{|\alpha)}}]}{2 p_2\cdot \bar k-i\epsilon}\left[1\textcolor{blue}{+\frac{\hbar \bar q^2-\hbar(\bar q-\bar k)^2}{2 p_2\cdot \bar k-i\epsilon}}\right]-
    \frac{4[ p_2{}_\mu p_2{}_\alpha\textcolor{orange}{-\hbar\bar k_{(\mu|} p_2{}_{|\alpha)}}]}{2 p_2\cdot \bar k+i\epsilon}\left[1\textcolor{blue}{+\frac{\hbar\bar k^2}{2 p_2\cdot \bar k+i\epsilon}}\right]\right\}\nonumber\\
    &&\times\left\{
    \frac{4[ p_1{}^\mu p_1{}^\alpha\textcolor{orange}{+\hbar\bar k^{[\mu} p_1{}^{\alpha]}}]}{2 p_1\cdot \bar k-i\epsilon}\left[1\textcolor{blue}{-\frac{\hbar\bar k^2}{2 p_1\cdot \bar k-i\epsilon}}\right]-\frac{4[ p_1{}^\mu p_1{}^\alpha\textcolor{orange}{-\hbar\bar k^{[\mu} p_1{}^{\alpha]}}]}{2 p_1\cdot \bar k+i\epsilon}\left[1\textcolor{blue}{-\frac{\hbar \bar q^2-\hbar(\bar q-\bar k)^2}{2 p_1\cdot \bar k+i\epsilon}}\right]\right\}\nonumber\\
    &&\times\frac{1}{\bar k^2-i\epsilon}\frac{1}{(\bar q-\bar k)^2-i\epsilon}+\order{\frac 1 \hbar^2}\nonumber\\
    &\equiv&i\mathcal{A}^{(-1)}_{1,\Box}+i\mathcal{A}^{(0),\text{\textcolor{blue}{blue}}}_{1,\Box}+i\mathcal{A}^{(0),\text{\textcolor{orange}{orange}}}_{1,\Box}\,,
\end{eqnarray}
where the round brackets placed around indices denote symmetrization, and the square brackets denote antisymmetrization. Above, we  split the subleading term into two contributions; we label the terms originating \textcolor{blue}{from the denominator of the propagator by blue}, while we label those \textcolor{orange}{from the numerator with orange}. We now separate the above into terms that are leading order in the $\hbar$ expansion, terms that are blue at subleading order, and terms that are orange at subleading order.

The superclassical subleading eikonal integral can be evaluated similarly to the triangle diagrams above. We note that the leading order goes like $\order{1/\hbar^4}$, thereby being at \textit{super}-classical order: 
\begin{eqnarray}
    \label{ap22}i\mathcal{A}_{1,\Box}^{(-1)}
    &=& i\frac{(ie)^4}{2i^4\hbar^4}\int \dfourk{\bar k} \, \frac{(4\fourdot{p_1}{p_2})^2}{ \bar k^2 (\bar q-\bar k)^2}\left\{\frac{1}{2\fourdot{p_1}{  \bar k} -i\epsilon}- \frac{1}{2\fourdot{p_1}{ \bar k}+i\epsilon}\right\} \left\{\frac{1}{2\fourdot{p_2}{ \bar  k}-i\epsilon}
     - \frac{1}{2\fourdot{p_2}{ \bar k}+i\epsilon}\right\}\nonumber\\
     &=&i\frac{(ie)^4}{8i^4\hbar^4}\int \dfourk{\bar k} \, \frac{(4\fourdot{p_1}{p_2})^2}{ \bar k^2 (\bar q-\bar k)^2}(2\pi i)^2\delta(\fourdot{p_1}{k})\delta(\fourdot{p_2}{k}),
\end{eqnarray}
where we have used the same replacement as the triangle diagrams where we subtract out principal parts of the propagator. Adapting to the rest frame of $p_1$, we can evaluate the time-component integration with the delta function and define the only non-zero spatial component of the momentum for the other particle, $p_2^{\mu=z}:=p_z$. This essentially follows from a combination of the transversality of $q$ to $p_i$ at this  order as well due to the assumption that the transfer momentum is small. This gives 
\begin{eqnarray}
    i\mathcal{A}_{1,\Box}^{(-1)} &=&\frac{i^3(ie)^4(4\fourdot{p_1}{p_2})^2}{8i^4\hbar^4m_1p_z}\int \frac{d^2\mathbf{\bar k_\perp}}{(2\pi)^2}\, \frac{1}{ \mathbf{\bar k_\perp}^2 (\mathbf{\bar q}-\mathbf{\bar k_\perp})^2},
\end{eqnarray}
which looks functionally the same as the integral in Eq.~\eqref{eq:ATriangleLO} except dimension of the loop-momentum integral. Thus, the main difference is the solution of the Gaussian integral performed after doing both the Feynman parameterization and the Schwinger trick. This is explicitly given by
\begin{eqnarray}
    i\mathcal{A}_{1,\Box}^{(-1)} 
    &=&\frac{i^3(ie)^4(4\fourdot{p_1}{p_2})^2}{8i^4\hbar^4m_1p_z}\frac{1}{4\pi}\int_0^1 dx\frac 1{x(1-x)\mathbf{\bar q}^2}.
\end{eqnarray}
The divergent Feynman parameter $x$-integral signals a threshold IR  divergence. We introduce an IR photon mass $\nu$ to regularize it, yielding the final result 
\begin{eqnarray}
    i\mathcal{A}_{1,\Box}^{(-1)} &=&\frac{i^3(ie)^4(4\fourdot{p_1}{p_2})^2}{8i^4\hbar^4m_1p_z}\frac{1}{4\pi}\int_0^1 dx\frac 1{x(1-x)\mathbf{\bar q}^2+\nu^2}=-i\frac{2e^4(\fourdot{p_1}{p_2})^2}{\hbar^4\pi m_1p_z}\frac{\mathrm{arctanh}\left[\frac{|\mathbf q|}{\sqrt{\mathbf q^2+4\nu^2}}\right]}{|\mathbf q|\sqrt{\mathbf q^2+4\nu^2}}.\nonumber\\
\end{eqnarray}
This the virtual photon soft contribution.  When computing observables such as the scattering probability (or cross-sections), this virtual IR divergence  cancels against the IR divergent contribution from the emission of real IR photons
~\cite{Weinberg:1965nx}. 

 An alternative form of $\mathcal{A}_{1,\Box}^{(-1)} $ which makes it clear that this superclassical contribution is the iteration of the leading order eikonal can be derived as follows. Begin by multiplying \eqref{ap22}
by $\delta(q\cdot p_1)\delta(q\cdot p_2)$, which ensures that the momentum transferred is appropriately constrained such that the outgoing momenta are on-shell, and then Fourier transforming to impact parameter space. The resulting expression $i \int d^4 \bar q \,e^{i\bar q\cdot b}\mathcal{A}_{1,\Box}^{(-1)}\delta(q\cdot p_1)\delta(q\cdot p_2) $ is the square  of the leading eikonal, $i \int d^4 \bar q \,e^{i\bar q\cdot b}
\mathcal A_0(\bar q)$.

For the classical subleading eikonal, the orange term can be easily shown to vanish. First, we collect such terms:
\begin{eqnarray}
    i\mathcal{A}^{(0),\text{\textcolor{orange}{orange}}}_{1,\Box}&=& \frac{i^5}{2i^4}\frac{e^4}{ \hbar ^{3}}\int_{\bar {k}} \,\left\{ \left(
    \frac{1}{2 p_2\cdot \bar k-i\epsilon}-\frac{1}{2 p_2\cdot \bar k+i\epsilon}\right)\left[
    \frac{4^2(\bar k\cdot p_2) (p_1\cdot p_2)}{2 p_1\cdot \bar k-i\epsilon}+\frac{4^2(\bar k\cdot p_2) (p_1\cdot p_2)}{2 p_1\cdot \bar k+i\epsilon}\right]\right.\nonumber\\
    &&+\left.\left[
    \frac{4^2(\bar k\cdot p_1) (p_1\cdot p_2)}{2 p_2\cdot \bar k-i\epsilon}+\frac{4^2(\bar k\cdot p_1) (p_1\cdot p_2)}{2 p_2\cdot \bar k+i\epsilon}\right]\left(
    \frac{1}{2 p_1\cdot \bar k-i\epsilon}-\frac{1}{2 p_1\cdot \bar k+i\epsilon}\right)\right\}.
\end{eqnarray}
Recognizing again the delta functions $\delta(2 p_{1,2}\cdot \bar k)$ factors and taking into account their action under the integral sign,  
we see that $\mathcal{A}^{(0),\text{\textcolor{orange}{orange}}}_{1,\Box}$
is zero.

Finally, we examine the blue classical subleading eikonal term for the box diagrams. It can be seen that the bracket term  with which we keep the leading order contribution yields a delta function, so we can immediately write 
\begin{eqnarray}
    \mathcal{A}^{(0),\text{\textcolor{blue}{blue}}}_{1,\Box}&=& -(2\pi)\frac{(ie)^4}{2i^4}\frac{16(p_1\cdot p_2)^2}{ \hbar ^{3}}\int_{\bar {k}} \,\left\{ \delta(2p_2\cdot \bar k)\left[
    \frac{ \bar q^2-(\bar q-\bar k)^2}{(2 p_1\cdot \bar k+i\epsilon)^2}-\frac{\bar k^2}{(2 p_1\cdot \bar k-i\epsilon)^2}\right]\right.\nonumber\\
    &&+\left.\delta(2p_1\cdot \bar k)\left[
    \frac{ \bar q^2-(\bar q-\bar k)^2}{(2 p_2\cdot \bar k-i\epsilon)^2}-\frac{\bar k^2}{(2 p_2\cdot \bar k+i\epsilon)^2}\right]\right\}\nonumber\\
    &=& -(2\pi)\frac{(ie)^4}{2i^4}\frac{16(p_1\cdot p_2)^2}{ \hbar ^{3}}\int_{\bar {k}} \,\left\{ \delta(2p_2\cdot \bar k)\left[
    \frac{ 2\bar k\cdot\bar q-\bar k^2}{(2 p_1\cdot \bar k+i\epsilon)^2}-\frac{(-\bar k)^2}{(2 p_1\cdot \bar k+i\epsilon)^2}\right]\right.\nonumber\\
    &&+\left.\delta(2p_1\cdot \bar k)\left[
    \frac{  2\bar k\cdot\bar q-\bar k^2}{(2 p_2\cdot \bar k-i\epsilon)^2}-\frac{(-\bar k)^2}{(2 p_2\cdot \bar k-i\epsilon)^2}\right]\right\}\nonumber\\
    &=& -(2\pi )\frac{(ie)^4}{i^4}\frac{16(p_1\cdot p_2)^2}{ \hbar ^{3}}\int_{\bar {k}} \,\left\{ \bar k\cdot(\bar q-\bar k) \left[
    \frac{\delta(2p_2\cdot \bar k)}{(2 p_1\cdot \bar k+i\epsilon)^2}+ \frac{\delta(2p_1\cdot \bar k)}{(2 p_2\cdot \bar k-i\epsilon)^2}\right]\right\}+\order{\frac 1 \hbar^2}.\nonumber\\
\end{eqnarray}
We must thus perform an integral of the type,
\begin{equation}
    I(\vb q) = \int \dfourk{k} \frac{1}{k^2 (q-k)^2}\frac{k \cdot (q-k)}{(k \cdot p_2 + i \epsilon)^2} (2 \pi)\delta(p_1 \cdot k).
\end{equation}
Let us first take care of the delta function. Since $p_1 \cdot k$ is Lorentz invariant, let us go to the rest frame of $p_1$ and evaluate it. 
\begin{equation}
    p_1 \cdot k \suchthat_{\text{rest frame of }p_1} = -m_1 k^0.
\end{equation}
If we assume the initial momenta $\vb p_1$ and $\vb p_2$ are along $z$ axis in the lab frame, then to go to the rest frame of $\vb p_1$, we need to boost along $z$ axis. This will change the magnitude of $\vb p_2$, but it will still stay in the $z$ direction. We have then
\begin{equation}
    p_2\suchthat_{\text{rest frame of }p_1} = (E_2, 0, 0, p_z).
\end{equation}
Then, once we evaluate the $k^0$ integral using the delta function, we obtain:
\begin{equation}
    I(\vb q) = \int \dthreek{k} \frac{1}{\vb k^2 (\vb q - \vb k)^2} \frac{\vb k \cdot (\vb q - \vb k)}{(k_z p_z + i\epsilon)^2}.\label{a30}
\end{equation}
We also make the assumption that the transfer momentum is purely spatial, $q =\vb q$. To regulate the IR divergence, let us add masses to the mediator propagators, denoted by $\nu$, and combine them with Feynman parameters. Then \eqref{a30} becomes:
\begin{equation}
    I(\vb q) = \int_0^1 dx \, \int \dthreek{k} \frac{1}{(\vb k^2 - 2 x \vb k \cdot \vb q + x \vb q^2 + \nu^2)^2}\frac{\vb k \cdot (\vb q - \vb k)}{(k_z p_z + i\epsilon)^2}.
\end{equation}
Shifting the integration variable by
\begin{equation}
    \vb k \to \vb k + x \vb q,
\end{equation}
we get,
\begin{equation}
    I(\vb q) = \int_0^1 dx \, \int \dthreek{k} \frac{1}{(\vb k^2 + x(1-x) \vb q^2 + \nu^2)^2}\frac{(\vb k + x \vb q) \cdot (\vb q - \vb k - x \vb q)}{(k_z p_z + i\epsilon)^2}.
\end{equation}
The $k_z$ term do not get affected by the change because, to this order in $\hbar$, $\vb q$ is in the transverse plane to the $z$ axis in the lab frame, and a boost along the $z$ axis will not change this. This dropped term corresponds to higher orders in $\hbar$, which shall only contribute quantum corrections at the next order.
Now, use the Schwinger trick to write,
\begin{equation}
    I(\vb q) = \int_0^1 dx \, \int \dthreek{k} \intinfT \frac{(\vb k + x \vb q) \cdot (\vb q - \vb k - x \vb q)}{(k_z p_z + i\epsilon)^2} T \exp{-T(\vb k^2 + x(1-x) \vb q^2 + \nu^2)}.
\end{equation}
Let us isolate the $\vb k$ integrals,
\begin{equation}
    I(\vb q) = \int_0^1 dx \, \intinfT T \exp{-T(\nu^2 + x(1-x)\vb q^2)} \underbrace{\int \dthreek{k} e^{-T \vb k^2} \frac{(\vb k + x \vb q) \cdot (\vb q - \vb k - x \vb q)}{(k_z p_z + i\epsilon)^2}}_{:=I_k},,
\end{equation}
where,
\begin{equation}\label{eq:IsubK}
    I_k=  \int \dthreek{k} e^{-T \vb k^2} \, \, \frac{-\vb k^2 + (1-2x)\vb k \cdot \vb q + x(1-x)\vb q^2}{(k_z p_z + i\epsilon)^2}=\int \dthreek{k} e^{-T \vb k^2} \, \, \frac{-\vb k^2+ x(1-x)\vb q^2}{(k_z p_z + i\epsilon)^2},
\end{equation}
where the second term in the numerator can be dropped for $k_x$ and $k_y$ and integrals because the integrand is odd, while $k_z$ integration vanishes since $q_z=0$ {at this order in $\hbar$}. 
 We perform the integrals for the remaining terms in Eq.~\eqref{eq:IsubK}. After examining the first term,
\begin{equation}
     I_1 = \int \dthreek{k} e^{-T \vb k^2} \, \, \frac{-\vb k^2}{(k_z p_z + i\epsilon)^2}
\end{equation}
we go to spherical coordinates, and write $ k_z = k \cos{\theta}$.
This gives:
\begin{align}
    I_1 &= \frac{2 \pi}{(2 \pi)^3} \intoinf{k} k^2 \int_{-1}^1 d(\cos{\theta}) \, \, e^{-T k^2} \frac{-k^2}{k^2 p_z^2 \cos^2\theta} \nn 
    &= \frac{1}{2 \pi^2} \frac{1}{p_z^2} \frac{\sqrt{\pi}}{4 T \sqrt{T}}.
\end{align}
Next, we evaluate the second term of Eq.~\eqref{eq:IsubK},
\begin{equation}
    I_2 = \int \dthreek{k} e^{-T \vb k^2} \, \, \frac{1}{(k_z p_z + i\epsilon)^2}.
\end{equation}
Again, going to spherical coordinates,
\begin{equation}
    I_2 = \frac{1}{(2\pi)^2} \frac{1}{p_z^2} \intoinf{k} k^2 \frac{1}{k^2 \cos^2\theta} e^{-T k^2} = -\frac{1}{4 \pi^2} \frac{1}{p_z^2}\sqrt{\frac{\pi}{T}}.
\end{equation}
We add both of the above expressions in order to reconstruct $I(\vb q)$:
\begin{equation}
    I(\vb q) = \frac{\sqrt{\pi}}{4 \pi^2 p_z^2}\int_0^1 dx \, \intinfT T \exp{-T(\nu^2 + x(1-x)\vb q^2)} \qty(\frac{1}{2T^{3/2}} - \frac{x(1-x)\vb q^2}{\sqrt{T}}).
\end{equation}
By performing the $T$ integration, we get 
\begin{equation}
    I(\vb q) = \frac{\sqrt{\pi}}{4 \pi^2 p_z^2}\int_0^1 dx \, \qty[\frac{\sqrt{\pi}}{2 \sqrt{\nu^2 + x(1-x)\vb q^2}} - \frac{x(1-x) \vb q^2 \sqrt{\pi}}{2 (\nu^2 + x(1-x)\vb q^2)^{3/2}}].
\end{equation}
Finally, by performing the $x$ integral, we obtain our final result
\begin{eqnarray}
    I(\vb q) &=& \frac{1}{8 \pi p_z^2} \qty[\dfrac{2\arcsin\left(\frac{\abs{\vb q}}{\sqrt{\vb q^2+4\nu^2}}\right)}{\abs{\vb q}} - \frac{1}{\abs{\vb q}} \qty(\dfrac{ \left(2\vb q^2+8\nu^2\right)\arcsin\left(\frac{\abs{\vb q}}{\sqrt{\vb q^2+4\nu^2}}\right)-4\nu\abs{\vb q}}{\vb q^2+4\nu^2})]\nonumber\\
    &=&\frac{1}{2 \pi p_z^2} \frac{\nu}{\vb q^2 + 4 \nu^2}.
\end{eqnarray}
We see that in the $\nu\to0$ limit, the blue terms from the subleading box integrals vanish as well. Thus, the only nonzero part of the box integrals is  superclassical. As argued before, in impact parameter space, the resummation of all the (classical and superclassical) leading eikonal terms at each loop order yields a phase, $e^{i\delta_0}$.
This concludes our next-to-leading order eikonal analysis.  Our approach is similar to that of \cite{Akhoury:2013yua} who studied the eikonal of scalars  interacting    gravitationally. We  differ in how we regularized the IR divergences: we employed a photon mass as an IR regulator, as opposed to the  dimensional regularization used  in  \cite{Akhoury:2013yua}. In this appendix we explicitly  evaluated the various one-loop  integrals with the goal of becoming familiar  with the structure of the various terms (the superclassical terms, the IR divergences, etc) and match them with the appropriate  WQFT diagrams.  Beyond one loop,  brute force calculations are less advisable, and instead one needs more clever ways of evaluating the integrals.
The next-to-next-to-leading (i.e. at 2-loop order) order  contributions to the eikonal  can be found in Refs.~\cite{Bern:2021xze}.

\bibliography{WQFTEikonal.bib}{}

\end{document}